\documentstyle[eqsecnum,aps,epsf,
feynmp]{revtex}
\newcommand{\nc}{\newcommand}
\nc{\fullg}{\mbox{\boldmath $G$}}
\nc{\fullm}{\mbox{\boldmath $M$}}
\nc{\scs}{\scriptstyle}
\nc{\beq}{\begin{eqnarray}}
\nc{\eeq}{\end{eqnarray}}
\nc{\la}{\label}
\nc{\r}{\ref}
\nc{\no}{\nonumber}
\nc{\ci}{\cite}
\nc{\meas}{{\cal D}\bar{\psi}{\cal D}\psi{\cal D}A\,}
\nc{\gmeas}{{\cal D}\psi^\dagger{\cal D}\psi{\cal D}A\,}
\nc{\qedaction}{{\cal A}_{\rm QED}[\bar{\psi},\psi,A_\mu]}
\nc{\crossout}[1]{#1\!\!\!/}
\nc{\threeint}{\int\!\!\int\!\!\int}
\nc{\twoint}{\int\!\!\int}
\nc{\freemean}[1]{\langle\,#1\,\rangle^{(0)}}
\nc{\mean}[1]{\langle\,#1\,\rangle}
\nc{\trace}{{\rm Tr\,}}
\nc{\setval}{\fmfset{wiggly_len}{1.5mm}\fmfset{arrow_len}{1.5mm}\fmfset{arrow_ang}{13}\fmfset{dash_len}{1.5mm}\fmfpen{0.125mm}\fmfset{dot_size}{1thick}}
\nc{\dphi}[3]{\frac{\delta #1}{\delta
\parbox{10mm}{\centerline{
\begin{fmfgraph*}(5,3)
\fmfpen{0.125mm}
\fmfleft{v1}
\fmfright{v2}
\fmf{plain}{v2,v1}
\fmfv{decor.size=0,label={\footnotesize #2},l.dist=0.5mm}{v1}
\fmfv{decor.size=0,label={\footnotesize #3},l.dist=0.5mm}{v2}
\end{fmfgraph*}
}}}}
\nc{\ddphi}[1]{\frac{\delta^2 #1}{\delta
\parbox{10mm}{\centerline{
\begin{fmfgraph*}(5,3)
\fmfpen{0.125mm}
\fmfleft{v1}
\fmfright{v2}
\fmf{plain}{v2,v1}
\fmfv{decor.size=0,label={\footnotesize 1},l.dist=0.5mm}{v1}
\fmfv{decor.size=0,label={\footnotesize 2},l.dist=0.5mm}{v2}
\end{fmfgraph*}
}}\,\delta
\parbox{10mm}{\centerline{
\begin{fmfgraph*}(5,3)
\fmfpen{0.125mm}
\fmfleft{v1}
\fmfright{v2}
\fmf{plain}{v2,v1}
\fmfv{decor.size=0,label={\footnotesize 3},l.dist=0.5mm}{v1}
\fmfv{decor.size=0,label={\footnotesize 4},l.dist=0.5mm}{v2}
\end{fmfgraph*}
}}}}
\nc{\dfermi}[3]{\frac{\delta #1}{\delta
\parbox{10mm}{\centerline{
\begin{fmfgraph*}(5,3)
\fmfset{arrow_len}{1.5mm}
\fmfpen{0.125mm}
\fmfleft{v1}
\fmfright{v2}
\fmf{fermion}{v2,v1}
\fmfv{decor.size=0,label={\footnotesize #2},l.dist=0.5mm}{v1}
\fmfv{decor.size=0,label={\footnotesize #3},l.dist=0.5mm}{v2}
\end{fmfgraph*}
}}}}
\nc{\ddfermi}[1]{\frac{\delta^2 #1}{\delta
\parbox{10mm}{\centerline{
\begin{fmfgraph*}(5,3)
\fmfset{arrow_len}{1.5mm}
\fmfpen{0.125mm}
\fmfleft{v1}
\fmfright{v2}
\fmf{fermion}{v2,v1}
\fmfv{decor.size=0,label={\footnotesize 1},l.dist=0.5mm}{v1}
\fmfv{decor.size=0,label={\footnotesize 2},l.dist=0.5mm}{v2}
\end{fmfgraph*}
}}\,\delta
\parbox{10mm}{\centerline{
\begin{fmfgraph*}(5,3)
\fmfset{arrow_len}{1.5mm}
\fmfpen{0.125mm}
\fmfleft{v1}
\fmfright{v2}
\fmf{fermion}{v2,v1}
\fmfv{decor.size=0,label={\footnotesize 3},l.dist=0.5mm}{v1}
\fmfv{decor.size=0,label={\footnotesize 4},l.dist=0.5mm}{v2}
\end{fmfgraph*}
}}
}}
\nc{\dbose}[3]{\frac{\delta #1}{\delta
\parbox{10mm}{\centerline{
\begin{fmfgraph*}(5,3)
\fmfset{wiggly_len}{1.5mm}
\fmfpen{0.125mm}
\fmfleft{v1}
\fmfright{v2}
\fmf{boson}{v2,v1}
\fmfv{decor.size=0,label={\footnotesize #2},l.dist=0.5mm}{v1}
\fmfv{decor.size=0,label={\footnotesize #3},l.dist=0.5mm}{v2}
\end{fmfgraph*}
}}}}
\begin{document}
\setlength{\unitlength}{1mm}
\begin{fmffile}{graph1}
\title{Recursive Graphical Construction of Feynman Diagrams\\
and Their Multiplicities in $\phi^4$- and in $\phi^2A$-Theory}
\author{Hagen Kleinert$^1$, Axel Pelster$^1$, Boris Kastening$^2$, 
and M. Bachmann$^1$}
\address{$^1$Institut f\"ur Theoretische Physik, Freie Universit\"at Berlin, 
Arnimallee 14, 14195 Berlin, Germany\\
$^2$Institut f\"ur Theoretische Physik, Universit\"at Heidelberg, 
Philosophenweg 16, 69120 Heidelberg, Germany}
%\date{}
\maketitle
\begin{abstract}
The free energy of a field theory can be considered as a functional
of the free correlation function. As such it obeys a 
nonlinear functional differential equation which can be turned into a 
recursion relation. This is solved 
order by order in the coupling constant to find all 
connected vacuum diagrams with their proper 
multiplicities. The procedure is  applied to
a multicomponent scalar field theory with a $\phi^4$-self-interaction
and then to a theory of two scalar fields $\phi$ and $A$
with an interaction  $\phi^2 A$.
All Feynman diagrams with external lines are
obtained
from functional derivatives of the connected vacuum diagrams
with respect to the free correlation function.
Finally, the recursive graphical construction 
is automatized by computer algebra with the help of a 
unique matrix notation for the Feynman diagrams.
\end{abstract}
\section{Introduction}
If one wants to draw all Feynman diagrams of higher orders
by hand, it becomes increasingly difficult
to identify all topologically different connections between the vertices. 
To count the corresponding
multiplicites is an even more tedious task. Fortunately,
there exist now various convenient computer programs,  
for instance {\it FeynArts} \ci{FeynArts1,FeynArts2,FeynArts3} or
{\it QGRAF} \ci{QGRAF1,QGRAF2},
for constructing and counting Feynman graphs in 
different field theories. These programs are based on a 
combinatorial enumeration of all possible ways of
connecting vertices by lines
according to Feynman's rules. \\

The purpose of this paper is to develop an alternative systematic 
approach to construct all Feynman diagrams of a field theory.
It relies on considering a
Feynman diagram as a functional of its graphical elements, i.e. its lines and
vertices. Functional derivatives with respect these elements
are represented by graphical operations which remove lines or vertices of a
Feynman diagram in all possible ways. With these operations, our approach
proceeds in two steps. 
First the connected vacuum diagrams
are constructed, together with their proper multiplicities, as solutions of a
graphical recursion relation derived from a
nonlinear functional differential equation. These relations have been
set up a long time ago \cite{Kleinert1,Kleinert2}, but so far they have
only been solved to all orders in the coupling strength in the trivial
case of zero-dimensional quantum field theories. 
The present paper extends the previous
work by developing an efficient graphical
algorithm for solving this equation for two simple scalar 
field theories, a multicomponent scalar
field theory with $\phi^4$-self-interaction and a theory with two 
scalar fields $\phi$ and $A$ with the interaction $\phi^2 A$.
In a second step, all connected diagrams with external lines are
obtained
from functional derivatives of the connected vacuum diagrams
with respect to the free correlation function. Finally, we  
automatize our construction method  
by computer algebra with the help of a
unique matrix notation for Feynman diagrams.
\section{Scalar $\phi^4$-Theory}
\la{PHI}
Consider a self-interacting scalar field $\phi$ with $N$ components 
in $d$ euclidean dimensions whose
thermal fluctuations are controlled by the energy functional
\beq
\la{EF}
E [ \phi ] = \frac{1}{2} \int_{12} G^{-1}_{12} \phi_1 \phi_2 
+ \frac{g}{4!} \int_{1234} V_{1234}  \phi_1 \phi_2 \phi_3 \phi_4
\eeq
with some coupling constant $g$.
In this short-hand notation, the spatial and tensorial
arguments of the
field $\phi$, the bilocal kernel $G^{-1}$, and the quartic 
interaction $V$ are indicated
by simple number indices, i.e.,
\beq
1 \equiv \{ x_1 , \alpha_1 \} \, , \,\, 
\int_1 \equiv \sum_{\alpha_1} \int d^d x_1 \, , \,\,
\phi_1 \equiv \phi_{\alpha_1} ( x_1 ) \, , \,\,
G^{-1}_{12} \equiv G^{-1}_{\alpha_1 , \alpha_2} ( x_1 , x_2 ) \, , \,\, 
V_{1234} \equiv V_{\alpha_1 , \alpha_2 , \alpha_3 , \alpha_4} 
( x_1 , x_2 , x_3 , x_4 ) \, .
\eeq
The kernel is a functional matrix $G^{-1}$, while $V$ is a functional 
tensor, both being symmetric in their
indices. The energy functional (\r{EF}) describes generically
$d$-dimensional euclidean $\phi^4$-theories. These are models for
a family of universality classes of continuous phase transitions,
such as the $O(N)$-symmetric $\phi^4$-theory which serves to derive
the critical phenomena in dilute polymer solutions ($N=0$), Ising- and
Heisenberg-like magnets ($N=1,3$), and superfluids ($N=2$).
In all these
cases, the energy functional (\r{EF}) is specified by
\beq
G_{\alpha_1 , \alpha_2}^{-1} ( x_1 , x_2 ) & = &
\delta_{\alpha_1 , \alpha_2} \, \left( - \partial_{x_1}^2 + m^2 
\right) \delta ( x_1 - x_2 ) \, , \la{PH1} \\ \la{PH2}
V_{\alpha_1,\alpha_2,\alpha_3,\alpha_4} ( x_1 , x_2 , x_3 , x_4 ) 
&=&  \frac{1}{3} \, \left\{ 
\delta_{\alpha_1 , \alpha_2} \delta_{\alpha_3 , \alpha_4} +
\delta_{\alpha_1 , \alpha_3} \delta_{\alpha_2 , \alpha_4} +
\delta_{\alpha_1 , \alpha_4} \delta_{\alpha_2 , \alpha_3} \right\} \,
\delta ( x_1 - x_2 ) \delta ( x_1 - x_3 ) \delta ( x_1 - x_4 ) \, ,
\eeq
where the mass $m^2$ is proportional to the temperature distance
from the critical point.
In the following we shall leave $G^{-1}$ and $V$ completely general, 
except for the symmetry with respect to their indices, and insert
the physical values (\r{PH1}) and (\r{PH2}) at the end.
By using natural units in which the Boltzmann constant $k_B$ times the
temperature $T$ equals unity, the partition function 
is determined as a functional integral over the Boltzmann
weight $e^{- E [ {\bf \phi} ]}$
\beq
\la{PF}
Z = \int {\cal D} {\bf \phi} \, e^{- E [ {\bf \phi} ]} 
\eeq
and may be evaluated perturbatively
as a power series in the coupling constant $g$. From this we obtain
the negative free energy $W = \ln Z$ as an expansion
\beq
\la{EX}
W = \sum_{p = 0}^{\infty} \frac{1}{p!} \left( \frac{- g}{4!} \right)^p
W^{(p)} \, .
\eeq
The coefficients $W^{(p)}$ may be displayed as
connected vacuum diagrams constructed from lines and vertices. Each
line represents a  free correlation function 
\beq
\la{PRO}
\parbox{20mm}{\centerline{
\begin{fmfgraph*}(8,3)
\setval
\fmfleft{v1}
\fmfright{v2}
\fmf{plain}{v1,v2}
\fmflabel{${\scs 1}$}{v1}
\fmflabel{${\scs 2}$}{v2}
\end{fmfgraph*}
}} \equiv G_{12} \, ,
\eeq
which is the functional inverse of the kernel $G^{-1}$ in the energy
functional (\r{EF}), defined by
\beq
\la{FP}
\int_{2} G_{12} \, G^{-1}_{23} = \delta_{13} \, .
\eeq
The vertices represent an integral over the interaction 
\beq
\la{VE}
\parbox{17mm}{\begin{center}
\begin{fmfgraph*}(6,5)
\setval
\fmfstraight
\fmfleft{i2,i1}
\fmfright{o2,o1}
\fmf{plain}{i1,v1}
\fmf{plain}{v1,i2}
\fmf{plain}{v1,o1}
\fmf{plain}{o2,v1}
\fmfdot{v1}
\end{fmfgraph*}
\end{center}}
\equiv \int_{1234} \, V_{1234} \, .
\eeq
To construct all 
connected vacuum diagrams contributing to
$W^{(p)}$ to each order $p$
in perturbation theory, one connects 
$p$ vertices with $4p$ legs  
in all possible ways according to Feynman's rules
which follow from Wick's expansion of correlation functions
into a sum of all pair contractions.
This yields an increasing number of Feynman diagrams, 
each with a certain
multiplicity which
follows from combinatorics. In total there are $4!^p p!$ ways of
ordering the $4p$ legs of the $p$ vertices. This number is reduced by
permutations of the legs and the vertices which leave a vacuum diagram 
invariant. Denoting the number of self-, double, triple and fourfold
connections with $S,D,T,F$, there are $2!^S, 2!^D, 3!^T, 4!^F$ leg
permutations. An additional reduction
arises from the  
number $N$ of identical vertex
permutations where the vertices remain attached to the lines
emerging from them in the same way as before. 
The resulting multiplicity of a 
connected vacuum diagram in the $\phi^4$-theory is therefore
given by the formula \ci{Neu,Verena}
\beq
\la{MU1}
M_{\phi^4}^{E=0} = \frac{4!^p \, p!}{2!^{S+D}\, 3!^T \, 4!^F \, N} \, .
\eeq
The superscript $E$ records that the number of external legs of
the connected vacuum diagrams is zero.
The diagrammatic representation of the
coefficients $W^{(p)}$ in the expansion
(\r{EX}) of the negative
free energy $W$ is displayed in Table \r{P} up to five loops
\ci{Kastening1,Kastening2,Dohm}. \\

For higher orders, the factorially increasing number of diagrams makes it
more and more difficult to construct all topologically different diagrams 
and to count their multiplicities. 
In particular, it becomes quite hard to identify by 
inspection
the number $N$ of identical vertex permutations. This identification
problem is solved by introducing a uniqued matrix notation
for the graphs, to be explained in detail in Section IV. \\

In the following, we shall generate iteratively all connected vacuum
diagrams. We start in Subsection II.A by identifying graphical
operations associated with functional derivatives with respect
to the kernel $G^{-1}$, or the propagator $G$. In Subsection II.B we show 
that these operations can be applied to the one-loop contribution of
the free partition function to
generate all perturbative contributions to the  
partition function (\r{PF}). In Subsection II.C we derive a nonlinear
functional differential equation for the negative free energy $W$, whose
graphical solution in Subsection II.D. yields all connected 
vacuum diagrams order by order in the coupling strength.
\subsection{Basic Graphical Operations} \la{NU}
Each Feynman diagram is composed of integrals over products of free 
correlation functions $G$ and 
may thus be considered as a functional of the kernel 
$G^{-1}$. The connected vacuum diagrams
satisfy a certain functional differential equation,
from which they will be constructed recursively. This will be done 
by a graphical procedure, for which we set up the necessary graphical
rules in this subsection. First we observe 
that functional derivatives with respect to the kernel
$G^{-1}$ or to the free propagator $G$
correspond to the graphical prescriptions of cutting or of removing a single
line of a diagram in all possible ways, respectively.
\subsubsection{Cutting Lines}
Since $\phi$ is a real scalar field, the kernel $G^{-1}$ is a symmetric
functional matrix.  This property has to be taken into account when performing
functional derivatives with respect to the kernel $G^{-1}$, whose basic rule 
is
\beq
\la{DR1}
\frac{\delta G^{-1}_{12}}{\delta G^{-1}_{34}} = \frac{1}{2} \left\{ 
\delta_{13} \delta_{42} + \delta_{14} \delta_{32} \right\} \, .
\eeq
>From the identity (\r{FP}) and the functional chain rule, 
we find the effect of this derivative
on the free propagator
\beq
\la{ACT}
- 2 \, \frac{\delta G_{12}}{\delta G^{-1}_{34}} = 
G_{13} G_{42} + G_{14} G_{32} \, .
\eeq
This has the graphical representation
\beq
- 2 \, \frac{\delta}{\delta G^{-1}_{34}}
\parbox{20mm}{\centerline{
\begin{fmfgraph*}(8,3)
\setval
\fmfleft{v1}
\fmfright{v2}
\fmf{plain}{v1,v2}
\fmflabel{${\scs 1}$}{v1}
\fmflabel{${\scs 2}$}{v2}
\end{fmfgraph*}
}}
=
\parbox{20mm}{\centerline{
\begin{fmfgraph*}(8,3)
\setval
\fmfleft{v1}
\fmfright{v2}
\fmf{plain}{v1,v2}
\fmflabel{${\scs 1}$}{v1}
\fmflabel{${\scs 3}$}{v2}
\end{fmfgraph*}
}}
\parbox{20mm}{\centerline{
\begin{fmfgraph*}(8,3)
\setval
\fmfleft{v1}
\fmfright{v2}
\fmf{plain}{v1,v2}
\fmflabel{${\scs 4}$}{v1}
\fmflabel{${\scs 2}$}{v2}
\end{fmfgraph*}
}}
+
\parbox{20mm}{\centerline{
\begin{fmfgraph*}(8,3)
\setval
\fmfleft{v1}
\fmfright{v2}
\fmf{plain}{v1,v2}
\fmflabel{${\scs 1}$}{v1}
\fmflabel{${\scs 4}$}{v2}
\end{fmfgraph*}
}}
\parbox{20mm}{\centerline{
\begin{fmfgraph*}(8,3)
\setval
\fmfleft{v1}
\fmfright{v2}
\fmf{plain}{v1,v2}
\fmflabel{${\scs 3}$}{v1}
\fmflabel{${\scs 2}$}{v2}
\end{fmfgraph*}
}} \, .
\eeq
Thus differentiating a propagator
with respect to the kernel $G^{-1}$
amounts to cutting the associated line
into two pieces. 
The differentiation rule (\r{DR1}) ensures that the spatial indices 
of the kernel are symmetrically
attached to the newly created line ends
in the two possible ways. When differentiating a general
Feynman integral 
with respect to $G^{-1}$, the product rule of functional
differentiation leads to a sum of diagrams in which
each line is cut once.\\

With this graphical operation, the
product of two fields can be rewritten
as a derivative of the energy functional with 
respect to the kernel
\beq
\label{SR}
\phi_1 \phi_2 = 2 \, \frac{\delta E [ \phi ]}{\delta G^{-1}_{12}} \, ,
\eeq
as follows directly from (\r{EF}) and (\r{DR1}). 
Applying the substitution rule (\r{SR})
to the functional integral for the fully interacting two-point function
\beq
\la{2P}
\fullg_{12} = \frac{1}{Z}\,
\int {\cal D} \phi \, \phi_1 \phi_2 \, e^{- E [ \phi ]} \, ,
\eeq
we obtain the fundamental identity
\beq
\la{GIT}
\fullg_{12}  = -2 \, \frac{\delta W}{\delta G^{-1}_{12}} \, .
\eeq
Thus, by cutting a line of the connected
vacuum diagrams in all possible ways, we obtain
all diagrams of the fully interacting 
two-point function. Analytically this
has a Taylor series expansion in powers of the 
coupling constant $g$ similar to (\r{EX})
\beq
\la{EX2}
\fullg_{12}  = \sum_{p = 0}^{\infty} 
\frac{1}{p!} \left( \frac{- g}{4!} \right)^p
\fullg_{12}^{(p)}
\eeq
with coefficients
\beq
\la{g12p}
\fullg_{12}^{(p)}=-2\, \frac{\delta W^{(p)}}{\delta G_{12}^{-1}}.
\eeq
The cutting prescription (\r{GIT}) converts the vacuum diagrams 
of $p$th order in the coefficients $W^{(p)}$
in Table \r{P} to the corresponding ones in the coefficients 
$\fullg_{12}^{(p)}$ of the two-point function. 
The results are shown in Table \r{TWO} up to
four loops. The numbering of diagrams used in Table \r{TWO}
reveals from which connected vacuum diagrams they are obtained by
cutting a line.
For instance, the diagrams \#15.1-\#15.5 and their 
multiplicities in Table \r{TWO} follow from the connected
vacuum diagram \#15 in Table \r{P}. 
We observe that the multiplicity of a diagram of a 
two-point function obeys a formula similar to (\r{MU1}):
\beq
\la{MU2}
M_{\phi^4}^{E=2} = \frac{4!^p \, p! \, 2!}{2!^{S+D}\, 3!^T \,N} \, .
\eeq
In the numerator, the $4!^p \,p!$ permutations of the $4p$ legs of the $p$
vertices are multiplied by a factor $2!$ for the permutations of the $E=2$ 
end points of
the two-point function. The number $N$ in the denominator
counts the identical permutations of both the $p$ vertices and the two
end points.\\

Performing a differentiation of the two-point function (\r{2P})
with respect to the kernel $G^{-1}$ yields
\beq
\la{FF}
-2 \,
\frac{\delta \fullg_{12}}{\delta G_{34}^{-1}}
= \fullg_{1234} - \fullg_{12} \fullg_{34} \, ,
\eeq
where $\fullg_{1234}$ denotes
the fully interacting four-point function
\beq
\la{4P}
\fullg_{1234} = \frac{1}{Z}\,
\int {\cal D} \phi \, \phi_1 \phi_2 \phi_3 \phi_4 \, e^{- E [ \phi ]} \, .
\eeq
The term $\fullg_{12} \fullg_{34}$ in (\r{FF}) subtracts a certain set of
disconnected diagrams from $\fullg_{1234}$.
By subtracting {\it all}  disconnected diagrams from $\fullg_{1234}$, we obtain
the connected four-point function
\beq
\la{C4P}
\fullg_{1234}^{\rm c}
\equiv \fullg_{1234} - \fullg_{12} \fullg_{34}
- \fullg_{13} \fullg_{24}- \fullg_{14} \fullg_{23}
\eeq
in the form
\beq
\label{GcfromG}
\fullg_{1234}^{\rm c}
= -2 \, \frac{\delta \fullg_{12}}{\delta G^{-1}_{34}}
- \fullg_{13} \fullg_{24}- \fullg_{14} \fullg_{23} \, .
\eeq
The first term contains all diagrams obtained by cutting a line
in the diagrams of the two-point-function $\fullg_{12}$.
The second and third terms
remove from these the disconnected diagrams.
In this way we obtain the perturbative expansion
\beq
\la{EX3}
\fullg_{1234}^{\rm c} =
\sum_{p = 1}^{\infty} 
\frac{1}{p!} \left( \frac{- g}{4!} \right)^p
\fullg_{1234}^{{\rm c},(p)}
\eeq
with coefficients
\beq
\la{g1234cp}
\fullg_{1234}^{c,(p)}=-2\,\frac{\delta\fullg_{12}^{(p)}}{\delta G_{34}^{-1}}
-\sum_{q=0}^p\left(\begin{array}{c}p\\q\end{array}\right)
\left(\fullg_{13}^{(p-q)}\fullg_{24}^{(q)}
+\fullg_{14}^{(p-q)}\fullg_{23}^{(q)}\right).
\eeq
They are listed diagrammatically in Table \r{FOUR} up to three loops.
As before in Table II,
the multiple numbering in Table \r{FOUR} indicates the origin of each
diagram of the connected four-point function.
For instance, the diagram \#11.2.2, \#11.4.3, \#14.1.2, \#14.3.3 
in Table \r{FOUR} stems together with its multiplicity from 
the diagrams \#11.2, \#11.4, \#14.1, \#14.3 in
Table \r{TWO}.\\

The multiplicity of each diagram of a connected four-point function obeys
a formula similar to (\r{MU2}):
\beq
\la{MU3}
M_{\phi^4}^{E=4} = \frac{4!^p \, p! \, 4!}{2!^{S+D}\, 3!^T \,N} \, .
\eeq
This multiplicity decomposes into equal parts if the spatial
indices $1,2,3,4$ are assigned to the $E=4$ end points of the connected
four-point function, for instance:
\beq
62208 \quad
\parbox{18mm}{\begin{center}
\begin{fmfgraph*}(15,5)
\setval
\fmfstraight
\fmfforce{0w,0h}{i1}
\fmfforce{0w,1h}{i2}
\fmfforce{1w,0h}{o1}
\fmfforce{1w,1h}{o2}
\fmfforce{1/6w,0.5h}{v1}
\fmfforce{1/2w,0.5h}{v2}
\fmfforce{5/6w,0.5h}{v3}
\fmf{plain}{i1,v1}
\fmf{plain}{v3,o1}
\fmf{plain}{i2,v1}
\fmf{plain}{v3,o2}
\fmf{plain,left=1}{v1,v2,v1}
\fmf{plain,left=1}{v2,v3,v2}
\fmfdot{v1,v2,v3}
\end{fmfgraph*}
\end{center}}
& \equiv &\,\,\, 20736 \quad
\parbox{18mm}{\begin{center}
\begin{fmfgraph*}(15,5)
\setval
\fmfstraight
\fmfforce{0w,0h}{i1}
\fmfforce{0w,1h}{i2}
\fmfforce{1w,0h}{o1}
\fmfforce{1w,1h}{o2}
\fmfforce{1/6w,0.5h}{v1}
\fmfforce{1/2w,0.5h}{v2}
\fmfforce{5/6w,0.5h}{v3}
\fmf{plain}{i1,v1}
\fmf{plain}{v3,o1}
\fmf{plain}{i2,v1}
\fmf{plain}{v3,o2}
\fmf{plain,left=1}{v1,v2,v1}
\fmf{plain,left=1}{v2,v3,v2}
\fmfv{decor.size=0, label=${\scs 2}$, l.dist=1mm, l.angle=-180}{i1}
\fmfv{decor.size=0, label=${\scs 1}$, l.dist=1mm, l.angle=-180}{i2}
\fmfv{decor.size=0, label=${\scs 4}$, l.dist=1mm, l.angle=0}{o1}
\fmfv{decor.size=0, label=${\scs 3}$, l.dist=1mm, l.angle=0}{o2}
\fmfdot{v1,v2,v3}
\end{fmfgraph*}
\end{center}}
\quad + \, 20736 \quad
\parbox{18mm}{\begin{center}
\begin{fmfgraph*}(15,5)
\setval
\fmfstraight
\fmfforce{0w,0h}{i1}
\fmfforce{0w,1h}{i2}
\fmfforce{1w,0h}{o1}
\fmfforce{1w,1h}{o2}
\fmfforce{1/6w,0.5h}{v1}
\fmfforce{1/2w,0.5h}{v2}
\fmfforce{5/6w,0.5h}{v3}
\fmf{plain}{i1,v1}
\fmf{plain}{v3,o1}
\fmf{plain}{i2,v1}
\fmf{plain}{v3,o2}
\fmf{plain,left=1}{v1,v2,v1}
\fmf{plain,left=1}{v2,v3,v2}
\fmfv{decor.size=0, label=${\scs 3}$, l.dist=1mm, l.angle=-180}{i1}
\fmfv{decor.size=0, label=${\scs 1}$, l.dist=1mm, l.angle=-180}{i2}
\fmfv{decor.size=0, label=${\scs 4}$, l.dist=1mm, l.angle=0}{o1}
\fmfv{decor.size=0, label=${\scs 2}$, l.dist=1mm, l.angle=0}{o2}
\fmfdot{v1,v2,v3}
\end{fmfgraph*}
\end{center}}
\quad + \, 20736 \quad
\parbox{18mm}{\begin{center}
\begin{fmfgraph*}(15,5)
\setval
\fmfstraight
\fmfforce{0w,0h}{i1}
\fmfforce{0w,1h}{i2}
\fmfforce{1w,0h}{o1}
\fmfforce{1w,1h}{o2}
\fmfforce{1/6w,0.5h}{v1}
\fmfforce{1/2w,0.5h}{v2}
\fmfforce{5/6w,0.5h}{v3}
\fmf{plain}{i1,v1}
\fmf{plain}{v3,o1}
\fmf{plain}{i2,v1}
\fmf{plain}{v3,o2}
\fmf{plain,left=1}{v1,v2,v1}
\fmf{plain,left=1}{v2,v3,v2}
\fmfv{decor.size=0, label=${\scs 4}$, l.dist=1mm, l.angle=-180}{i1}
\fmfv{decor.size=0, label=${\scs 1}$, l.dist=1mm, l.angle=-180}{i2}
\fmfv{decor.size=0, label=${\scs 3}$, l.dist=1mm, l.angle=0}{o1}
\fmfv{decor.size=0, label=${\scs 2}$, l.dist=1mm, l.angle=0}{o2}
\fmfdot{v1,v2,v3}
\end{fmfgraph*}
\end{center}}
\quad . 
\eeq
Generalizing the multiplicities (\r{MU1}), (\r{MU2}), and (\r{MU3}) 
for connected
vacuum diagrams, two- and four-point functions to an arbitrary 
connected correlation
function with an even number $E$ of end points, we see that 
\beq
\la{MU4}
M_{\phi^4}^{E} = \frac{4!^p \, p! \, E!}{2!^{S+D}\, 3!^T\,4!^F \,N} \, ,
\eeq
where $N$ counts the number of permutations of vertices and external lines
which leave the diagram unchanged.
\subsubsection{Removing Lines}
We now study the graphical effect of functional derivatives
with respect to the free propagator $G$,
where the basic differentiation rule (\r{DR1}) becomes
\beq
\la{DR2}
\frac{\delta G_{12}}{\delta G_{34}} = \frac{1}{2} \left\{ 
\delta_{13} \delta_{42} + \delta_{14} \delta_{32} \right\} \, .
\eeq
We represent this graphically by
extending the elements of Feynman diagrams by an open dot with two labeled
line ends representing the delta function:
\beq
{\scs 1}
\parbox{6mm}{\centerline{
\begin{fmfgraph}(4,3)
\setval
\fmfforce{0w,0.5h}{i1}
\fmfforce{1w,0.5h}{o1}
\fmfforce{0.5w,0.5h}{v1}
\fmf{plain}{i1,v1}
\fmf{plain}{v1,o1}
\fmfv{decor.size=0, label={\scs 1}, l.dist=1mm, l.angle=-180}{i1}
\fmfv{decor.size=0, label={\scs 2}, l.dist=1mm, l.angle=0}{o1}
\fmfv{decor.shape=circle,decor.filled=empty,decor.size=0.6mm}{v1}
\end{fmfgraph}}} {\scs 2}
= \quad \delta_{12} \, .
\eeq
Thus we can write
the differentiation (\r{DR2}) graphically as follows:
\beq
\dphi{}{3}{4} 
\parbox{20mm}{\centerline{
\begin{fmfgraph*}(8,3)
\setval
\fmfleft{v1}
\fmfright{v2}
\fmf{plain}{v1,v2}
\fmflabel{${\scs 1}$}{v1}
\fmflabel{${\scs 2}$}{v2}
\end{fmfgraph*}
}} = \frac{1}{2} \Bigg\{
{\scs 1}
\parbox{6mm}{\centerline{
\begin{fmfgraph}(4,3)
\setval
\fmfforce{0w,0.5h}{i1}
\fmfforce{1w,0.5h}{o1}
\fmfforce{0.5w,0.5h}{v1}
\fmf{plain}{i1,v1}
\fmf{plain}{v1,o1}
\fmfv{decor.size=0, label=${\scs 1}$, l.dist=1mm, l.angle=-180}{i1}
\fmfv{decor.size=0, label=${\scs 2}$, l.dist=1mm, l.angle=0}{o1}
\fmfv{decor.shape=circle,decor.filled=empty,decor.size=0.6mm}{v1}
\end{fmfgraph}}} {\scs 3} \quad
{\scs 4}
\parbox{6mm}{\centerline{
\begin{fmfgraph}(4,3)
\setval
\fmfforce{0w,0.5h}{i1}
\fmfforce{1w,0.5h}{o1}
\fmfforce{0.5w,0.5h}{v1}
\fmf{plain}{i1,v1}
\fmf{plain}{v1,o1}
\fmfv{decor.size=0, label=${\scs 1}$, l.dist=1mm, l.angle=-180}{i1}
\fmfv{decor.size=0, label=${\scs 2}$, l.dist=1mm, l.angle=0}{o1}
\fmfv{decor.shape=circle,decor.filled=empty,decor.size=0.6mm}{v1}
\end{fmfgraph}}} {\scs 2} \quad + \quad
{\scs 1}
\parbox{6mm}{\centerline{
\begin{fmfgraph}(4,3)
\setval
\fmfforce{0w,0.5h}{i1}
\fmfforce{1w,0.5h}{o1}
\fmfforce{0.5w,0.5h}{v1}
\fmf{plain}{i1,v1}
\fmf{plain}{v1,o1}
\fmfv{decor.size=0, label=${\scs 1}$, l.dist=1mm, l.angle=-180}{i1}
\fmfv{decor.size=0, label=${\scs 2}$, l.dist=1mm, l.angle=0}{o1}
\fmfv{decor.shape=circle,decor.filled=empty,decor.size=0.6mm}{v1}
\end{fmfgraph}}} {\scs 4} \quad
{\scs 3}
\parbox{6mm}{\centerline{
\begin{fmfgraph}(4,3)
\setval
\fmfforce{0w,0.5h}{i1}
\fmfforce{1w,0.5h}{o1}
\fmfforce{0.5w,0.5h}{v1}
\fmf{plain}{i1,v1}
\fmf{plain}{v1,o1}
\fmfv{decor.size=0, label=${\scs 1}$, l.dist=1mm, l.angle=-180}{i1}
\fmfv{decor.size=0, label=${\scs 2}$, l.dist=1mm, l.angle=0}{o1}
\fmfv{decor.shape=circle,decor.filled=empty,decor.size=0.6mm}{v1}
\end{fmfgraph}}} {\scs 2}
\Bigg\} \, .
\eeq
Differentiating a line with respect to the free correlation function
removes the line, leaving in a symmetrized way the spatial indices of
the free correlation function on the vertices to which the line was
connected.\\

The effect of this derivative is illustrated by studying
the diagrammatic effect of the operator
\beq
\la{OO}
\hat{L} = \int_{12} G_{12} \, 
\frac{\delta}{\delta G_{12}} \, .
\eeq
Applying $\hat{L}$ to a connected vacuum diagram in $W^{(p)}$, 
the functional derivative $\delta / \delta G_{12}$ removes successively
each of its $2p$ lines. Subsequently, the
removed lines are again reinserted, so that the connected vacuum diagrams
$W^{(p)}$ are eigenfunctions of $\hat{L}$, whose eigenvalues $2p$ count
the lines of the diagrams:
\beq
\la{EVP}
\hat{L} \, W^{(p)}  = 2 p \, W^{(p)} \, .
\eeq
As an example, take
the explicit first-order expression for the vacuum diagrams, i.e.
\beq
\la{FO}
W^{(1)} = 3 \int_{1234} V_{1234} \,G_{12} G_{34} \, ,
\eeq
and apply the basic rule (\r{DR2}), leading to the desired eigenvalue $2$.
\subsection{Perturbation Theory}
Field theoretic perturbation expressions are usually derived by  
introducing an external current $J$ into the energy functional (\r{EF})
which is linearly coupled to the field $\phi$
\ci{Amit,Zuber,Zinn}. 
Thus the partition function
(\r{PF}) becomes in the presence of $J$ the generating functional $Z [ J ]$
which allows us to find all free
$n$-point functions from functional derivatives with respect to this
external current $J$. In the normal phase of a $\phi^4$-theory, the 
expectation value of the field $\phi$ is zero and only correlation functions
of an even number of fields are nonzero. To calculate all of these, 
it is possible to 
substitute two functional derivatives with respect to the current $J$
by one functional derivative
with respect to the kernel $G^{-1}$. This reduces 
the number of functional derivatives in each order 
of perturbation theory by one half and has the additional advantage that
the introduction of the current $J$ becomes superfluous.
\subsubsection{Current Approach}
Recall briefly the standard perturbative treatment, in which
the energy functional
(\r{EF}) is artificially extended by a source term
\beq
E [ \phi , J ] = E [ \phi ] - \int_1 J_1 \phi_1 \, .
\eeq
The functional integral for the generating functional
\beq
\la{GF}
Z [J]= \int {\cal D} {\bf \phi} \, e^{- E [ {\bf \phi} ,J]} 
\eeq
is first explicitly calculated for a vanishing coupling constant $g$,
yielding
\beq
\la{GFF}
Z^{(0)} [ J ] = \exp \left\{ - \frac{1}{2} \, \mbox{Tr} \ln G^{-1} + 
\frac{1}{2} \int_{12} \, G_{12} \, J_1 J_2 \right\} \, ,
\eeq
where the trace of the logarithm of the kernel is defined by the series
\cite[p.~16]{Kleinert4}
\beq
\la{LOG}
\mbox{Tr} \ln G^{-1} = \sum_{n = 1}^{\infty} \frac{(-1)^{n + 1}}{n}
\int_{1 \ldots n} \left\{ G^{-1}_{12} - \delta_{12} \right\} \cdots
\left\{ G^{-1}_{n1} - \delta_{n1} \right\} \, .
\eeq
If the coupling constant $g$ does not vanish, one expands the generating
functional $Z [ J ]$ in powers of the quartic interaction  $V$,
and reexpresses the resulting
powers of the field
within the functional integral (\r{GF}) as functional
derivatives with respect to the current $J$. The original partition
function (\r{PF}) can thus be obtained from the free generating functional
(\r{GFF}) by the formula
\beq
\la{CP}
Z = \left.\exp \left\{ - \frac{g}{4!} \, \int_{1234} V_{1234} \, 
\frac{\delta^4}{\delta J_1 \delta J_2 \delta J_3 \delta J_4} \right\}
Z^{(0)} [ J ]\right|_{J=0} \, . 
\eeq
Expanding the exponential in a power series, we arrive at the perturbation
expansion
\beq
Z &=& \left\{ 1 + \frac{-g}{4!} \, \int_{1234} V_{1234} \, 
\frac{\delta^4}{\delta J_1 \delta J_2 \delta J_3 \delta J_4} \right.\no \\
&& \left. \left.
+  \frac{1}{2} \left(  \frac{-g}{4!} \right)^2 \int_{12345678}
V_{1234} V_{5678} \,
\frac{\delta^8}{\delta J_1 \delta J_2 \delta J_3 \delta J_4
\delta J_5 \delta J_6 \delta J_7 \delta J_8} + \ldots \right\}
Z^{(0)} [ J ]\right|_{J=0} \, , \la{EJ}
\eeq
in which 
the $p$th order contribution
for the partition function requires the evaluation of $4p$ functional
derivatives with respect to the current $J$.
\subsubsection{Kernel Approach}
The derivation of the perturbation expansion simplifies,
if we use functional derivatives
with respect to the kernel $G^{-1}$ in the energy functional (\r{EF})
rather than with respect to the
current $J$. This allows us to
substitute the previous expression (\r{CP}) for the partition function by
\beq
\la{CPN}
Z = \exp \left\{ - \frac{g}{6} \, \int_{1234} V_{1234} \, 
\frac{\delta^2}{\delta G^{-1}_{12} \delta G^{-1}_{34}} \right\}
e^{W^{(0)}} \, ,
\eeq
where the zeroth order of the negative free energy has the diagrammatic 
representation
\beq
\la{FPF}
W^{(0)} = - \frac{1}{2}\, \mbox{Tr} \ln G^{-1} \equiv
\frac{1}{2}\;
\parbox{8mm}{\centerline{
\begin{fmfgraph}(5,5)
\setval
\fmfi{plain}{reverse fullcircle scaled 1w shifted (0.5w,0.5h)}
\end{fmfgraph}
}}
\, .
\eeq
Expanding again the exponential in a power series, we obtain
\beq
\la{NANA}
Z =  \left\{ 1 + \frac{-g}{6} \, \int_{1234} V_{1234} \, 
\frac{\delta^2}{\delta G^{-1}_{12} \delta G^{-1}_{34}} + 
\frac{1}{2} \left(  \frac{-g}{6} \right)^2 \int_{12345678}
V_{1234} V_{5678} 
\frac{\delta^4}{\delta G^{-1}_{12} \delta G^{-1}_{34}
\delta G^{-1}_{56} \delta G^{-1}_{78}} + \ldots \right\}
e^{W^{(0)}} \, .
\eeq
Thus we need only half as many 
functional derivatives than in (\r{EJ}).
Taking into account (\r{DR1}), (\ref{ACT}), and (\r{LOG}), we obtain
\beq
\la{PROP}
\frac{\delta W^{(0)}}{\delta G^{-1}_{12}} = - \,\frac{1}{2} \, G_{12} \, , 
\quad \frac{\delta^2 W^{(0)}}{\delta G^{-1}_{12} \delta G^{-1}_{34}} =
\frac{1}{4} \left\{ G_{13} G_{24} + G_{14} G_{23} \right\} \, ,
\eeq
such that the partition function $Z$ becomes
\beq
Z & = & \left\{ 1 + \frac{-g}{4!} \, 3 \int_{1234} V_{1234} \, G_{12} G_{34} 
+ \, \frac{1}{2} \, \left( \frac{-g}{4!} \right)^2 \int_{12345678}
V_{1234} V_{5678}\right. \no \\
&&\,\times \, \left. \Bigg[
9 \, G_{12} G_{34} G_{56} G_{78} 
 + 24 \, G_{15} G_{26} G_{37} G_{48} 
+ 72 \, G_{12} G_{35} G_{46} G_{78} \Bigg] \, + \, \ldots \right\}
\, e^{W^{(0)}} \, . 
\eeq
This has the diagrammatic representation
\beq
\la{ZEX}
Z = \left\{ 1 + \frac{-g}{4!} \, 3 
\parbox{13mm}{\begin{center}
\begin{fmfgraph*}(10,5)
\setval
\fmfleft{i1}
\fmfright{o1}
\fmf{plain,left=1}{i1,v1,i1}
\fmf{plain,left=1}{o1,v1,o1}
\fmfdot{v1}
\end{fmfgraph*}\end{center}}
+ \, \frac{1}{2} \, \left( \frac{-g}{4!} \right)^2 \left[ 
9 \, 
\parbox{13mm}{\begin{center}
\begin{fmfgraph*}(10,5)
\setval
\fmfleft{i1}
\fmfright{o1}
\fmf{plain,left=1}{i1,v1,i1}
\fmf{plain,left=1}{o1,v1,o1}
\fmfdot{v1}
\end{fmfgraph*}\end{center}}
\parbox{13mm}{\begin{center}
\begin{fmfgraph*}(10,5)
\setval
\fmfleft{i1}
\fmfright{o1}
\fmf{plain,left=1}{i1,v1,i1}
\fmf{plain,left=1}{o1,v1,o1}
\fmfdot{v1}
\end{fmfgraph*}\end{center}}
+ 24 \,
\parbox{10.5mm}{\begin{center}
\begin{fmfgraph*}(7.5,5)
\setval
\fmfforce{0w,0.5h}{v1}
\fmfforce{1w,0.5h}{v2}
%\fmfleft{v1}
%\fmfright{v2}
\fmf{plain,left=1}{v1,v2,v1}
\fmf{plain,left=0.4}{v1,v2,v1}
\fmfdot{v1,v2}
\end{fmfgraph*}\end{center}} 
+ 72 \, 
\parbox{18mm}{\begin{center}
\begin{fmfgraph*}(15,5)
\setval
\fmfleft{i1}
\fmfright{o1}
\fmf{plain,left=1}{i1,v1,i1}
\fmf{plain,left=1}{v1,v2,v1}
\fmf{plain,left=1}{o1,v2,o1}
\fmfdot{v1,v2}
\end{fmfgraph*}\end{center}} 
\right] \, + \, \ldots \right\}
\, e^{W^{(0)}} \, .
\eeq
All diagrams in this expansion follow directly by successively
cutting lines of the basic one-loop
vacuum diagram (\r{FPF}) according to (\r{NANA}).
By going to the logarithm of the partition function $Z$, we find a 
diagrammatic expansion for the negative free energy $W$ 
\beq
W = 
\frac{1}{2}\;
\parbox{8mm}{\centerline{
\begin{fmfgraph}(5,5)
\setval
\fmfi{plain}{reverse fullcircle scaled 1w shifted (0.5w,0.5h)}
\end{fmfgraph}
}}
+ \,\frac{-g}{4!} \,\, 3\, 
\parbox{13mm}{\begin{center}
\begin{fmfgraph*}(10,5)
\setval
\fmfleft{i1}
\fmfright{o1}
\fmf{plain,left=1}{i1,v1,i1}
\fmf{plain,left=1}{o1,v1,o1}
\fmfdot{v1}
\end{fmfgraph*}\end{center}}
+ \, \frac{1}{2} \, \left( \frac{-g}{4!} \right)^2 \left\{ 
\,24 \,
\parbox{10.5mm}{\begin{center}
\begin{fmfgraph*}(7.5,5)
\setval
\fmfforce{0w,0.5h}{v1}
\fmfforce{1w,0.5h}{v2}
%\fmfleft{v1}
%\fmfright{v2}
\fmf{plain,left=1}{v1,v2,v1}
\fmf{plain,left=0.4}{v1,v2,v1}
\fmfdot{v1,v2}
\end{fmfgraph*}\end{center}} 
+ 72 \, 
\parbox{18mm}{\begin{center}
\begin{fmfgraph*}(15,5)
\setval
\fmfleft{i1}
\fmfright{o1}
\fmf{plain,left=1}{i1,v1,i1}
\fmf{plain,left=1}{v1,v2,v1}
\fmf{plain,left=1}{o1,v2,o1}
\fmfdot{v1,v2}
\end{fmfgraph*}\end{center}} 
\right\} \, + \, \ldots \, ,
\eeq
which turns out to contain precisely all connected diagrams in 
(\r{ZEX}) with the same multiplicities. In the next section we show that
this diagrammatic expansion for the negative free energy can be derived more
efficiently by solving a differential equation.
\subsection{Functional Differential Equation for $W = \ln Z$}
Regarding the partition function $Z$ as a functional of the kernel
$G^{-1}$, we derive a
functional differential equation for $Z$. We start with the
trivial identity
\beq
\int {\cal D} \phi \frac{\delta}{\delta \phi_1} \left\{ \phi_2 
\, e^{- E [ \phi ]} \right\} = 0 \, ,
\eeq
which follows via functional integration by parts from the vanishing of 
the exponential at infinite fields. Taking
into account the explicit form of
the energy functional (\r{EF}), we perform
the functional derivative with respect to the field
and obtain
\beq
\la{FD1}
\int {\cal D} \phi \left\{ \delta_{12} 
- \int_3 G^{-1}_{13} \phi_2 \phi_3
- \frac{g}{6} \int_{345} V_{1345} \, \phi_2 \phi_3 \phi_4 \phi_5 
\right\} e^{- E [ \phi ]} = 0 \, .
\eeq
Applying the substitution rule (\r{SR}), this equation 
can be expressed in terms of the partition function (\r{PF})
and its derivatives with respect to the kernel $G^{-1}$:
\beq
\la{FD2}
\delta_{12} \, Z + 2 \int_3 G_{13}^{-1} 
\frac{\delta Z}{\delta G^{-1}_{23}} = \frac{2}{3} \, g\,\int_{345} V_{1345}
\frac{\delta^2 Z}{\delta G^{-1}_{23} \delta G^{-1}_{45}} \, .
\eeq
Note that this linear functional differential equation for the partition
function $Z$ is, indeed, solved by (\r{CPN}) due to the commutation
relation
\beq
\la{CR1}
&&\exp \left\{ - \frac{g}{6} \int_{1234} V_{1234} \, 
\frac{\delta^2}{\delta G^{-1}_{12} \delta G^{-1}_{34}}\right\}
\, G^{-1}_{56} - G^{-1}_{56} \, 
\exp \left\{ - \frac{g}{6} \int_{1234} V_{1234} \, 
\frac{\delta^2}{\delta G^{-1}_{12} \delta G^{-1}_{34}}\right\} 
\nonumber \\
&&\hspace*{1cm}
= - \frac{g}{3} \int_{78} V_{5678} \, \frac{\delta}{\delta G^{-1}_{78}}\, 
\exp \left\{ - \frac{g}{6} \int_{1234} V_{1234} \, 
\frac{\delta^2}{\delta G^{-1}_{12} \delta G^{-1}_{34}}\right\}
\eeq
which follows from the canonical one
\beq
\la{CR2}
\frac{\delta}{\delta G^{-1}_{12}} \, G^{-1}_{34} - 
 G^{-1}_{34} \,\frac{\delta}{\delta G^{-1}_{12}} = 
\frac{1}{2} \, \left\{ \delta_{13} \delta_{24} + \delta_{14} \delta_{23} 
\right\} \, .
\eeq
Going over from $Z$ to $W = \ln Z$,
the linear functional differential equation (\r{FD2}) turns
into a nonlinear one:
\beq
\la{FD3}
\delta_{12} + 2 \int_3 G_{13}^{-1} 
\frac{\delta W}{\delta G^{-1}_{23}} = \frac{2}{3}\,g\, \int_{345} V_{1345} 
\left\{ \frac{\delta^2 W}{\delta G^{-1}_{23} \delta G^{-1}_{45}} 
+\frac{\delta W}{\delta G^{-1}_{23}} \, \frac{\delta W}{\delta G^{-1}_{45}}
\right\} \, .
\eeq
If the coupling constant $g$ vanishes, this is immediately 
solved by (\r{FPF}). For a non-vanishing coupling constant $g$, the
right-hand side in (\r{FD3}) produces corrections to (\r{FPF})
which we shall denote with $W^{({\rm int})}$. Thus
the negative free energy $W$ decomposes according to
\beq
\la{DEC}
W = W^{(0)} + W^{({\rm int})} \, .
\eeq
Inserting this into (\r{FD3}) and taking into account (\ref{PROP}),
we obtain the following functional differential equation for the interaction
negative 
free energy $W^{({\rm int})}$:
\beq
\la{FD4}
\int_{12} G_{12}^{-1} \,\frac{\delta W^{({\rm int})}}{\delta G^{-1}_{12}} &=&
\frac{g}{4} \int_{1234} V_{1234} \,G_{12} G_{34} - \frac{g}{3} 
\int_{1234} V_{1234} \,G_{12} \, \frac{\delta W^{({\rm int})}}{\delta 
G^{-1}_{34}} \no \\
&&+
\frac{g}{3} \int_{1234} V_{1234} \, \left\{ \frac{\delta^2
W^{({\rm int})}}{\delta G^{-1}_{12} \delta G^{-1}_{34}} + 
\frac{\delta W^{({\rm int})}}{\delta G^{-1}_{12}}\, 
\frac{\delta W^{({\rm int})}}{\delta G^{-1}_{34}} \right\} \, . 
\eeq
With the help of the functional chain rule, the first and second 
derivatives with respect to the kernel $G^{-1}$ are rewritten as
\beq
\la{NR1}
\frac{\delta}{\delta G^{-1}_{12}} = - \int_{34} G_{13} G_{24} 
\frac{\delta}{\delta G_{34}} 
\eeq
and
\beq
\label{NR2}
\frac{\delta^2}{\delta G^{-1}_{12} \delta G^{-1}_{34}} & = & \int_{5678}
G_{15} G_{26} G_{37} G_{48} \frac{\delta^2}{\delta G_{56} \delta G_{78}} 
\no \\
&&+ \frac{1}{2} \int_{56} \left\{ G_{13} G_{25} G_{46} +
G_{14} G_{25} G_{36} + G_{23} G_{15} G_{46} + G_{24} G_{15} G_{36} 
\right\} \frac{\delta}{\delta G_{56}} \, ,
\eeq
respectively, so that
the functional differential equation (\r{FD4}) for  
$W^{({\rm int})}$ takes the form
(compare Eq.~(51) in Ref.~\cite{Kleinert2})
\beq
\la{FDD}
\int_{12} G_{12} \,\frac{\delta W^{({\rm int})}}{\delta G_{12}} &=&
- \frac{g}{4} \int_{1234} V_{1234} \,G_{12} G_{34} 
- g \, \int_{123456} V_{1234} \, G_{12} G_{35} G_{46} \frac{\delta
W^{({\rm int})}}{\delta G_{56}} \no \\
&&- \frac{g}{3}
\int_{12345678} V_{1234} \, G_{15} G_{26} G_{37} G_{48}
\left\{ \frac{\delta^2 W^{({\rm int})}}{\delta G_{56} \delta G_{78}}
+ \frac{\delta W^{({\rm int})}}{\delta G_{56}} 
\,\frac{\delta W^{({\rm int})}}{\delta G_{78}} \right\} \, .
\eeq
\subsection{Recursion Relation and Graphical Solution}
We now convert the functional differential equation (\r{FDD}) into a recursion
relation by expanding $W^{({\rm int})}$ into a power series in $G$:
\beq
\label{EXX}
W^{({\rm int})} = 
\sum_{p = 1}^{\infty} \frac{1}{p!} \left( \frac{- g}{4!} \right)^p
W^{(p)} \, .
\eeq
Using the property (\r{EVP}) that the coefficient $W^{(p)}$ satisfies the 
eigenvalue problem of the line numbering operator (\ref{OO}), 
we obtain the recursion relation
\beq
\label{RR1}
W^{({\rm p+1})}& =& 12 \int_{123456} V_{1234} \, G_{12} G_{35} G_{46} 
\, \frac{\delta W^{({\rm p})}}{\delta G_{56}} + 4
\int_{12345678} V_{1234} \,G_{15} G_{26} G_{37} G_{48} 
\frac{\delta^2 W^{({\rm p})}}{\delta G_{56} \delta G_{78}} \no \\
&&+ 4 
\sum\limits_{q=1}^{p-1} 
\left( \begin{array}{c} p \\ q \end{array} \right) 
\int_{12345678} V_{1234}\, G_{15} G_{26} G_{37} G_{48}\, 
\frac{\delta W^{({\rm p-q})}}{\delta G_{56}} \,
\frac{\delta W^{({\rm q})}}{\delta G_{78}} 
\eeq
and the initial condition (\r{FO}). With the help
of the graphical rules of Subsection \r{NU}, the
recursion relation (\r{RR1}) can be written diagrammatically as follows
\beq
W^{(p+1)} & = &\quad  4 \quad\ddphi{W^{(p)}}\quad
\parbox{14mm}{\begin{center}
\begin{fmfgraph*}(3,10)
\setval
\fmfstraight
\fmfleft{i1,i2,i3,i4}
\fmfright{o1}
\fmf{plain}{o1,i1}
\fmf{plain}{o1,i2}
\fmf{plain}{i3,o1}
\fmf{plain}{i4,o1}
\fmfdot{o1}
\fmfv{decor.size=0, label=${\scs 4}$, l.dist=1mm, l.angle=-180}{i1}
\fmfv{decor.size=0, label=${\scs 3}$, l.dist=1mm, l.angle=-180}{i2}
\fmfv{decor.size=0, label=${\scs 2}$, l.dist=1mm, l.angle=-180}{i3}
\fmfv{decor.size=0, label=${\scs 1}$, l.dist=1mm, l.angle=-180}{i4}
\end{fmfgraph*}
\end{center}}
+ \,\, 12 \quad\dphi{W^{(p)}}{1}{2}\quad
\parbox{17mm}{\begin{center}
\begin{fmfgraph*}(8,5)
\setval
\fmfstraight
\fmfforce{0w,0h}{i1}
\fmfforce{0w,1h}{i2}
\fmfforce{3/8w,0.5h}{v1}
\fmfforce{1w,0.5h}{v2}
\fmf{plain}{i1,v1}
\fmf{plain}{v1,i2}
\fmf{plain,left}{v1,v2,v1}
\fmfdot{v1}
\fmfv{decor.size=0, label=${\scs 2}$, l.dist=1mm, l.angle=-180}{i1}
\fmfv{decor.size=0, label=${\scs 1}$, l.dist=1mm, l.angle=-180}{i2}
\end{fmfgraph*}
\end{center}}
\no \\
&&\quad+\quad4\quad\sum\limits_{q=1}^{p-1} 
\left( \begin{array}{c} p \\ q \end{array} \right) 
\quad\dphi{W^{(p-q)}}{1}{2}\quad
\parbox{17mm}{\begin{center}
\begin{fmfgraph*}(6,5)
\setval
\fmfstraight
\fmfleft{i2,i1}
\fmfright{o2,o1}
\fmf{plain}{i1,v1}
\fmf{plain}{v1,i2}
\fmf{plain}{v1,o1}
\fmf{plain}{o2,v1}
\fmfdot{v1}
\fmfv{decor.size=0, label=${\scs 1}$, l.dist=1mm, l.angle=-180}{i1}
\fmfv{decor.size=0, label=${\scs 2}$, l.dist=1mm, l.angle=-180}{i2}
\fmfv{decor.size=0, label=${\scs 3}$, l.dist=1mm, l.angle=0}{o1}
\fmfv{decor.size=0, label=${\scs 4}$, l.dist=1mm, l.angle=0}{o2}
\end{fmfgraph*}
\end{center}}
\quad\dphi{W^{(q)}}{3}{4} \, , \qquad p \ge 1 \, .
\la{GRR}
\eeq
This is iterated starting from
\beq
\la{STTT}
W^{(1)} \, = \, 3 \,\,\,
\parbox{12mm}{\begin{center}
\begin{fmfgraph*}(10,5)
\setval
\fmfleft{i1}
\fmfright{o1}
\fmf{plain,left=1}{i1,v1,i1}
\fmf{plain,left=1}{o1,v1,o1}
\fmfdot{v1}
\end{fmfgraph*}\end{center}} \, .
\eeq
The right-hand side of (\r{GRR}) contains
three different graphical
operations. The first two are linear and involve
one or two line amputations of the previous perturbative order. 
The third operation is nonlinear and mixes two 
different one-line amputations of lower orders.\\

An alternative way of formulating the above recursion relation may be
based on the graphical rules
\beq
\la{ALTER}
W^{(p)} \hspace*{0.1cm} = 
\parbox{13mm}{\begin{center}
\begin{fmfgraph*}(10,10)
\setval
\fmfforce{0w,1/2h}{v1}
\fmfforce{1w,1/2h}{v2}
\fmfforce{1/2w,1/2h}{v3}
\fmf{plain,left=1}{v2,v1,v2}
\fmfv{decor.size=0, label=$p$, l.dist=0mm, l.angle=0}{v3}
\end{fmfgraph*}\end{center}} 
\, , \hspace*{1cm}
\frac{\delta W^{(p)}}{\delta G_{12}} \hspace*{0.1cm} = \hspace*{0.3cm}
\parbox{13mm}{\begin{center}
\begin{fmfgraph*}(10,10)
\setval
\fmfforce{0w,1/2h}{v1}
\fmfforce{1w,1/2h}{v2}
\fmfforce{1/2w,1/2h}{v3}
\fmfforce{0.12w,0.8h}{v4}
\fmfforce{0.12w,0.2h}{v5}
\fmf{plain,left=1}{v2,v1,v2}
\fmfv{decor.size=0, label=$p$, l.dist=0mm, l.angle=0}{v3}
\fmfv{decor.size=0, label=${\scs 1}$, l.dist=2mm, l.angle=-180}{v4}
\fmfv{decor.size=0, label=${\scs 2}$, l.dist=2mm, l.angle=-180}{v5}
\fmfdot{v4,v5}
\end{fmfgraph*}\end{center}} 
\, , \hspace*{1cm}
\frac{\delta^2 W^{(p)}}{\delta G_{12} \delta G_{34}} \hspace*{0.1cm}
= \hspace*{0.3cm}
\parbox{13mm}{\begin{center}
\begin{fmfgraph*}(10,10)
\setval
\fmfforce{0w,1/2h}{v1}
\fmfforce{1w,1/2h}{v2}
\fmfforce{1/2w,1/2h}{v3}
\fmfforce{0.12w,0.8h}{v4}
\fmfforce{0.12w,0.2h}{v5}
\fmfforce{0.3w,0.95h}{v6}
\fmfforce{0.3w,0.05h}{v7}
\fmf{plain,left=1}{v2,v1,v2}
\fmfv{decor.size=0, label=$p$, l.dist=0mm, l.angle=0}{v3}
\fmfv{decor.size=0, label=${\scs 2}$, l.dist=2mm, l.angle=-180}{v4}
\fmfv{decor.size=0, label=${\scs 3}$, l.dist=2mm, l.angle=-180}{v5}
\fmfv{decor.size=0, label=${\scs 1}$, l.dist=1.5mm, l.angle= 120}{v6}
\fmfv{decor.size=0, label=${\scs 4}$, l.dist=1.5mm, l.angle=-120}{v7}
\fmfdot{v4,v5,v6,v7}
\end{fmfgraph*}\end{center}} \, .
\eeq
With these, the recursion relation (\r{GRR}) reads
\beq
\la{graphrec}
\parbox{13mm}{\begin{center}
\begin{fmfgraph*}(10,10)
\setval
\fmfforce{0w,1/2h}{v1}
\fmfforce{1w,1/2h}{v2}
\fmfforce{1/2w,1/2h}{v3}
\fmf{plain,left=1}{v2,v1,v2}
\fmfv{decor.size=0, label=$p+1$, l.dist=0mm, l.angle=0}{v3}
\end{fmfgraph*}\end{center}} \hspace*{0.1cm} = \hspace*{0.1cm} 4 
\hspace*{0.1cm}
\parbox{18mm}{\begin{center}
\begin{fmfgraph*}(15,10)
\setval
\fmfforce{1/3w,1/2h}{v1}
\fmfforce{1w,1/2h}{v2}
\fmfforce{2/3w,1/2h}{v3}
\fmfforce{0.413333333333w,0.8h}{v4}
\fmfforce{0.413333333333w,0.2h}{v5}
\fmfforce{0.533333333333w,0.95h}{v6}
\fmfforce{0.533333333333w,0.05h}{v7}
\fmfforce{0w,1/2h}{v8}
\fmf{plain,right=0.2}{v4,v8,v5}
\fmf{plain,right=0.4}{v6,v8}
\fmf{plain,right=0.4}{v8,v7}
\fmf{plain,left=1}{v2,v1,v2}
\fmfv{decor.size=0, label=$p$, l.dist=0mm, l.angle=0}{v3}
\fmfdot{v4,v5,v6,v7,v8}
\end{fmfgraph*}\end{center}}  \hspace*{0.1cm} + \hspace*{0.1cm} 12 
\hspace*{0.1cm}
\parbox{23mm}{\begin{center}
\begin{fmfgraph*}(20,10)
\setval
\fmfforce{1/2w,1/2h}{v1}
\fmfforce{1w,1/2h}{v2}
\fmfforce{3/4w,1/2h}{v3}
\fmfforce{0.56w,0.8h}{v4}
\fmfforce{0.56w,0.2h}{v5}
\fmfforce{1/4w,1/2h}{v8}
\fmfforce{0w,1/2h}{v9}
\fmf{plain,right=0.2}{v4,v8,v5}
\fmf{plain,right=1}{v8,v9,v8}
\fmf{plain,left=1}{v2,v1,v2}
\fmfv{decor.size=0, label=$p$, l.dist=0mm, l.angle=0}{v3}
\fmfdot{v4,v5,v8}
\end{fmfgraph*}\end{center}}
\hspace*{0.1cm}+\hspace*{0.1cm} 4\hspace*{0.1cm}\sum\limits_{q=1}^{p-1} 
\hspace*{0.1cm}\left( \begin{array}{c} p \\ q \end{array} \right) 
\hspace*{0.1cm}
\parbox{33mm}{\begin{center}
\begin{fmfgraph*}(30,10)
\setval
\fmfforce{0w,1/2h}{v1}
\fmfforce{1/3w,1/2h}{v2}
\fmfforce{1/2w,1/2h}{v3}
\fmfforce{2/3w,1/2h}{v4}
\fmfforce{1w,1/2h}{v5}
\fmfforce{1/6w,1/2h}{v6}
\fmfforce{5/6w,1/2h}{v7}
\fmfforce{0.2933333w,0.8h}{v8}
\fmfforce{0.2933333w,0.2h}{v9}
\fmfforce{0.7066666w,0.8h}{v10}
\fmfforce{0.7066666w,0.2h}{v11}
\fmfv{decor.size=0, label=$p-q$, l.dist=0mm, l.angle=0}{v6}
\fmfv{decor.size=0, label=$q$, l.dist=0mm, l.angle=0}{v7}
\fmf{plain,right=1}{v1,v2,v1}
\fmf{plain,right=1}{v4,v5,v4}
\fmf{plain,left=0.2}{v8,v3}
\fmf{plain,left=0.2}{v3,v9}
\fmf{plain,right=0.2}{v10,v3}
\fmf{plain,right=0.2}{v3,v11}
\fmfdot{v3,v8,v9,v10,v11}
\end{fmfgraph*}\end{center}} 
\, , \qquad p \ge 1 \, .
\eeq
To demonstrate the working of (\r{GRR}), we
calculate the connected vacuum diagrams up to five loops.
Applying the linear
operations to (\r{RR1}), we obtain immediately
\beq
\la{O1}
\dphi{W^{(1)}}{1}{2} \, = \, 6\,\,
\parbox{17mm}{\begin{center}
\begin{fmfgraph*}(8,5)
\setval
\fmfstraight
\fmfforce{0w,0h}{i1}
\fmfforce{0w,1h}{i2}
\fmfforce{3/8w,0.5h}{v1}
\fmfforce{1w,0.5h}{v2}
\fmf{plain}{i1,v1}
\fmf{plain}{v1,i2}
\fmf{plain,left}{v1,v2,v1}
\fmfdot{v1}
\fmfv{decor.size=0, label=${\scs 2}$, l.dist=1mm, l.angle=-180}{i1}
\fmfv{decor.size=0, label=${\scs 1}$, l.dist=1mm, l.angle=-180}{i2}
\end{fmfgraph*}
\end{center}}
\, , \quad 
\ddphi{W^{(1)}} = \, 6 
\parbox{24mm}{\begin{center}
\begin{fmfgraph*}(6,5)
\setval
\fmfstraight
\fmfleft{i2,i1}
\fmfright{o2,o1}
\fmf{plain}{i1,v1}
\fmf{plain}{v1,i2}
\fmf{plain}{v1,o1}
\fmf{plain}{o2,v1}
\fmfdot{v1}
\fmfv{decor.size=0, label=${\scs 1}$, l.dist=1mm, l.angle=-180}{i1}
\fmfv{decor.size=0, label=${\scs 2}$, l.dist=1mm, l.angle=-180}{i2}
\fmfv{decor.size=0, label=${\scs 3}$, l.dist=1mm, l.angle=0}{o1}
\fmfv{decor.size=0, label=${\scs 4}$, l.dist=1mm, l.angle=0}{o2}
\end{fmfgraph*}
\end{center}} \, .
\eeq
Inserted into (\r{GRR}), these lead to the three-loop vacuum diagrams
\beq
\la{O2}
W^{(2)} \, = \,  24 \,\,\,
\parbox{10.5mm}{\begin{center}
\begin{fmfgraph*}(7.5,5)
\setval
\fmfforce{0w,0.5h}{v1}
\fmfforce{1w,0.5h}{v2}
%\fmfleft{v1}
%\fmfright{v2}
\fmf{plain,left=1}{v1,v2,v1}
\fmf{plain,left=0.4}{v1,v2,v1}
\fmfdot{v1,v2}
\end{fmfgraph*}\end{center}} 
+ \, 72 \,\,\,
\parbox{20mm}{\begin{center}
\begin{fmfgraph*}(15,5)
\setval
\fmfleft{i1}
\fmfright{o1}
\fmf{plain,left=1}{i1,v1,i1}
\fmf{plain,left=1}{v1,v2,v1}
\fmf{plain,left=1}{o1,v2,o1}
\fmfdot{v1,v2}
\end{fmfgraph*}\end{center}} 
\, .
\eeq
Proceeding to the next order, we have to perform one- and 
two-line amputations on the vacuum graphs in (\r{O2}), leading to
\beq
\la{O31}
\dphi{W^{(2)}}{1}{2} = 
96\, 
\parbox{20mm}{\begin{center}
\begin{fmfgraph*}(10,5)
\setval
\fmfforce{0w,0.5h}{i1}
\fmfforce{1/4w,1/2h}{v1}
\fmfforce{3/4w,0.5h}{v2}
\fmfforce{1w,0.5h}{o1}
\fmf{plain}{i1,o1}
\fmf{plain,left=1}{v1,v2,v1}
\fmfdot{v1,v2}
\fmfv{decor.size=0, label=${\scs 1}$, l.dist=1mm, l.angle=-180}{i1}
\fmfv{decor.size=0, label=${\scs 2}$, l.dist=1mm, l.angle=0}{o1}
\end{fmfgraph*}
\end{center}}
+ \, 144 \,
\parbox{15mm}{\begin{center}
\begin{fmfgraph*}(5,10)
\setval
\fmfforce{0w,0h}{i1}
\fmfforce{0.5w,0h}{v1}
\fmfforce{1w,0h}{o1}
\fmfforce{0.5w,0.5h}{v2}
\fmfforce{0.5w,1h}{v3}
\fmf{plain}{i1,v1}
\fmf{plain}{v1,o1}
\fmf{plain,left=1}{v1,v2,v1}
\fmf{plain,left=1}{v2,v3,v2}
\fmfdot{v1,v2}
\fmfv{decor.size=0, label=${\scs 1}$, l.dist=1mm, l.angle=-180}{i1}
\fmfv{decor.size=0, label=${\scs 2}$, l.dist=1mm, l.angle=0}{o1}
\end{fmfgraph*}
\end{center}}
+ \, 144 \,
\parbox{20mm}{\begin{center}
\begin{fmfgraph*}(12,5)
\setval
\fmfforce{0w,0h}{i1}
\fmfforce{1/5w,0h}{v1}
\fmfforce{1/5w,1h}{v2}
\fmfforce{4/5w,0h}{v3}
\fmfforce{4/5w,1h}{v4}
\fmfforce{1w,0h}{o1}
\fmf{plain}{i1,o1}
\fmf{plain,left=1}{v1,v2,v1}
\fmf{plain,left=1}{v3,v4,v3}
\fmfdot{v1,v3}
\fmfv{decor.size=0, label=${\scs 1}$, l.dist=1mm, l.angle=-180}{i1}
\fmfv{decor.size=0, label=${\scs 2}$, l.dist=1mm, l.angle=0}{o1}
\end{fmfgraph*}
\end{center}}
\, ,
\eeq
and subsequently to
\beq
\ddphi{W^{(2)}} &= &\, 288 \, 
\parbox{20mm}{\begin{center}
\begin{fmfgraph*}(10,5)
\setval
\fmfforce{0w,0h}{i1}
\fmfforce{0w,1h}{i2}
\fmfforce{1w,0h}{o1}
\fmfforce{1w,1h}{o2}
\fmfforce{1/4w,1/2h}{v1}
\fmfforce{3/4w,1/2h}{v2}
\fmf{plain}{i1,v1}
\fmf{plain}{i2,v1}
\fmf{plain}{o1,v2}
\fmf{plain}{o2,v2}
\fmf{plain,left=1}{v1,v2,v1}
\fmfdot{v1,v2}
\fmfv{decor.size=0, label=${\scs 3}$, l.dist=1mm, l.angle=-180}{i1}
\fmfv{decor.size=0, label=${\scs 4}$, l.dist=1mm, l.angle=0}{o1}
\fmfv{decor.size=0, label=${\scs 1}$, l.dist=1mm, l.angle=-180}{i2}
\fmfv{decor.size=0, label=${\scs 2}$, l.dist=1mm, l.angle=0}{o2}
\end{fmfgraph*}
\end{center}}
+ \, 144 \, 
\parbox{20mm}{\begin{center}
\begin{fmfgraph*}(10,5)
\setval
\fmfforce{0w,0h}{i1}
\fmfforce{0w,1h}{i2}
\fmfforce{1w,0h}{o1}
\fmfforce{1w,1h}{o2}
\fmfforce{1/4w,1/2h}{v1}
\fmfforce{3/4w,1/2h}{v2}
\fmf{plain}{i1,v1}
\fmf{plain}{i2,v1}
\fmf{plain}{o1,v2}
\fmf{plain}{o2,v2}
\fmf{plain,left=1}{v1,v2,v1}
\fmfdot{v1,v2}
\fmfv{decor.size=0, label=${\scs 2}$, l.dist=1mm, l.angle=-180}{i1}
\fmfv{decor.size=0, label=${\scs 4}$, l.dist=1mm, l.angle=0}{o1}
\fmfv{decor.size=0, label=${\scs 1}$, l.dist=1mm, l.angle=-180}{i2}
\fmfv{decor.size=0, label=${\scs 3}$, l.dist=1mm, l.angle=0}{o2}
\end{fmfgraph*}
\end{center}}
+ \, 288 \,
\parbox{20mm}{\begin{center}
\begin{fmfgraph*}(10,5)
\setval
\fmfforce{0w,0.5h}{i1}
\fmfforce{1/4w,0h}{i2}
\fmfforce{1/4w,1h}{i3}
\fmfforce{1w,0.5h}{o1}
\fmfforce{1/4w,1/2h}{v1}
\fmfforce{3/4w,1/2h}{v2}
\fmfforce{3/4w,3/2h}{v3}
\fmf{plain}{i1,o1}
\fmf{plain}{i2,v1}
\fmf{plain}{i3,v1}
\fmf{plain,left=1}{v2,v3,v2}
\fmfdot{v1,v2}
\fmfv{decor.size=0, label=${\scs 2}$, l.dist=1mm, l.angle=-180}{i1}
\fmfv{decor.size=0, label=${\scs 3}$, l.dist=1mm, l.angle=-90}{i2}
\fmfv{decor.size=0, label=${\scs 1}$, l.dist=1mm, l.angle=90}{i3}
\fmfv{decor.size=0, label=${\scs 4}$, l.dist=1mm, l.angle=0}{o1}
\end{fmfgraph*}
\end{center}}
\no \\
\la{O32}
&&+ \, 144 \,
\parbox{20mm}{\begin{center}
\begin{fmfgraph*}(10,5)
\setval
\fmfforce{0w,0.5h}{i1}
\fmfforce{1/4w,0h}{i2}
\fmfforce{1/4w,1h}{i3}
\fmfforce{1w,0.5h}{o1}
\fmfforce{1/4w,1/2h}{v1}
\fmfforce{3/4w,1/2h}{v2}
\fmfforce{3/4w,3/2h}{v3}
\fmf{plain}{i1,o1}
\fmf{plain}{i2,v1}
\fmf{plain}{i3,v1}
\fmf{plain,left=1}{v2,v3,v2}
\fmfdot{v1,v2}
\fmfv{decor.size=0, label=${\scs 3}$, l.dist=1mm, l.angle=-180}{i1}
\fmfv{decor.size=0, label=${\scs 4}$, l.dist=1mm, l.angle=-90}{i2}
\fmfv{decor.size=0, label=${\scs 1}$, l.dist=1mm, l.angle=90}{i3}
\fmfv{decor.size=0, label=${\scs 2}$, l.dist=1mm, l.angle=0}{o1}
\end{fmfgraph*}
\end{center}}
+ \, 144 \,
\parbox{20mm}{\begin{center}
\begin{fmfgraph*}(10,5)
\setval
\fmfforce{0w,0.5h}{i1}
\fmfforce{1/4w,0h}{i2}
\fmfforce{1/4w,1h}{i3}
\fmfforce{1w,0.5h}{o1}
\fmfforce{1/4w,1/2h}{v1}
\fmfforce{3/4w,1/2h}{v2}
\fmfforce{3/4w,3/2h}{v3}
\fmf{plain}{i1,o1}
\fmf{plain}{i2,v1}
\fmf{plain}{i3,v1}
\fmf{plain,left=1}{v2,v3,v2}
\fmfdot{v1,v2}
\fmfv{decor.size=0, label=${\scs 1}$, l.dist=1mm, l.angle=-180}{i1}
\fmfv{decor.size=0, label=${\scs 3}$, l.dist=1mm, l.angle=-90}{i2}
\fmfv{decor.size=0, label=${\scs 2}$, l.dist=1mm, l.angle=90}{i3}
\fmfv{decor.size=0, label=${\scs 4}$, l.dist=1mm, l.angle=0}{o1}
\end{fmfgraph*}
\end{center}}
+ \, 144 \,
\parbox{15mm}{\begin{center}
\begin{fmfgraph*}(5,12.5)
\setval
\fmfforce{0w,1/10h}{i11}
\fmfforce{1/2w,1/10h}{v11}
\fmfforce{1/2w,1/2h}{v21}
\fmfforce{1w,1/10h}{o11}
\fmfforce{0w,7/10h}{i12}
\fmfforce{1/2w,7/10h}{v12}
\fmfforce{1/2w,11/10h}{v22}
\fmfforce{1w,7/10h}{o12}
\fmf{plain}{i11,o11}
\fmf{plain}{i12,o12}
\fmf{plain,left=1}{v11,v21,v11}
\fmf{plain,left=1}{v12,v22,v12}
\fmfdot{v11,v12}
\fmfv{decor.size=0, label=${\scs 2}$, l.dist=1mm, l.angle=-180}{i11}
\fmfv{decor.size=0, label=${\scs 4}$, l.dist=1mm, l.angle=0}{o11}
\fmfv{decor.size=0, label=${\scs 1}$, l.dist=1mm, l.angle=-180}{i12}
\fmfv{decor.size=0, label=${\scs 3}$, l.dist=1mm, l.angle=0}{o12}
\end{fmfgraph*}
\end{center}} \, .
\eeq
Inserting (\r{O31}) and (\r{O32}) into (\r{GRR}) and
taking into account (\r{O1}),
we find the connected vacuum diagrams of order $p=3$
with their multiplicities as shown in Table \r{P}. We observe
that the nonlinear operation in (\ref{GRR}) does not lead to 
topologically new diagrams. It only corrects the multiplicities of 
the diagrams
generated from the first two operations. This is true also in higher orders.
The connected vacuum diagrams of the 
subsequent order $p=4$ and their multiplicities are listed in
Table \r{P}.\\

As a cross-check we can also determine the total multiplicities $M^{(p)}$ 
of all connected vacuum diagrams contributing to $W^{(p)}$. 
To this end we recall that
each of the $M^{(p)}$ diagrams in $W^{(p)}$ consists of $2p$ lines. 
The amputation of one or two lines therefore leads to 
$2p M^{(p)}$ and $2p (2p -1) M^{(p)}$
diagrams with $2p-1$ and $2p-2$ lines, respectively. Considering only the
total multiplicities, the graphical recursion relations (\r{GRR}) reduce to
the form derived before in Ref.~\cite{Kleinert2}
\beq
\la{NRR}
M^{(p+1)} = 16 p ( p + 1 ) M^{(p)} + 16 \sum_{q = 1}^{p -1} \frac{p!}{(p-q-1)! 
(q-1)!} M^{(q)} M^{(p-q)} \, ;  \hspace*{1cm} p \ge 1 \,.
\eeq
These are solved starting with the initial value 
\begin{eqnarray}
M^{(1)}=3\, ,
\end{eqnarray}
leading to the total multiplicities
\beq
M^{(2)}  =  96 \, , \quad M^{(3)} = 9504 \, , \quad M^{(4)}
= 1880064 \, ,
\eeq
which agree with the results listed in Table \r{P}. In addition
we note that the next orders would contain
\beq
M^{(5)}  =  616108032 \, , \quad
M^{(6)} =  301093355520 \, , \quad
M^{(7)} = 205062331760640
\eeq
connected vacuum diagrams.
\section{Scalar $\phi^2 A$-Theory}
For the sake of generality, let us also
study the situation where the quartic interaction of the
$\phi^4$-theory is generated by a scalar field $A$ from a cubic 
$\phi^2 A$-interaction. The associated energy functional
\beq
\la{YU}
E [ \phi, A ] =
E^{(0)} [ \phi, A ] +
E^{({\rm int})} [ \phi, A ]
\eeq
decomposes into the free part
\beq
E^{(0)} [ \phi, A ] = \frac{1}{2} \int_{12} G^{-1}_{12} \,\phi_1 \phi_2
+ \frac{1}{2} \int_{12} H^{-1}_{12} \,A_1 A_2
\eeq
and the interaction
\beq
E^{({\rm int})} [ \phi, A ]= \frac{\sqrt{g}}{2} \int_{123} V_{123} \,
\phi_1 \phi_2 A_3 \, .
\eeq
Indeed, as the field $A$ appears only quadratically in (\r{YU}),
the functional integral for the partition function
\beq
\la{YPF}
Z = \int {\cal D} \,\phi {\cal D} A \, e^{- E [ \phi , A ]}
\eeq
can be exactly evaluated with respect to the field $A$, yielding
\beq
\la{EFF}
Z = \int {\cal D} \, e^{- E^{({\rm eff})} [ \phi ]}
\eeq
with the effective energy functional
\beq
\la{EFE}
E^{({\rm eff})} [ \phi ] = - \frac{1}{2} \mbox{Tr} \ln H^{-1}
+ \frac{1}{2} \int_{12} G_{12}^{-1} \,\phi_1 \phi_2 -
\frac{g}{8} \int_{123456} V_{125} V_{346} H_{56} \,\phi_1 
\phi_2 \phi_3 \phi_4 \, .
\eeq
Apart from a trivial shift due to the negative free energy of the field $A$,
the effective energy functional (\r{EFE}) coincides with one (\ref{EF})
of a $\phi^4$-theory with the quartic interaction
\beq
\la{REL}
V_{1234} = -3 \, \int_{56} V_{125} V_{346} H_{56} \, .
\eeq
If we supplement the previous Feynman rules (\r{PRO}), (\r{VE})
by the free correlation function of the field $A$
\beq
\label{vac08}
\parbox{20mm}{\centerline{
\begin{fmfgraph*}(7,3)
\setval
\fmfleft{v1}
\fmfright{v2}
\fmf{boson}{v1,v2}
\fmflabel{$1$}{v1}
\fmflabel{$2$}{v2}
\end{fmfgraph*}
}}&\equiv& \quad H_{12}
\eeq
and the cubic interaction
\beq
\label{vac09}
\parbox{20mm}{\centerline{
\begin{fmfgraph}(6,5)
\setval
\fmfleft{v2,v1}
\fmfright{v3}
\fmfforce{0.5w,0.5h}{v4}
\fmf{boson}{v4,v3}
\fmf{plain}{v4,v1}
\fmf{plain}{v2,v4}
\fmfdot{v4}
\end{fmfgraph}
}}&\equiv& \quad \int_{123}V_{123} \, ,
\eeq
the intimate relation (\r{REL})
between the $\phi^4$-theory and the $\phi^2 A$-
theory can be graphically illustrated by
\beq
\la{INTI}
\parbox{17mm}{\begin{center}
\begin{fmfgraph*}(6,5)
\setval
\fmfstraight
\fmfleft{i2,i1}
\fmfright{o2,o1}
\fmf{plain}{i1,v1}
\fmf{plain}{v1,i2}
\fmf{plain}{v1,o1}
\fmf{plain}{o2,v1}
\fmfdot{v1}
\end{fmfgraph*}
\end{center}}
= \, - \, 
\parbox{20mm}{\centerline{
\begin{fmfgraph}(14,5)
\setval
\fmfforce{0w,0h}{i1}
\fmfforce{0w,1h}{i2}
\fmfforce{1w,0h}{o1}
\fmfforce{1w,1h}{o2}
\fmfforce{3/14w,0.5h}{v1}
\fmfforce{11/14w,0.5h}{v2}
\fmf{boson}{v1,v2}
\fmf{plain}{i1,v1}
\fmf{plain}{i2,v1}
\fmf{plain}{o1,v2}
\fmf{plain}{o2,v2}
\fmfdot{v1,v2}
\end{fmfgraph}}} 
\, - \, 
\parbox{10mm}{\centerline{
\begin{fmfgraph}(5,14)
\setval
\fmfforce{0w,0h}{i1}
\fmfforce{1w,0h}{i2}
\fmfforce{0w,1h}{o1}
\fmfforce{1w,1h}{o2}
\fmfforce{1/2w,3/14h}{v1}
\fmfforce{1/2w,11/14h}{v2}
\fmf{boson}{v1,v2}
\fmf{plain}{i1,v1}
\fmf{plain}{i2,v1}
\fmf{plain}{o1,v2}
\fmf{plain}{o2,v2}
\fmfdot{v1,v2}
\end{fmfgraph}}}
\, - \, 
\parbox{15mm}{\centerline{
\begin{fmfgraph}(10,14)
\setval
\fmfforce{1/4w,1h}{i1}
\fmfforce{3/4w,1h}{i2}
\fmfforce{0w,0h}{o1}
\fmfforce{1w,0h}{o2}
\fmfforce{1/8w,1/6h}{v1}
\fmfforce{7/8w,1/6h}{v2}
\fmf{boson}{v1,v2}
\fmf{plain}{i1,o2}
\fmf{plain}{i2,o1}
\fmfdot{v1,v2}
\end{fmfgraph}}} \, .
\eeq
This corresponds to a photon exchange in the so-called $s$-, $t$- 
and $u$-channels
of Mandelstam's theory of the scattering matrix. Their infinite
repetitions yield the relevant forces in the Hartree, Fock and Bogoliubov
approximations of many-body physics.
In the following we analyze the $\phi^2 A$-theory along similar lines 
as before the
$\phi^4$-theory. 
\subsection{Perturbation Theory}
Expanding the exponential in the partition function (\r{YPF}) in powers
of the coupling constant $g$, the resulting perturbation series reads
\beq
\la{YP1}
Z = \sum_{p = 0}^{\infty} \frac{1}{(2 p)!}
\left( \frac{g}{4} \right)^p \int {\cal D} \phi
\, {\cal D} A \, \left( \int_{123456} V_{123} V_{456} \phi_1 \phi_2
\phi_4 \phi_5 A_3 A_6 \right)^p \, e^{- E^{(0)} [ \phi,A]} \, .
\eeq
Substituting the product of two fields $\phi$ or $A$ by a functional
derivative with respect to the kernels $G^{-1}$ or $H^{-1}$, we conclude
from (\r{YP1})
\beq
\la{YP2}
Z = \sum_{p = 0}^{\infty} \frac{(-2 g)^p}{(2p)!} \, \left( \int_{123456} 
V_{123} V_{456} \, \frac{\delta^3}{\delta G^{-1}_{12} \delta G^{-1}_{45}
\delta H^{-1}_{36}} \right)^p \, e^{W^{(0)}} \, ,
\eeq
where the zeroth order of the negative free energy reads
\beq
\la{YYPF}
W^{(0)} = - \frac{1}{2} \mbox{Tr} \ln G^{-1}
- \frac{1}{2} \mbox{Tr} \ln H^{-1} \equiv
\frac{1}{2}\;
\parbox{8mm}{\centerline{
\begin{fmfgraph}(5,5)
\setval
\fmfi{plain}{reverse fullcircle scaled 1w shifted (0.5w,0.5h)}
\end{fmfgraph}
}} + 
\frac{1}{2}\;
\parbox{8mm}{\centerline{
\begin{fmfgraph}(5,5)
\setval
\fmfi{boson}{reverse fullcircle scaled 1w shifted (0.5w,0.5h)}
\end{fmfgraph}
}}
\, .
\eeq
Inserting (\r{YYPF}) in (\r{YP2}), the first-order contribution to the 
negative free energy yields
\beq
\la{FOY1}
W^{(1)} = 
2 \, \int_{123456} V_{123} V_{456} \, H_{36} G_{14} G_{25} + 
 \int_{123456} V_{123} V_{456} \, H_{36} G_{12} G_{45} \, ,
\eeq
which corresponds to the Feynman diagrams
\beq
\la{FOY2}
W^{(1)} = 
2 \quad
\parbox{10mm}{\centerline{
\begin{fmfgraph}(7,5)
\setval
\fmfleft{v1}
\fmfright{v2}
\fmf{boson}{v1,v2}
\fmf{plain,left=0.7}{v2,v1}
\fmf{plain,left=0.7}{v1,v2}
\fmfdot{v1,v2}
\end{fmfgraph}
}} \,\, + \,\,
\parbox{19mm}{\centerline{
\begin{fmfgraph}(15,7)
\setval
\fmfforce{0.33w,0.5h}{v1}
\fmfforce{0.66w,0.5h}{v2}
\fmf{boson}{v1,v2}
\fmfi{plain}{reverse fullcircle scaled 0.33w shifted (0.165w,0.5h)}
\fmfi{plain}{fullcircle rotated 180 scaled 0.33w shifted (0.825w,0.5h)}
\fmfdot{v1,v2}
\end{fmfgraph}
}} \, .
\eeq
\subsection{Functional Differential Equation for $W = \ln Z$}
The derivation of a functional differential equation for the negative
free energy
$W$ requires the combination of two independent steps. Consider first the
identity
\beq
\la{F10}
\int {\cal D} \,\phi {\cal D} A \, \frac{\delta}{\delta \phi_1} \,
\left\{ \phi_2 \, e^{- E[\phi,A]} \right\} = 0 \, ,
\eeq
which immediately yields with the energy functional (\r{YU})
\beq
\la{F11}
\delta_{12} Z + 2 \int_3 G^{-1}_{13} \, \frac{\delta Z}{\delta G^{-1}_{23}}
+ 2 \sqrt{g} \,
\int_{34} V_{134} \, \frac{\delta \left\{\langle A_4 \rangle Z
\right\}}{\delta G^{-1}_{23}} = 0 \, ,
\eeq
where $\langle A \rangle$ denotes the expectation value of the field $A$.
In order to close the functional differential equation, we 
consider the second identity
\beq
\la{F12}
\int {\cal D} \,\phi {\cal D} A \, \frac{\delta}{\delta A_1} \,
e^{- E[\phi,A]}  = 0 \, ,
\eeq
which leads to
\beq
\la{F13}
\langle A_1 \rangle Z =  \sqrt{g} \,
\int_{234} V_{234} H_{14} \, \frac{\delta 
Z}{\delta G^{-1}_{23}} \, .
\eeq
Inserting (\r{F13}) in (\r{F11}), we result in the desired
functional differential equation for the negative free energy $W = \ln Z$:
\beq
\delta_{12} + 2\, \int_2 G^{-1}_{13} \, \frac{\delta W}{\delta G_{23}^{-1}}
= - 2 g \, \int_{34567} V_{134} V_{567} H_{47} 
\left\{ \frac{\delta^2 W}{\delta G^{-1}_{23} \delta G^{-1}_{56}} + 
\frac{\delta W}{\delta G^{-1}_{23}} 
\frac{\delta W}{\delta G^{-1}_{56}} \right\}\, .
\eeq
A subsequent separation (\r{DEC}) of the zeroth order (\r{YYPF})
leads to a functional differential equation for the interaction part of the
free energy $W^{({\rm int})}$:
\beq 
\la{D1}
\int_{12} G^{-1}_{12} \, \frac{\delta W^{({\rm int})}}{\delta G^{-1}_{12}}
&=& - \frac{g}{4} \, \int_{123456} V_{123} V_{456} H_{36} \left\{
G_{12} G_{45} + 2 G_{14} G_{25} \right\} 
+ g\,\int_{123456} V_{123} V_{456} G_{12} H_{36}
\, \frac{\delta W^{({\rm int})}}{\delta
G^{-1}_{45}} \no \\
&&-  g\,\int_{123456} V_{123} V_{456} H_{36}
\left\{ \frac{\delta^2 W^{({\rm int})}}{\delta G^{-1}_{12} \delta G^{-1}_{45}}
+ \frac{\delta W^{({\rm int})}}{\delta G^{-1}_{12}} \, 
\frac{\delta W^{({\rm int})}}{\delta G^{-1}_{45}} \right\} \, .
\eeq
Taking into account the functional chain rules (\r{NR1}), (\r{NR2}),
the functional derivatives with respect to $G^{-1}$ in (\r{D1}) can be
rewritten in terms of $G$:
\beq 
\la{D2}
\int_{12} G_{12} \, \frac{\delta W^{({\rm int})}}{\delta G_{12}}
= \frac{g}{4} \, \int_{123456} V_{123} V_{456} H_{36} \left\{
G_{12} G_{45}
+ 2 G_{14} G_{25} \right\}
+ g\,\int_{123456} V_{123} V_{456} H_{36} \, \left\{ G_{12} G_{47} G_{58}
\no \right.
\eeq
\beq
\left. + 2 G_{14} G_{27} G_{58} \right\} \, 
\frac{\delta W^{({\rm int})}}{\delta G_{78}} + g\,
\int_{123456789\bar{1}} V_{123} V_{456} H_{36} G_{17} G_{28} 
G_{39} G_{4\bar{1}} \, 
\left\{ \frac{\delta^2 W^{({\rm int})}}{\delta G_{78} \delta G_{9\bar{1}}}
+ \frac{\delta W^{({\rm int})}}{\delta G_{78}} \, 
\frac{\delta W^{({\rm int})}}{\delta G_{9\bar{1}}} 
\right\} \, .
\eeq
\subsection{Recursion Relation And Graphical Solution}
The functional differential
equation (\r{D2}) is now solved by the power series
\beq
\la{EXPN}
W^{({\rm int})} = \sum_{p=1}^{\infty} \frac{1}{(2p)!}
\, \left( \frac{g}{4} \right)^p \, W^{(p)} \, .
\eeq
Using the property (\r{EVP}) that the coefficients $W^{(p)}$ satisfy the
eigenvalue condition of the operator (\r{OO}), we obtain both the
recursion relation
\beq
W^{(p+1)} &= &4 ( 2 p + 1 ) \left\{
\int_{12345678} V_{123} V_{456} \, 
H_{36} \left( G_{12} G_{47} G_{58} + 2 G_{14} G_{27} G_{58} \right)
\, \frac{\delta W^{(p)}}{\delta G_{78}} \right. \no
\eeq
\beq
\la{RRA}
\left. 
+ \int_{123456789\bar{1}} V_{123} V_{456} \,
H_{36} G_{17} G_{28} G_{39} G_{4\bar{1}} \, \left(
\frac{\delta^2 W^{(p)}}{\delta G_{78} \delta  G_{9\bar{1}}} 
+ \sum_{q = 1}^{p-1} \left( \begin{array}{@{}c}
2 p \\ 2 q \end{array} \right) 
\frac{\delta W^{(p-q)}}{\delta G_{78}} 
\, \frac{\delta W^{(q)}}{\delta G_{9\bar{1}}} \right)  \right\}
\eeq
and the initial value (\r{FOY1}). Using the Feynman rules (\r{PRO}),
(\r{vac08}) and (\r{vac09}), the recursion relation (\r{RRA}) reads
graphically
\beq
W^{(p+1)} &= &4 ( 2 p + 1 ) \Bigg\{
\quad\dphi{W^{(p)}}{1}{2} \quad
\parbox{20mm}{\begin{center}
\begin{fmfgraph*}(16,5)
\setval
\fmfstraight
\fmfforce{0w,0h}{i1}
\fmfforce{0w,1h}{i2}
\fmfforce{3/16w,0.5h}{v1}
\fmfforce{11/16w,0.5h}{v2}
\fmfforce{1w,0.5h}{v3}
\fmf{plain}{i1,v1}
\fmf{plain}{v1,i2}
\fmf{boson}{v1,v2}
\fmf{plain,left}{v3,v2,v3}
\fmfdot{v1,v2}
\fmfv{decor.size=0, label=${\scs 2}$, l.dist=1mm, l.angle=-180}{i1}
\fmfv{decor.size=0, label=${\scs 1}$, l.dist=1mm, l.angle=-180}{i2}
\end{fmfgraph*}
\end{center}}
+ 2 
\quad\dphi{W^{(p)}}{1}{2}\quad
\parbox{11mm}{\begin{center}
\begin{fmfgraph*}(7,8)
\setval
\fmfstraight
\fmfforce{0w,0h}{i1}
\fmfforce{0w,1h}{i2}
\fmfforce{3/7w,0h}{v1}
\fmfforce{3/7w,1h}{v2}
\fmf{plain}{i1,v1}
\fmf{plain}{i2,v2}
\fmf{boson}{v1,v2}
\fmf{plain,right=1}{v1,v2}
\fmfdot{v1,v2}
\fmfv{decor.size=0, label=${\scs 2}$, l.dist=1mm, l.angle=-180}{i1}
\fmfv{decor.size=0, label=${\scs 1}$, l.dist=1mm, l.angle=-180}{i2}
\end{fmfgraph*}
\end{center}} 
\no \\
\la{RRG} 
&&+ \quad\ddphi{W^{(p)}}\quad
\parbox{7mm}{\begin{center}
\begin{fmfgraph*}(3,12)
\setval
\fmfstraight
\fmfforce{0w,0h}{i1}
\fmfforce{0w,1/4h}{i2}
\fmfforce{0w,3/4h}{i3}
\fmfforce{0w,1h}{i4}
\fmfforce{1w,1/6h}{o1}
\fmfforce{1w,5/6h}{o2}
\fmf{plain}{o1,i1}
\fmf{plain}{o1,i2}
\fmf{plain}{i3,o2}
\fmf{plain}{i4,o2}
\fmf{boson}{o1,o2}
\fmfdot{o1,o2}
\fmfv{decor.size=0, label=${\scs 4}$, l.dist=1mm, l.angle=-180}{i1}
\fmfv{decor.size=0, label=${\scs 3}$, l.dist=1mm, l.angle=-180}{i2}
\fmfv{decor.size=0, label=${\scs 2}$, l.dist=1mm, l.angle=-180}{i3}
\fmfv{decor.size=0, label=${\scs 1}$, l.dist=1mm, l.angle=-180}{i4}
\end{fmfgraph*}
\end{center}}
\,\,+\,\,\quad\sum\limits_{q=1}^{p-1} \quad 
\left( \begin{array}{c} 2p \\ 2q \end{array} \right) 
\quad\dphi{W^{(p-q)}}{1}{2}\quad
\parbox{18mm}{\centerline{
\begin{fmfgraph*}(14,5)
\setval
\fmfforce{0w,0h}{i1}
\fmfforce{0w,1h}{i2}
\fmfforce{1w,0h}{o1}
\fmfforce{1w,1h}{o2}
\fmfforce{3/14w,0.5h}{v1}
\fmfforce{11/14w,0.5h}{v2}
\fmf{boson}{v1,v2}
\fmf{plain}{i1,v1}
\fmf{plain}{i2,v1}
\fmf{plain}{o1,v2}
\fmf{plain}{o2,v2}
\fmfdot{v1,v2}
\fmfv{decor.size=0, label=${\scs 1}$, l.dist=1mm, l.angle=-180}{i1}
\fmfv{decor.size=0, label=${\scs 2}$, l.dist=1mm, l.angle=-180}{i2}
\fmfv{decor.size=0, label=${\scs 3}$, l.dist=1mm, l.angle=0}{o1}
\fmfv{decor.size=0, label=${\scs 4}$, l.dist=1mm, l.angle=0}{o2}
\end{fmfgraph*}
}} 
\quad\dphi{W^{(q)}}{3}{4} \Bigg\} \, , \qquad p \ge 1 \, ,
\eeq
which is iterated starting from (\r{FOY2}). In analogy to (\r{graphrec}),
this recursion relation can be cast in a closed diagrammatic way
by using the alternative graphical rules (\r{ALTER}):
\beq
\parbox{13mm}{\begin{center}
\begin{fmfgraph*}(10,10)
\setval
\fmfforce{0w,1/2h}{v1}
\fmfforce{1w,1/2h}{v2}
\fmfforce{1/2w,1/2h}{v3}
\fmf{plain,left=1}{v2,v1,v2}
\fmfv{decor.size=0, label=$p+1$, l.dist=0mm, l.angle=0}{v3}
\end{fmfgraph*}\end{center}} \hspace*{0.1cm} 
&=&\hspace*{1mm}4 ( 2 p + 1 ) \hspace*{1mm}\Bigg\{
\parbox{31mm}{\begin{center}
\begin{fmfgraph*}(28,10)
\setval
\fmfforce{0w,1/2h}{v1}
\fmfforce{5/28w,1/2h}{v2}
\fmfforce{13/28w,1/2h}{v3}
\fmfforce{18/28w,1/2h}{v4}
\fmfforce{1w,1/2h}{v5}
\fmfforce{19.2/28w,0.8h}{v6}
\fmfforce{19.2/28w,0.2h}{v7}
\fmfforce{23/28w,0.5h}{v8}
\fmf{plain,left=1}{v2,v1,v2}
\fmf{boson}{v2,v3}
\fmf{plain,left=1}{v4,v5,v4}
\fmf{plain,left=0.2}{v3,v6}
\fmf{plain,right=0.2}{v3,v7}
\fmfv{decor.size=0, label=$p$, l.dist=0mm, l.angle=0}{v8}
\fmfdot{v2,v3,v6,v7}
\end{fmfgraph*}\end{center}} \hspace*{1mm} + \hspace*{1mm} 2\hspace*{1mm}
\parbox{22mm}{\begin{center}
\begin{fmfgraph*}(19,10)
\setval
\fmfforce{4/19w,0.9h}{v1}
\fmfforce{4/19w,0.1h}{v2}
\fmfforce{9/19w,0.5h}{v3}
\fmfforce{1w,0.5h}{v4}
\fmfforce{10.2/19w,0.8h}{v6}
\fmfforce{10.2/19w,0.2h}{v7}
\fmfforce{14/19w,0.5h}{v8}
\fmf{boson}{v1,v2}
\fmf{plain,right=1}{v1,v2}
\fmf{plain,left=1}{v4,v3,v4}
\fmf{plain,left=0.2}{v1,v6}
\fmf{plain,right=0.2}{v2,v7}
\fmfv{decor.size=0, label=$p$, l.dist=0mm, l.angle=0}{v8}
\fmfdot{v2,v1,v6,v7}
\end{fmfgraph*}\end{center}} 
\no \\ && + 
\parbox{18mm}{\begin{center}
\begin{fmfgraph*}(15,10)
\setval
\fmfforce{0w,0.9h}{v1}
\fmfforce{0w,0.1h}{v2}
\fmfforce{5/15w,0.5h}{v3}
\fmfforce{1w,0.5h}{v4}
\fmfforce{6.2/15w,0.8h}{v6}
\fmfforce{6.2/15w,0.2h}{v7}
\fmfforce{10/15w,0.5h}{v8}
\fmfforce{8/15w,0.95h}{v9}
\fmfforce{8/15w,0.05h}{v10}
\fmf{boson}{v1,v2}
\fmf{plain,left=1}{v4,v3,v4}
\fmf{plain,left=0.2}{v1,v6}
\fmf{plain,right=0.2}{v2,v7}
\fmf{plain,left=0.4}{v1,v9}
\fmf{plain,right=0.4}{v2,v10}
\fmfv{decor.size=0, label=$p$, l.dist=0mm, l.angle=0}{v8}
\fmfdot{v2,v1,v6,v7,v9,v10}
\end{fmfgraph*}\end{center}} 
\,\,+\,\,\quad\sum\limits_{q=1}^{p-1} \quad 
\left( \begin{array}{c} 2p \\ 2q \end{array} \right) 
\parbox{41mm}{\begin{center}
\begin{fmfgraph*}(38,10)
\setval
\fmfforce{0w,0.5h}{v1}
\fmfforce{10/38w,0.5h}{v2}
\fmfforce{15/38w,0.5h}{v3}
\fmfforce{23/38w,0.5h}{v4}
\fmfforce{28/38w,0.5h}{v5}
\fmfforce{1w,0.5h}{v6}
\fmfforce{33/38w,0.5h}{v71}
\fmfforce{5/38w,0.5h}{v70}
\fmfforce{8.8/38w,0.8h}{v8}
\fmfforce{8.8/38w,0.2h}{v9}
\fmfforce{29.2/38w,0.8h}{v10}
\fmfforce{29.2/38w,0.2h}{v11}
\fmf{plain,left=1}{v1,v2,v1}
\fmf{plain,left=1}{v5,v6,v5}
\fmf{plain,left=0.2}{v8,v3}
\fmf{plain,right=0.2}{v9,v3}
\fmf{boson}{v3,v4}
\fmf{plain,right=0.2}{v10,v4}
\fmf{plain,left=0.2}{v11,v4}
\fmfv{decor.size=0, label=$p-q$, l.dist=0mm, l.angle=0}{v70}
\fmfv{decor.size=0, label=$q$, l.dist=0mm, l.angle=0}{v71}
\fmfdot{v3,v4,v8,v9,v10,v11}
\end{fmfgraph*}\end{center}} 
\, \Bigg\} \, , \qquad p \ge 1 \, ,
\eeq
We illustrate the procedure of solving the recursion relation (\r{RRG}) by
construction the three-loop vacuum diagrams. Applying one or two
functional derivatives to (\r{FOY2}), we have
\beq
\dphi{W^{(1)}}{1}{2} = 2 \quad
\parbox{20mm}{\begin{center}
\begin{fmfgraph*}(16,5)
\setval
\fmfstraight
\fmfforce{0w,0h}{i1}
\fmfforce{0w,1h}{i2}
\fmfforce{3/16w,0.5h}{v1}
\fmfforce{11/16w,0.5h}{v2}
\fmfforce{1w,0.5h}{v3}
\fmf{plain}{i1,v1}
\fmf{plain}{v1,i2}
\fmf{boson}{v1,v2}
\fmf{plain,left}{v3,v2,v3}
\fmfdot{v1,v2}
\fmfv{decor.size=0, label=${\scs 2}$, l.dist=1mm, l.angle=-180}{i1}
\fmfv{decor.size=0, label=${\scs 1}$, l.dist=1mm, l.angle=-180}{i2}
\end{fmfgraph*}
\end{center}}
+\,\, 4 \,\,
\parbox{11mm}{\begin{center}
\begin{fmfgraph*}(7,8)
\setval
\fmfstraight
\fmfforce{0w,0h}{i1}
\fmfforce{0w,1h}{i2}
\fmfforce{3/7w,0h}{v1}
\fmfforce{3/7w,1h}{v2}
\fmf{plain}{i1,v1}
\fmf{plain}{i2,v2}
\fmf{boson}{v1,v2}
\fmf{plain,right=1}{v1,v2}
\fmfdot{v1,v2}
\fmfv{decor.size=0, label=${\scs 2}$, l.dist=1mm, l.angle=-180}{i1}
\fmfv{decor.size=0, label=${\scs 1}$, l.dist=1mm, l.angle=-180}{i2}
\end{fmfgraph*}
\end{center}} 
\, , \quad 
\ddphi{W^{(p)}} \, = \,\,2 \quad
\parbox{18mm}{\centerline{
\begin{fmfgraph*}(14,5)
\setval
\fmfforce{0w,0h}{i1}
\fmfforce{0w,1h}{i2}
\fmfforce{1w,0h}{o1}
\fmfforce{1w,1h}{o2}
\fmfforce{3/14w,0.5h}{v1}
\fmfforce{11/14w,0.5h}{v2}
\fmf{boson}{v1,v2}
\fmf{plain}{i1,v1}
\fmf{plain}{i2,v1}
\fmf{plain}{o1,v2}
\fmf{plain}{o2,v2}
\fmfdot{v1,v2}
\fmfv{decor.size=0, label=${\scs 1}$, l.dist=1mm, l.angle=-180}{i1}
\fmfv{decor.size=0, label=${\scs 2}$, l.dist=1mm, l.angle=-180}{i2}
\fmfv{decor.size=0, label=${\scs 3}$, l.dist=1mm, l.angle=0}{o1}
\fmfv{decor.size=0, label=${\scs 4}$, l.dist=1mm, l.angle=0}{o2}
\end{fmfgraph*}
}}
\quad + \,\,4 \quad\quad
\parbox{18mm}{\centerline{
\begin{fmfgraph*}(14,5)
\setval
\fmfforce{0w,0h}{i1}
\fmfforce{0w,1h}{i2}
\fmfforce{1w,0h}{o1}
\fmfforce{1w,1h}{o2}
\fmfforce{3/14w,0.5h}{v1}
\fmfforce{11/14w,0.5h}{v2}
\fmf{boson}{v1,v2}
\fmf{plain}{i1,v1}
\fmf{plain}{i2,v1}
\fmf{plain}{o1,v2}
\fmf{plain}{o2,v2}
\fmfv{decor.size=0, label=${\scs 3}$, l.dist=1mm, l.angle=-180}{i1}
\fmfv{decor.size=0, label=${\scs 1}$, l.dist=1mm, l.angle=-180}{i2}
\fmfv{decor.size=0, label=${\scs 4}$, l.dist=1mm, l.angle=0}{o1}
\fmfv{decor.size=0, label=${\scs 2}$, l.dist=1mm, l.angle=0}{o2}
\fmfdot{v1,v2}
\end{fmfgraph*}}} \,\,\,\, .
\eeq
This is inserted into (\r{RRG}) to yield the three-loop diagrams shown
in Table \r{Y} with their multiplicities. The table also contains
the subsequent 4-loop results which we shall not derive here in
detail. Observe that
the multiplicity of a connected vacuum diagram ${\rm D}$
in the $\phi^2 A$-theory
is given by a formula similar to (\r{MU1}) in the $\phi^4$-theory:
\beq 
M_{\phi^2 A}^{\rm D} = \frac{(2p)! 4^p}{2!^{S+D}N} \, . 
\eeq
Here $S$ and $D$ denote the number of self- and double connections, whereas
$N$ represents the number of identical vertex permutations.\\

The  connected vacuum diagrams of the 
$\phi^2 A$-theory in Table \r{Y}
can, of course, be converted to corresponding ones of the 
$\phi^4$-theory in Table \r{P}, by shrinking wiggly lines to a point and
dividing the resulting multiplicity by 3 in accordance with 
(\r{INTI}). This relation between connected
vacuum diagrams in $\phi^4$- and $\phi^2 A$-theory is emphasized by the
numbering used in Table \r{Y}. For instance, the shrinking
converts the five diagrams \#4.1-\#4.5 in Table \r{Y} to the diagram
\#4 in Table \r{P}.
Taking into account the different combinatorial factors in
the expansion (\r{EX}) and (\r{EXPN}) as well as the factor 3 in the 
shrinkage (\r{INTI}), the multiplicity $M_{\phi^4}^{E=0}$ 
of a $\phi^4$-diagram
results from the corresponding one $M_{\phi^2 A}^{E=0}$ of the 
$\phi^2 A$-partner diagrams via the rule
\beq
M_{\phi^4}^{E=0} = \frac{1}{(2p-1)!!}\, M_{\phi^2 A}^{E=0} \, .
\eeq
\section{Computer Generation of Diagrams}
Continuing the solution of the graphical recursion relations (\r{GRR}) 
and (\r{RRG}) to higher loops becomes an arduous task. We therefore
automatize the procedure by computer algebra. Here we restrict
ourselves to the $\phi^4$-theory because of its relevance for critical 
phenomena.
\subsection{Matrix Representation of Diagrams}
\label{matrepdiag}
To implement the procedure on a computer
we must represent Feynman diagrams in the $\phi^4$-theory
by algebraic symbols. For this we 
use matrices as defined in Refs.~\ci{Neu,Verena}. 
Let $p$ be the number of vertices of a given diagram and label them by
indices from $1$ to $p$.
Set up a matrix $\fullm$ whose elements $M_{ij}$ ($0\leq i,j\leq p$)
specify the number of lines joining the vertices $i$ and $j$.
The diagonal elements $M_{ii}$ ($i>0$) count the number of self-connections of
the $i$th vertex. External lines of a diagram are labeled as if they were
connected to a single additional dummy vertex with number $0$. 
As matrix element $M_{00}$ is set to zero as a convention.
The off-diagonal elements lie in the interval $0\leq M_{ij}\leq 4$, while
the diagonal elements for $i>0$ are restricted by
$0\leq M_{ii}\leq 2$. We observe that the sum of
the matrix elements $M_{ij}$ in each row or column equals 4, where the 
diagonal elements count twice, 

\beq
\sum_{j=0}^p M_{ij} + M_{ii} = 
\sum_{i=0}^p M_{ij} + M_{jj} = 4 \, .
\eeq
The matrix $\fullm$ is symmetric
and is thus specified by $(p+1)(p+2)/2-1$ elements.
Each matrix characterizes a unique diagram and determines its multiplicity
via formula (\r{MU4}). From the matrix $\fullm$ we read offdirectly 
the number of self-, double, triple,
fourfold connections $S,D,T,F$ and the number of external legs
$E = \sum_{i=1}^p M_{0i}$. It also permits us to calculate the number $N$
of identical vertex permutations. For this we observe
that the matrix $\fullm$ is not unique,
since so far the vertex numbering is arbitrary.
In fact, $N$ is the number of permutations of vertices and external lines
which leave the matrix $\fullm$ unchanged [compare to the statement after
(\ref{MU4})].
If $n_M$ denotes the number of different matrices representing the same 
diagram, the number $N$ is given by
\beq
\la{N}
N=\frac{p!}{n_M}\,\prod_{i=1}^p M_{0i} ! \, .
\eeq
The matrix elements
$M_{0i}$ count the number of external legs connected to the $i$th
vertex. Inserting Eq.~(\r{N}) into the formula (\r{MU4}) for $E=0$, we obtain
the multiplicity of the diagram represented by $\fullm$. This may be used to
cross-check the multiplicities obtained before when
solving the recursion relation (\r{GRR}).\\

As an example, consider the following diagram of the four-point function
with $p=3$ vertices:
\beq
\la{BD}
\parbox{13mm}{\begin{center}
\begin{fmfgraph}(10,5)
\setval
\fmfstraight
\fmfforce{0w,1/2h}{i1}
\fmfforce{1w,1/2h}{o1}
\fmfforce{1/4w,1/2h}{v1}
\fmfforce{3/4w,1/2h}{v2}
\fmfforce{1/2w,1h}{v3}
\fmfforce{0.75w,1.5h}{i2}
\fmfforce{0.25w,1.5h}{o2}
\fmf{plain}{i1,o1}
\fmf{plain}{i2,v3}
\fmf{plain}{o2,v3}
\fmf{plain,left=1}{v1,v2,v1}
\fmfdot{v1,v2,v3}
\end{fmfgraph}
\end{center}} \, .
\eeq
This diagram can be represented  by altogether  $n_M = 3$ different matrices,
depending on the labeling of the top 
vertex with two external legs by $1$, $2$, or $3$:
\beq
\la{UM}
\begin{array}{@{}c|cccc}
  & 0 & 1 & 2 & 3 \\ \hline
0 & 0 & 2 & 1 & 1 \\
1 & 2 & 0 & 1 & 1 \\
2 & 1 & 1 & 0 & 2 \\
3 & 1 & 1 & 2 & 0 
\end{array}
\hspace*{2cm}
\begin{array}{@{}c|cccc}
  & 0 & 1 & 2 & 3 \\ \hline
0 & 0 & 1 & 2 & 1 \\
1 & 1 & 0 & 1 & 2 \\
2 & 2 & 1 & 0 & 1 \\
3 & 1 & 2 & 1 & 0 
\end{array}
\hspace*{2cm}
\begin{array}{@{}c|cccc}
  & 0 & 1 & 2 & 3 \\ \hline
0 & 0 & 1 & 1 & 2 \\
1 & 1 & 0 & 2 & 1 \\
2 & 1 & 2 & 0 & 1 \\
3 & 2 & 1 & 1 & 0 
\end{array}
\, .
\eeq
>From the zeroth row or column of these matrices or by inspecting the 
diagram (\r{BD}), we read off that there exist two one-connections and
one two-connection between external legs and vertices. Thus 
Eq.~(\r{N}) states
that the number of identical vertex permutations of the
diagram (\r{BD}) is 4 (compare the corresponding entry in Table \r{FOUR}).\\

The matrix $\fullm$ contains of course all information on the
topological properties of a diagram \ci{Neu,Verena}. For this we
define the submatrix $\tilde{\fullm}$
by removing the zeroth row and column from $\fullm$. This allows us to
recognize the connectedness of a diagram:
A diagram is disconnected if there is a vertex numbering for which
$\tilde{\fullm}$ is a block matrix.
Furthermore a vertex is a cutvertex, e.g. a vertex which links two 
otherwise disconnected
parts of a diagram, if the matrix $\tilde{\fullm}$ has an almost block-form
for an appropriate numbering of vertices in which
the blocks overlap only on some diagonal element $\tilde{M}_{ii}$, i.e.
the matrix $\tilde{\fullm}$ takes block form if the $i$th row and
column are removed.
Similarly, the matrix $\tilde{\fullm}$ allows us to recognize a  
one-particle-reducible diagram, which falls into two pieces
by cutting a certain line.
Removing a line amounts to
reducing the associated matrix elements in the submatrix $\tilde{\fullm}$
by one unit. If the resulting matrix
$\tilde{\fullm}$ has block form, the diagram is one-particle-reducible.\\ 

So far, the vertex numbering has been arbitrary, making the matrix 
representation of a diagram non-unique. To achieve uniqueness, we 
proceed as follows. We associate with each matrix a number whose digits
are composed of the matrix elements $M_{ij}$ ($0\leq j \leq i\leq p$), 
i.e. we form the number with the $(p+1)(p+2)/2-1$ elements
\beq
M_{10} M_{11} M_{20} M_{21} M_{22} M_{30} M_{31} M_{32} M_{33} \,
\ldots M_{pp} \, .
\eeq
The smallest of these numbers is chosen to represent the diagram unique.
For instance, the three matrices (\r{UM}) carry the numbers
\beq
\label{UN}
201101120 \, , \hspace*{2cm} 
102101210 \, , \hspace*{2cm}
101202110 \, ,
\eeq
the smallest one being the last.
Thus we uniquely represent the diagram (\r{BD}) by this number.
\subsection{Practical Generation}
We are now prepared for the computer
generation of Feynman diagrams.
First the vacuum diagrams are generated from the recursion relation
(\r{GRR}). From these the diagrams of the
connected two- and four-point functions are obtained
by cutting or removing lines. A MATHEMATICA program performs this task.
The resulting unique matrix representations of the diagrams up
to the order $p=4$ are listed in Table \r{DATA1}--\r{DATA3}.
They are the same as those derived before by hand in Table \r{P}--\r{FOUR}. 
Higher-order results up to $p=6$, containing all diagrams which are
relevant for the five loop renormalization of the $\phi^4$-theory
in $d=4-\epsilon$ dimensions \ci{Verena,FIVE}, are made available on
internet \ci{CODE}, where also
the program can be found. 
\subsubsection{Connected Vacuum Diagrams}
\la{vac}
The computer solution of the recursion relation (\r{GRR}) necessitates
to keep an exact record of the labeling of external legs of intermediate
diagrams which arise from differentiating a vacuum diagram with respect to
a line once or twice. To this end we have to extend our previous matrix
representation of diagrams where the external legs are labeled as if they
were connected to a simple additional vertex with number 0. For each matrix
representing a diagram we define an associated vector which contains the
labels of the external legs connected to each vertex. This vector has the
length of the dimension of the matrix and will be prepended  to the 
matrix. By doing so, it is understood that the rows and columns of the
matrix are labeled from 0 to the number of vertices as explained in
Subsection IV.A, so that we may omit these labels from now on.
Consider, as an example, the diagram (\r{BD}) of the four-point function
with $p=3$ vertices, where the spatial indices $1,2,3,4$ are assigned in a
particular order:\\

\beq
\la{BDEX}
\parbox{13mm}{\begin{center}
\begin{fmfgraph*}(10,5)
\setval
\fmfstraight
\fmfforce{0w,1/2h}{i1}
\fmfforce{1w,1/2h}{o1}
\fmfforce{1/4w,1/2h}{v1}
\fmfforce{3/4w,1/2h}{v2}
\fmfforce{1/2w,1h}{v3}
\fmfforce{0.75w,1.5h}{i2}
\fmfforce{0.25w,1.5h}{o2}
\fmf{plain}{i1,o1}
\fmf{plain}{i2,v3}
\fmf{plain}{o2,v3}
\fmf{plain,left=1}{v1,v2,v1}
\fmfv{decor.size=0, label=${\scs 2}$, l.dist=1.2mm, l.angle=90}{i2}
\fmfv{decor.size=0, label=${\scs 1}$, l.dist=1.2mm, l.angle=90}{o2}
\fmfv{decor.size=0, label=${\scs 3}$, l.dist=1mm, l.angle=-180}{i1}
\fmfv{decor.size=0, label=${\scs 4}$, l.dist=1mm, l.angle=0}{o1}
\fmfdot{v1,v2,v3}
\end{fmfgraph*}
\end{center}} \quad .
\eeq
In our extended matrix notation, such a diagram can be represented in total
by six matrices:
\beq
\left(\begin{array}{c}\{\}\\\{1,2\}\\\{3\}\\\{4\}\end{array}\left|
\begin{array}{cccc}
0&2&1&1\\
2&0&1&1\\
1&1&0&2\\
1&1&2&0
\end{array}\right)\right. \, , \quad
\left(\begin{array}{c}\{\}\\\{3\}\\\{1,2\}\\\{4\}\end{array}\left|
\begin{array}{cccc}
0&1&2&1\\
1&0&1&2\\
2&1&0&1\\
1&2&1&0
\end{array}\right)\right. \, , \quad
\left(\begin{array}{c}\{\}\\\{3\}\\\{4\}\\\{1,2\}\end{array}\left|
\begin{array}{cccc}
0&1&1&2\\
1&0&2&1\\
1&2&0&1\\
2&1&1&0
\end{array}\right)\right. \, , \quad \no\\ \hspace*{-1cm}
\left(\begin{array}{c}\{\}\\\{1,2\}\\\{4\}\\\{3\}\end{array}\left|
\begin{array}{cccc}
0&2&1&1\\
2&0&1&1\\
1&1&0&2\\
1&1&2&0
\end{array}\right)\right. \, , \quad
\left(\begin{array}{c}\{\}\\\{4\}\\\{1,2\}\\\{3\}\end{array}\left|
\begin{array}{cccc}
0&1&2&1\\
1&0&1&2\\
2&1&0&1\\
1&2&1&0
\end{array}\right)\right. \, , \quad
\left(\begin{array}{c}\{\}\\\{4\}\\\{3\}\\\{1,2\}\end{array}\left|
\begin{array}{cccc}
0&1&1&2\\
1&0&2&1\\
1&2&0&1\\
2&1&1&0
\end{array}\right)\right. \,.
\eeq
In the calculation of the vacuum diagrams from the recursion relation 
(\r{GRR}), starting from the two-loop diagram (\ref{STTT}), we have
to represent three different elementary operations in our extended matrix
notation:
\begin{enumerate}
\item 
Taking one or two derivatives of a vacuum diagram
with respect to a line. For example, we apply
this operation to the vacuum diagram \#2 in Table \ref{P}
\beq
\frac{\delta^2}{\delta G_{12}\delta G_{34}} \hspace*{1mm}
\parbox{10.5mm}{\begin{center}
\begin{fmfgraph*}(7.5,5)
\setval
\fmfforce{0w,0.5h}{v1}
\fmfforce{1w,0.5h}{v2}
\fmf{plain,left=1}{v1,v2,v1}
\fmf{plain,left=0.4}{v1,v2,v1}
\fmfdot{v1,v2}
\end{fmfgraph*}\end{center}} = 2 \, \frac{\delta}{\delta G_{12}} \,
\left[
\parbox{18mm}{\begin{center}
\begin{fmfgraph*}(5,5)
\setval
\fmfforce{-1/2w,0.5h}{i1}
\fmfforce{0w,0.5h}{v1}
\fmfforce{1w,0.5h}{v2}
\fmfforce{1.5w,0.5h}{o1}
\fmfforce{0.5w,0h}{v3}
\fmfforce{0.5w,1h}{v4}
\fmf{plain}{i1,o1}
\fmf{plain,left=1}{v3,v4,v3}
\fmfv{decor.size=0, label=${\scs 3}$, l.dist=1mm, l.angle=-180}{i1}
\fmfv{decor.size=0, label=${\scs 4}$, l.dist=1mm, l.angle=0}{o1}
\fmfdot{v1,v2}
\end{fmfgraph*} 
\end{center}}
+
\parbox{18mm}{\begin{center}
\begin{fmfgraph*}(5,5)
\setval
\fmfforce{-1/2w,0.5h}{i1}
\fmfforce{0w,0.5h}{v1}
\fmfforce{1w,0.5h}{v2}
\fmfforce{1.5w,0.5h}{o1}
\fmfforce{0.5w,0h}{v3}
\fmfforce{0.5w,1h}{v4}
\fmf{plain}{i1,o1}
\fmf{plain,left=1}{v3,v4,v3}
\fmfv{decor.size=0, label=${\scs 4}$, l.dist=1mm, l.angle=-180}{i1}
\fmfv{decor.size=0, label=${\scs 3}$, l.dist=1mm, l.angle=0}{o1}
\fmfdot{v1,v2}
\end{fmfgraph*} 
\end{center}} \right] \hspace*{2cm} \no \\
\hspace*{2cm} = 3 \left[ 
\parbox{18mm}{\begin{center}
\begin{fmfgraph*}(10,5)
\setval
\fmfstraight
\fmfforce{0w,0h}{i1}
\fmfforce{0w,1h}{i2}
\fmfforce{1w,0h}{o1}
\fmfforce{1w,1h}{o2}
\fmfforce{1/4w,0.5h}{v1}
\fmfforce{3/4w,0.5h}{v2}
\fmf{plain}{i1,v1}
\fmf{plain}{v2,o1}
\fmf{plain}{i2,v1}
\fmf{plain}{v2,o2}
\fmf{plain,left=1}{v1,v2,v1}
\fmfv{decor.size=0, label=${\scs 3}$, l.dist=1mm, l.angle=-180}{i1}
\fmfv{decor.size=0, label=${\scs 4}$, l.dist=1mm, l.angle=0}{o1}
\fmfv{decor.size=0, label=${\scs 2}$, l.dist=1mm, l.angle=0}{o2}
\fmfv{decor.size=0, label=${\scs 1}$, l.dist=1mm, l.angle=-180}{i2}
\fmfdot{v1,v2}
\end{fmfgraph*}
\end{center}}
+ 
\parbox{18mm}{\begin{center}
\begin{fmfgraph*}(10,5)
\setval
\fmfstraight
\fmfforce{0w,0h}{i1}
\fmfforce{0w,1h}{i2}
\fmfforce{1w,0h}{o1}
\fmfforce{1w,1h}{o2}
\fmfforce{1/4w,0.5h}{v1}
\fmfforce{3/4w,0.5h}{v2}
\fmf{plain}{i1,v1}
\fmf{plain}{v2,o1}
\fmf{plain}{i2,v1}
\fmf{plain}{v2,o2}
\fmf{plain,left=1}{v1,v2,v1}
\fmfv{decor.size=0, label=${\scs 3}$, l.dist=1mm, l.angle=-180}{i1}
\fmfv{decor.size=0, label=${\scs 4}$, l.dist=1mm, l.angle=0}{o1}
\fmfv{decor.size=0, label=${\scs 1}$, l.dist=1mm, l.angle=0}{o2}
\fmfv{decor.size=0, label=${\scs 2}$, l.dist=1mm, l.angle=-180}{i2}
\fmfdot{v1,v2}
\end{fmfgraph*}
\end{center}}
+
\parbox{18mm}{\begin{center}
\begin{fmfgraph*}(10,5)
\setval
\fmfstraight
\fmfforce{0w,0h}{i1}
\fmfforce{0w,1h}{i2}
\fmfforce{1w,0h}{o1}
\fmfforce{1w,1h}{o2}
\fmfforce{1/4w,0.5h}{v1}
\fmfforce{3/4w,0.5h}{v2}
\fmf{plain}{i1,v1}
\fmf{plain}{v2,o1}
\fmf{plain}{i2,v1}
\fmf{plain}{v2,o2}
\fmf{plain,left=1}{v1,v2,v1}
\fmfv{decor.size=0, label=${\scs 4}$, l.dist=1mm, l.angle=-180}{i1}
\fmfv{decor.size=0, label=${\scs 3}$, l.dist=1mm, l.angle=0}{o1}
\fmfv{decor.size=0, label=${\scs 2}$, l.dist=1mm, l.angle=0}{o2}
\fmfv{decor.size=0, label=${\scs 1}$, l.dist=1mm, l.angle=-180}{i2}
\fmfdot{v1,v2}
\end{fmfgraph*}
\end{center}}
+
\parbox{18mm}{\begin{center}
\begin{fmfgraph*}(10,5)
\setval
\fmfstraight
\fmfforce{0w,0h}{i1}
\fmfforce{0w,1h}{i2}
\fmfforce{1w,0h}{o1}
\fmfforce{1w,1h}{o2}
\fmfforce{1/4w,0.5h}{v1}
\fmfforce{3/4w,0.5h}{v2}
\fmf{plain}{i1,v1}
\fmf{plain}{v2,o1}
\fmf{plain}{i2,v1}
\fmf{plain}{v2,o2}
\fmf{plain,left=1}{v1,v2,v1}
\fmfv{decor.size=0, label=${\scs 4}$, l.dist=1mm, l.angle=-180}{i1}
\fmfv{decor.size=0, label=${\scs 3}$, l.dist=1mm, l.angle=0}{o1}
\fmfv{decor.size=0, label=${\scs 1}$, l.dist=1mm, l.angle=0}{o2}
\fmfv{decor.size=0, label=${\scs 2}$, l.dist=1mm, l.angle=-180}{i2}
\fmfdot{v1,v2}
\end{fmfgraph*}
\end{center}}
\right] \, . \la{NANU}
\eeq
This has the matrix representation
\beq
\lefteqn{\frac{\delta^2}{\delta G_{12}\delta G_{34}}
\left(\begin{array}{c}\{\}\\\{\}\\\{\}\end{array}\left|
\begin{array}{ccc}0&0&0\\0&0&4\\0&4&0\end{array}\right)\right.
=
2\,\frac{\delta}{\delta G_{12}}
\left[\left(\begin{array}{c}\{\}\\\{3\}\\\{4\}\end{array}\left|
\begin{array}{ccc}0&1&1\\1&0&3\\1&3&0\end{array}\right)\right.
+\left(\begin{array}{c}\{\}\\\{4\}\\\{3\}\end{array}\left|
\begin{array}{ccc}0&1&1\\1&0&3\\1&3&0\end{array}\right)\right.\right]}
\no\\
&=&
3\left[\left(\begin{array}{c}\{\}\\\{1,3\}\\\{2,4\}\end{array}\left|
\begin{array}{ccc}0&2&2\\2&0&2\\2&2&0\end{array}\right)\right.
+\left(\begin{array}{c}\{\}\\\{2,3\}\\\{1,4\}\end{array}\left|
\begin{array}{ccc}0&2&2\\2&0&2\\2&2&0\end{array}\right)\right.
+\left(\begin{array}{c}\{\}\\\{1,4\}\\\{2,3\}\end{array}\left|
\begin{array}{ccc}0&2&2\\2&0&2\\2&2&0\end{array}\right)\right.
+\left(\begin{array}{c}\{\}\\\{2,4\}\\\{1,3\}\end{array}\left|
\begin{array}{ccc}0&2&2\\2&0&2\\2&2&0\end{array}\right)\right.\right] \, .
\eeq
The first and fourth matrix as well as the second and third matrix represent
the same diagram in (\r{NANU}),
as can be seen bu permutating rows and columns of either matrix.
\item
Combining two or three diagrams to one. We perform this operation by
creating a block matrix of internal lines from the submatrices
representing the internal lines of the original diagrams.  Then the zeroth
row or column is added to represent the respective original external spatial 
arguments. Let us illustrate the combination of two diagrams by the example
\beq
\parbox{18mm}{\begin{center}
\begin{fmfgraph*}(5,5)
\setval
\fmfforce{-1/2w,0.5h}{i1}
\fmfforce{0w,0.5h}{v1}
\fmfforce{1w,0.5h}{v2}
\fmfforce{1.5w,0.5h}{o1}
\fmfforce{0.5w,0h}{v3}
\fmfforce{0.5w,1h}{v4}
\fmf{plain}{i1,o1}
\fmf{plain,left=1}{v3,v4,v3}
\fmfv{decor.size=0, label=${\scs 1}$, l.dist=1mm, l.angle=-180}{i1}
\fmfv{decor.size=0, label=${\scs 2}$, l.dist=1mm, l.angle=0}{o1}
\fmfdot{v1,v2}
\end{fmfgraph*} 
\end{center}}
\parbox{17mm}{\begin{center}
\begin{fmfgraph*}(8,5)
\setval
\fmfstraight
\fmfforce{0w,0h}{i1}
\fmfforce{0w,1h}{i2}
\fmfforce{3/8w,0.5h}{v1}
\fmfforce{1w,0.5h}{v2}
\fmf{plain}{i1,v1}
\fmf{plain}{v1,i2}
\fmf{plain,left}{v1,v2,v1}
\fmfdot{v1}
\fmfv{decor.size=0, label=${\scs 2}$, l.dist=1mm, l.angle=-180}{i1}
\fmfv{decor.size=0, label=${\scs 1}$, l.dist=1mm, l.angle=-180}{i2}
\end{fmfgraph*}
\end{center}} \quad \equiv \quad
\left(\begin{array}{c}\{\}\\\{1\}\\\{2\}\end{array}\left|
\begin{array}{ccc}
0&1&1\\
1&0&3\\
1&3&0\end{array}\right)\right. \quad , \quad
\left(\begin{array}{c}\{\}\\\{1,2\}\end{array}\left|
\begin{array}{cc}
0&2\\
2&1\\ \end{array}\right)\right.  \quad \rightarrow \quad
\left(\begin{array}{c}\{\}\\\{1,2\}\\\{1\}\\\{2\}\end{array}\left|
\begin{array}{cccc}
0&2&1&1\\
2&1&0&0\\
1&0&0&3\\
1&0&3&0
\end{array}\right)\right.
\la{X1}
\eeq
and the combination of three diagrams by
\beq
\parbox{17mm}{\begin{center}
\begin{fmfgraph*}(8,5)
\setval
\fmfstraight
\fmfforce{0w,0.5h}{v1}
\fmfforce{5/8w,0.5h}{v2}
\fmfforce{1w,0h}{i1}
\fmfforce{1w,1h}{i2}
\fmf{plain}{i1,v2}
\fmf{plain}{v2,i2}
\fmf{plain,left}{v1,v2,v1}
\fmfdot{v2}
\fmfv{decor.size=0, label=${\scs 2}$, l.dist=1mm, l.angle=0}{i1}
\fmfv{decor.size=0, label=${\scs 1}$, l.dist=1mm, l.angle=0}{i2}
\end{fmfgraph*}
\end{center}} 
\parbox{17mm}{\begin{center}
\begin{fmfgraph*}(6,5)
\setval
\fmfstraight
\fmfleft{i2,i1}
\fmfright{o2,o1}
\fmf{plain}{i1,v1}
\fmf{plain}{v1,i2}
\fmf{plain}{v1,o1}
\fmf{plain}{o2,v1}
\fmfdot{v1}
\fmfv{decor.size=0, label=${\scs 1}$, l.dist=1mm, l.angle=-180}{i1}
\fmfv{decor.size=0, label=${\scs 2}$, l.dist=1mm, l.angle=-180}{i2}
\fmfv{decor.size=0, label=${\scs 3}$, l.dist=1mm, l.angle=0}{o1}
\fmfv{decor.size=0, label=${\scs 4}$, l.dist=1mm, l.angle=0}{o2}
\end{fmfgraph*}
\end{center}}
\parbox{17mm}{\begin{center}
\begin{fmfgraph*}(8,5)
\setval
\fmfstraight
\fmfforce{0w,0h}{i1}
\fmfforce{0w,1h}{i2}
\fmfforce{3/8w,0.5h}{v1}
\fmfforce{1w,0.5h}{v2}
\fmf{plain}{i1,v1}
\fmf{plain}{v1,i2}
\fmf{plain,left}{v1,v2,v1}
\fmfdot{v1}
\fmfv{decor.size=0, label=${\scs 4}$, l.dist=1mm, l.angle=-180}{i1}
\fmfv{decor.size=0, label=${\scs 3}$, l.dist=1mm, l.angle=-180}{i2}
\end{fmfgraph*}
\end{center}} \quad \equiv \hspace*{5cm} \no \\
\la{X2}
\left(\begin{array}{c}\{\}\\\{1,2\}\end{array}\left|
\begin{array}{cc}
0&2\\
2&1\\ \end{array}\right)\right. \, , \, 
\left(\begin{array}{c}\{\}\\\{1,2,3,4\}\end{array}\left|
\begin{array}{cc}
0&4\\
4&0\\ \end{array}\right)\right. \, , \, 
\left(\begin{array}{c}\{\}\\\{3,4\}\end{array}\left|
\begin{array}{cc}
0&2\\
2&1\\ \end{array}\right)\right. \quad \rightarrow \quad
\left(\begin{array}{c}\{\}\\\{1,2\}\\\{1,2,3,4\}\\\{3,4\}\end{array}\left|
\begin{array}{cccc}
0&2&4&2\\
2&1&0&0\\
4&0&0&0\\
2&0&0&1
\end{array}\right)\right. \, .
\eeq
We observe that the ordering of the submatrices in the block matrix is 
arbitrary at this point, we just have to make sure to distribute the 
spatial labels of the external legs correctly.
\item
Connecting external legs with the same label and creating an internal
line. This is achieved in our extended matrix notation by eliminating the 
spatial labels of external legs which appear twice, and by performing
an appropriate entry in the matrix for the additional line. Thus we
obtain, for instance, from (\r{X1}) 
\beq
\parbox{10.5mm}{\begin{center}
\begin{fmfgraph*}(7.5,12.5)
\setval
\fmfforce{0w,0.3h}{v1}
\fmfforce{1w,0.3h}{v2}
\fmfforce{0.5w,0.6h}{v3}
\fmfforce{0.5w,1h}{v4}
\fmf{plain,left=1}{v1,v2,v1}
\fmf{plain,left=0.4}{v1,v2,v1}
\fmf{plain,left=1}{v3,v4,v3}
\fmfdot{v1,v2,v3}
\end{fmfgraph*}\end{center}} 
\quad \equiv \quad
\left(\begin{array}{c}\{\}\\\{\}\\\{\}\\\{\}\end{array}\left|
\begin{array}{cccc}
0&0&0&0\\
0&1&1&1\\
0&1&0&3\\
0&1&3&0
\end{array}\right)\right.
\eeq
and similarily from (\r{X2})
\beq
\parbox{23mm}{\begin{center}
\begin{fmfgraph*}(20,5)
\setval
\fmfleft{i1}
\fmfright{o1}
\fmf{plain,left=1}{i1,v1,i1}
\fmf{plain,left=1}{v1,v2,v1}
\fmf{plain,left=1}{v2,v3,v2}
\fmf{plain,left=1}{o1,v3,o1}
\fmfdot{v1,v2,v3}
\end{fmfgraph*}\end{center}}
\quad \equiv \quad
\left(\begin{array}{c}\{\}\\\{\}\\\{\}\\\{\}\end{array}\left|
\begin{array}{cccc}
0&0&0&0\\
0&1&2&0\\
0&2&0&2\\
0&0&2&1
\end{array}\right)\right. \, .
\eeq
As we reobtain at this stage connected vacuum diagrams where there are no
more external legs to be labeled, we may 
omit the prepended vector.
\end{enumerate}
The selection of a unique matrix representation for the resulting 
vacuum diagrams obtained at each stage of the recursion relation proceeds
as explained in detail in Subsection IV.A. By comparing we find out which
of the vacuum diagrams are topologically identical
and sum up their individual 
multiplicities.
Along these lines, the recursion relation (\r{GRR}) is solved by 
a MATHEMATICA program up to the order $p=6$. The results are shown in
Table \r{DATA1} and in Ref.~\ci{CODE}. To each order $p$, the
numbers $n^{(0)}_p$ of topologically different connected vacuum
diagrams are\\

\beq
\begin{array}{|c||c|c|c|c|c|c|}\hline
p & \,1\, & \,2 \,& \,3\, &\, 4\, &\, 5\, &\, 6 \\ \hline
\,n^{(0)}_p\, &\, 1\, &\, 2\, &\, 4\, &\, 10\, &\, 28\,& \,97\, \\ \hline
\end{array}
\eeq
\subsubsection{Two- and Four-Point Functions 
$\fullg_{12}$ and $\fullg_{1234}^c$ from Cutting Lines}
Having found all connected vacuum diagrams, we derive from these
the diagrams of the
connected two- and four-point functions by using the relations
(\ref{g12p}) and (\ref{g1234cp}).
In the matrix representation, cutting a line is essentially identical
to removing a line as explained above, except that we now interpret
the labels which represent the external spatial labels as sitting on the
end of lines.
Since we are not going to distinguish between trivially
``crossed'' graphs which are related
by exchanging external labels in our computer implementation,
we need no longer carry around
external spatial labels.
Thus we omit the additional vector prepended to the matrix
representing a diagram when generating vacuum diagrams.
As an example, consider cutting a line in diagram \#3 in Table \ref{P}
\beq
-\frac{\delta}{\delta G^{-1}}
\parbox{20mm}{\begin{center}
\begin{fmfgraph*}(15,5)
\setval
\fmfleft{i1}
\fmfright{o1}
\fmf{plain,left=1}{i1,v1,i1}
\fmf{plain,left=1}{v1,v2,v1}
\fmf{plain,left=1}{o1,v2,o1}
\fmfdot{v1,v2}
\end{fmfgraph*}\end{center}} \hspace*{2mm} =  \hspace*{2mm} 2 \hspace*{2mm}
\parbox{15mm}{\begin{center}
\begin{fmfgraph}(12,5)
\setval
\fmfforce{0w,0h}{i1}
\fmfforce{1/5w,0h}{v1}
\fmfforce{1/5w,1h}{v2}
\fmfforce{4/5w,0h}{v3}
\fmfforce{4/5w,1h}{v4}
\fmfforce{1w,0h}{o1}
\fmf{plain}{i1,o1}
\fmf{plain,left=1}{v1,v2,v1}
\fmf{plain,left=1}{v3,v4,v3}
\fmfdot{v1,v3}
\end{fmfgraph}
\end{center}}
+ 
\parbox{18mm}{\begin{center}
\begin{fmfgraph*}(13,5)
\setval
\fmfforce{0w,0h}{i1}
\fmfforce{0w,1h}{i2}
\fmfforce{3/13w,0.5h}{v1}
\fmfforce{8/13w,0.5h}{v2}
\fmfforce{1w,0.5h}{v3}
\fmf{plain}{i1,v1}
\fmf{plain}{i2,v1}
\fmf{plain,left=1}{v1,v2,v1}
\fmf{plain,left=1}{v2,v3,v2}
\fmfdot{v1,v2}
\end{fmfgraph*}\end{center}} 
+
\parbox{18mm}{\begin{center}
\begin{fmfgraph*}(13,5)
\setval
\fmfforce{1w,0h}{i1}
\fmfforce{1w,1h}{i2}
\fmfforce{0w,0.5h}{v1}
\fmfforce{5/13w,0.5h}{v2}
\fmfforce{10/13w,0.5h}{v3}
\fmf{plain}{i1,v3}
\fmf{plain}{i2,v3}
\fmf{plain,left=1}{v1,v2,v1}
\fmf{plain,left=1}{v2,v3,v2}
\fmfdot{v3,v2}
\end{fmfgraph*}\end{center}} \, ,
\la{TT}
\eeq
which has the matrix representation
\beq
\la{cut1}
-\frac{\delta}{\delta G^{-1}}
\left(\begin{array}{ccc}0&0&0\\0&1&2\\0&2&1\end{array}\right)
&=&
2\left(\begin{array}{ccc}0&1&1\\1&1&1\\1&1&1\end{array}\right)
+\left(\begin{array}{ccc}0&2&0\\2&0&2\\0&2&1\end{array}\right)
+\left(\begin{array}{ccc}0&0&2\\0&1&2\\2&2&0\end{array}\right) \, .
\eeq
Here the plus signs and multiplication by $2$ have a set-theoretical
meaning and are not to be understood as matrix algebra operations.
The last two matrices represent, incidentally, the same graph in (\r{TT})
as can be seen  
by exchanging the last two rows and columns
of either matrix.\\

To create the connected four-point function, we also have to consider
second derivatives of vacuum diagrams with respect to $G^{-1}$.
If an external line is cut, an additional external line will be created which
is not
connected to any vertex. It can be interpreted as a self-connection
of the zeroth vertex which collects the external lines. This may be
accomodated in the matrix notation by letting
the matrix element $M_{00}$  
count the number of lines not connected to any vertex.
For example, taking the derivative of the first diagram
in Eq.~(\ref{TT}) gives
\beq
-\frac{\delta}{\delta G^{-1}} 
\parbox{18mm}{\begin{center}
\begin{fmfgraph}(12,5)
\setval
\fmfforce{0w,0h}{i1}
\fmfforce{1/5w,0h}{v1}
\fmfforce{1/5w,1h}{v2}
\fmfforce{4/5w,0h}{v3}
\fmfforce{4/5w,1h}{v4}
\fmfforce{1w,0h}{o1}
\fmf{plain}{i1,o1}
\fmf{plain,left=1}{v1,v2,v1}
\fmf{plain,left=1}{v3,v4,v3}
\fmfdot{v1,v3}
\end{fmfgraph}
\end{center}}
= 
\parbox{13mm}{\begin{center}
\begin{fmfgraph}(10,10)
\setval
\fmfstraight
\fmfforce{0w,1/2h}{i1}
\fmfforce{1/4w,3/4h}{i2}
\fmfforce{1/4w,1/4h}{i3}
\fmfforce{1w,1/2h}{o1}
\fmfforce{3/4w,1/2h}{v1}
\fmfforce{3/4w,1h}{v2}
\fmfforce{1/4w,1/2h}{v3}
\fmf{plain}{i1,o1}
\fmf{plain}{i2,v3}
\fmf{plain}{i3,v3}
\fmf{plain,left=1}{v1,v2,v1}
\fmfdot{v1,v3}
\end{fmfgraph}
\end{center}}
+
\parbox{13mm}{\begin{center}
\begin{fmfgraph}(10,10)
\setval
\fmfstraight
\fmfforce{1w,1/2h}{i1}
\fmfforce{3/4w,3/4h}{i2}
\fmfforce{3/4w,1/4h}{i3}
\fmfforce{0w,1/2h}{o1}
\fmfforce{1/4w,1/2h}{v1}
\fmfforce{1/4w,1h}{v2}
\fmfforce{3/4w,1/2h}{v3}
\fmf{plain}{i1,o1}
\fmf{plain}{i2,v3}
\fmf{plain}{i3,v3}
\fmf{plain,left=1}{v1,v2,v1}
\fmfdot{v1,v3}
\end{fmfgraph}
\end{center}}
+
\parbox{15mm}{\begin{center}
\begin{fmfgraph*}(5,12.5)
\setval
\fmfforce{0w,1/10h}{i11}
\fmfforce{1/2w,1/10h}{v11}
\fmfforce{1/2w,1/2h}{v21}
\fmfforce{1w,1/10h}{o11}
\fmfforce{0w,7/10h}{i12}
\fmfforce{1/2w,7/10h}{v12}
\fmfforce{1/2w,11/10h}{v22}
\fmfforce{1w,7/10h}{o12}
\fmf{plain}{i11,o11}
\fmf{plain}{i12,o12}
\fmf{plain,left=1}{v11,v21,v11}
\fmf{plain,left=1}{v12,v22,v12}
\fmfdot{v11,v12}
\end{fmfgraph*}
\end{center}}
+ \hspace*{1mm} 2 
\parbox{18mm}{\begin{center}
\begin{fmfgraph}(12,7.5)
\setval
\fmfforce{0w,0h}{i1}
\fmfforce{1/5w,0h}{v1}
\fmfforce{1/5w,3/4h}{v2}
\fmfforce{4/5w,0h}{v3}
\fmfforce{4/5w,3/4h}{v4}
\fmfforce{0w,1h}{v5}
\fmfforce{1w,1h}{v6}
\fmfforce{1w,0h}{o1}
\fmf{plain}{i1,o1}
\fmf{plain}{v5,v6}
\fmf{plain,left=1}{v1,v2,v1}
\fmf{plain,left=1}{v3,v4,v3}
\fmfdot{v1,v3}
\end{fmfgraph}
\end{center}} \, ,
\la{SS}
\eeq
with the matrix notation
\beq
\la{cut2}
-\frac{\delta}{\delta G^{-1}}
\left(\begin{array}{ccc}0&1&1\\1&1&1\\1&1&1\end{array}\right)
=
\left(\begin{array}{ccc}0&3&1\\3&0&1\\1&1&1\end{array}\right)
+\left(\begin{array}{ccc}0&1&3\\1&1&1\\3&1&0\end{array}\right)
+\left(\begin{array}{ccc}0&2&2\\2&1&0\\2&0&1\end{array}\right)
+\hspace*{1mm} 
2\left(\begin{array}{ccc}1&1&1\\1&1&1\\1&1&1\end{array}\right).
\la{MM}
\eeq
The first two matrices represent the same diagram as can be seen from
Eq.~(\r{SS}). 
The last two matrices in Eq.~(\r{MM}) correspond to
disconnected diagrams:
the first because of the absence of a connection between the
two vertices, the second because of the disconnected line
represented by the entry $M_{00}=1$.
In the full expression for the two loop contribution $G_{1234}^{c,(2)}$ 
to the four-point function in Eq.~(\r{g1234cp}) all
disconnected diagrams arising from cutting a line in $G_{12}^{(2)}$
are canceled by diagrams resulting from the sum.
Therefore we may omit the sum, 
take only the first term and discard all
disconnected graphs it creates.
This is particularly useful for treating low orders by hand.
If we include the sum, we use the prescription of combining diagrams
into one as described above in Subsection IV.B.1, except that we now omit
the extra vector with the labels of spatial arguments.
\subsubsection{Two- and Four-Point Function 
$\fullg_{12}$ and $\fullg_{1234}^c$ from Removing Lines}
Instead of cutting lines of connected
vacuum graphs once or twice, 
the perturbative coefficients of $\fullg_{12}$ and $\fullg_{1234}^c$
can also be obtained graphically by removing lines. Indeed,
from (\ref{GIT}), (\ref{PROP}), (\ref{DEC}) and (\ref{NR1}) we get for
the two-point function
\beq
\label{g12}
\fullg_{12}
=G_{12}+2\int_{34}G_{13}G_{24}\frac{\delta W^{({\rm int})}}{\delta G_{34}}\,,
\eeq
so that we have for $p>0$ 
\beq
\fullg_{12}^{(p)}
=2\int_{34}G_{13}G_{24}\frac{\delta W^{(p)}}{\delta G_{34}}
\eeq
at our disposal to compute the coefficients $\fullg_{12}^{(p)}$
from removing one line in the connected vacuum diagrams $W^{(p)}$ in
all possible ways.
The corresponding matrix operations are identical to the ones for cutting
a line
so that in this respect there is no difference between both
procedures to obtain $\fullg_{12}$.\\

Combining (\ref{g12}) with (\ref{ACT}), (\ref{GcfromG}) and
(\ref{NR1}), we get for the connected four-point function
\beq
\fullg_{1234}^c
&=&
4\int_{5678}G_{15}G_{26}G_{37}G_{48}
\frac{\delta^2W^{({\rm int})}}{\delta G_{56}\delta G_{78}}
-4\int_{5678}G_{15}G_{27}(G_{36}G_{48}+G_{46}G_{38})
\frac{\delta W^{({\rm int})}}{\delta G_{56}}
\frac{\delta W^{({\rm int})}}{\delta G_{78}}
\eeq
which is equivalent to
\beq
\fullg_{1234}^{c,(p)}
&=&
4\int_{5678}G_{15}G_{26}G_{37}G_{48}
\frac{\delta^2W^{(p)}}{\delta G_{56}\delta G_{78}}
-4\sum_{q=1}^{p-1}\left(\begin{array}{c}p\\q\end{array}\right)
\int_{5678}G_{15}G_{27}(G_{36}G_{48}+G_{46}G_{38})
\frac{\delta W^{(q)}}{\delta G_{56}}
\frac{\delta W^{(p-q)}}{\delta G_{78}}\,.
\eeq
Again, the sum serves only to subtract disconnected diagrams which
are created by
the first term, so we may choose to discard both in the first term.\\

Now the problem of generating diagrams is reduced to the generation
of vacuum diagrams and subsequently taking functional 
derivatives with respect to $G_{12}$.
An advantage of this approach is that 
external lines do not appear at intermediate steps.
So when one uses the cancellation of disconnected terms as a cross check,
there are less operations to be performed than with cutting.
At the end one just interprets external labels 
as sitting on external lines.
Since all necessary operations on matrices have already been introduced,
we omit examples here and just note that we can again omit external labels
if we are not distinguishing between trivially ``crossed'' graphs.\\

The generation of diagrams of the connected two- and four-point functions
has been implemented in both possible ways. Cutting or removing one or
two lines in the connected vacuum diagrams up to the order $p=6$ lead to
the following numbers $n^{(2)}_p$ and $n^{(2)}_p$ of topologically different
diagrams of $\fullg^{(p)}_{12}$ and $\fullg_{1234}^{c,(p)}$:\\

\beq
\begin{array}{|c||c|c|c|c|c|c|}\hline
p & \,1\, & \,2 \,& \,3\, &\, 4\, &\, 5\, &\, 6 \\ \hline
\,n^{(2)}_p\, &\, 1\, &\, 3\, &\, 8\, &\, 30\, &\, 118\,& \,548\, \\ \hline
\,n^{(4)}_p\, &\, 1\, &\, 2\, &\, 8\, &\, 37\, &\, 181\,& \,1010\, \\ \hline
\end{array} 
\eeq
\section{Outlook}
In this work we have shown f
that all Feynman diagrams can be efficiently determined from a recursive
graphical solution of a nonlinear 
functional differential equation order by order
in the coupling strength. 
In separate publications, this method will be 
applied to the ordered phase of $\phi^4$-theory, where the energy functional 
contains a mixture of a cubic and a quartic interaction, 
and to quantum electrodynamics \ci{QED}. It is hoped that our method
will eventually be combined with
efficient numerical algorithms for actually evaluating these Feynman
diagrams.
\section*{Acknowledgements}
We are thankful to
Dr.~Bruno~van~den~Bossche and Florian~Jasch
for contributing various useful comments. 
One of us (M.B.) acknowledges support by the Studienstiftung 
des deutschen Volkes.
\newpage
\end{fmffile}
\newpage
\begin{fmffile}{graph2}
\begin{table}[t]
\begin{center}
\begin{tabular}{|c|c|}
\,\,\,$p$\,\,\,
&
\begin{tabular}{@{}c}
$\mbox{}$\\
$\mbox{}$
\end{tabular}
$W^{(p)}$
\\
\hline
%$0$ &
%\hspace{-10pt}
%\rule[-10pt]{0pt}{26pt}
%\begin{tabular}{@{}c}
%$1 / 2$ \\ ${\scs ( 1, 0, 0 , 0 , 1 )}$
%\end{tabular}
%
%\parbox{8mm}{\begin{center}
%\begin{fmfgraph}(5,5)
%\setval
%\fmfleft{i1}
%\fmfright{o1}
%\fmf{plain,left=1}{i1,o1,i1}
%\end{fmfgraph}
%\end{center}}
%
%\\
%\hline
$1$ &
\hspace{-10pt}
\rule[-10pt]{0pt}{26pt}
\begin{tabular}{@{}c}
$\mbox{}$\\
${\scs \mbox{\#1}}$ \\
$3$ \\ ${\scs ( 2, 1, 0 , 0 ; 1 )}$\\
$\mbox{}$
\end{tabular}
\parbox{13mm}{\begin{center}
\begin{fmfgraph*}(10,5)
\setval
\fmfleft{i1}
\fmfright{o1}
\fmf{plain,left=1}{i1,v1,i1}
\fmf{plain,left=1}{o1,v1,o1}
\fmfdot{v1}
\end{fmfgraph*}\end{center}}
\\
\hline
$2$ &
\hspace{-10pt}
\rule[-10pt]{0pt}{26pt}
\begin{tabular}{@{}c}
$\mbox{}$\\
${\scs \mbox{\#2}}$ \\
$24$ \\ 
${\scs ( 0, 0, 0 , 1 ; 2 )}$\\
$\mbox{}$
\end{tabular}
\parbox{10.5mm}{\begin{center}
\begin{fmfgraph*}(7.5,5)
\setval
\fmfforce{0w,0.5h}{v1}
\fmfforce{1w,0.5h}{v2}
%\fmfleft{v1}
%\fmfright{v2}
\fmf{plain,left=1}{v1,v2,v1}
\fmf{plain,left=0.4}{v1,v2,v1}
\fmfdot{v1,v2}
\end{fmfgraph*}\end{center}} 
\quad
\begin{tabular}{@{}c}
${\scs \mbox{\#3}}$ \\
$72$ \\ 
${\scs ( 2, 1, 0 , 0 ; 2 )}$
\end{tabular}
\parbox{18mm}{\begin{center}
\begin{fmfgraph*}(15,5)
\setval
\fmfleft{i1}
\fmfright{o1}
\fmf{plain,left=1}{i1,v1,i1}
\fmf{plain,left=1}{v1,v2,v1}
\fmf{plain,left=1}{o1,v2,o1}
\fmfdot{v1,v2}
\end{fmfgraph*}\end{center}}
\\
\hline
$3$&
\hspace{-10pt}
\rule[-10pt]{0pt}{26pt}
\begin{tabular}{@{}c}
${\scs \mbox{\#4}}$ \\
$1728$ \\ 
${\scs ( 0, 3, 0 , 0 ; 6 )}$
\end{tabular}
\parbox{11mm}{\begin{center}
\begin{fmfgraph*}(8,8)
\setval
\fmfforce{0.5w,0h}{v1}
\fmfforce{0.5w,1h}{v2}
\fmfforce{0.066987w,0.25h}{v3}
\fmfforce{0.93301w,0.25h}{v4}
\fmf{plain,left=1}{v1,v2,v1}
\fmf{plain}{v2,v3}
\fmf{plain}{v3,v4}
\fmf{plain}{v2,v4}
\fmfdot{v2,v3,v4}
\end{fmfgraph*}
\end{center}} 
\quad
\begin{tabular}{@{}c}
${\scs \mbox{\#5}}$ \\
$3456$\\ 
${\scs ( 1, 0, 1 , 0 ; 2 )}$
\end{tabular}
\parbox{10.5mm}{\begin{center}
\begin{fmfgraph*}(7.5,12.5)
\setval
\fmfforce{0w,0.3h}{v1}
\fmfforce{1w,0.3h}{v2}
\fmfforce{0.5w,0.6h}{v3}
\fmfforce{0.5w,1h}{v4}
\fmf{plain,left=1}{v1,v2,v1}
\fmf{plain,left=0.4}{v1,v2,v1}
\fmf{plain,left=1}{v3,v4,v3}
\fmfdot{v1,v2,v3}
\end{fmfgraph*}\end{center}} 
\quad
\begin{tabular}{@{}c}
${\scs \mbox{\#6}}$ \\
$1728$\\ 
${\scs ( 3, 0, 0 , 0 ; 6 )}$
\end{tabular}
\parbox{18mm}{\begin{center}
\begin{fmfgraph*}(15,15)
\setval
\fmfforce{1/2w,1/3h}{v1}
\fmfforce{1/2w,2/3h}{v2}
\fmfforce{1/2w,1h}{v3}
\fmfforce{0.355662432w,0.416666666h}{v4}
\fmfforce{0.64433568w,0.416666666h}{v5}
\fmfforce{0.067w,1/4h}{v6}
\fmfforce{0.933w,1/4h}{v7}
\fmf{plain,left=1}{v1,v2,v1}
\fmf{plain,left=1}{v2,v3,v2}
\fmf{plain,left=1}{v4,v6,v4}
\fmf{plain,left=1}{v5,v7,v5}
\fmfdot{v2,v4,v5}
\end{fmfgraph*}\end{center}} 
\quad
\begin{tabular}{@{}c}
${\scs \mbox{\#7}}$ \\
$2592$\\ 
${\scs ( 2, 2, 0 , 0 ; 2 )}$
\end{tabular}
\parbox{23mm}{\begin{center}
\begin{fmfgraph*}(20,5)
\setval
\fmfleft{i1}
\fmfright{o1}
\fmf{plain,left=1}{i1,v1,i1}
\fmf{plain,left=1}{v1,v2,v1}
\fmf{plain,left=1}{v2,v3,v2}
\fmf{plain,left=1}{o1,v3,o1}
\fmfdot{v1,v2,v3}
\end{fmfgraph*}\end{center}}
\\
\hline
$4$&
\rule[-10pt]{0pt}{26pt}
\begin{tabular}{@{}c}
${\scs \mbox{\#8}}$ \\
$62208$ \\ 
${\scs ( 0, 4 , 0 , 0 ; 8 )}$
\end{tabular} 
\parbox{13mm}{\begin{center}
\begin{fmfgraph*}(10,10)
\setval
\fmfforce{0.1464466w,0.1464466h}{v1}
\fmfforce{0.1464466w,0.8535534h}{v2}
\fmfforce{0.8535534w,0.8535534h}{v3}
\fmfforce{0.8535534w,0.1464466h}{v4}
\fmfforce{1/2w,0h}{v5}
\fmfforce{1/2w,1h}{v6}
\fmf{plain,left=1}{v5,v6,v5}
\fmf{plain}{v1,v2}
\fmf{plain}{v2,v3}
\fmf{plain}{v3,v4}
\fmf{plain}{v4,v1}
\fmfdot{v1,v2,v3,v4}
\end{fmfgraph*}
\end{center}}
\quad
\begin{tabular}{@{}c}
${\scs \mbox{\#9}}$ \\
$248832$ \\ 
${\scs ( 0, 2 , 0 , 0 ; 8 )}$
\end{tabular}
\parbox{13mm}{\begin{center}
\begin{fmfgraph*}(10,10)
\setval
\fmfforce{0w,1/2h}{v1}
\fmfforce{1w,1/2h}{v2}
\fmfforce{1/2w,1/4h}{v3}
\fmfforce{1/2w,3/4h}{v4}
\fmf{plain,left=1}{v1,v2,v1}
\fmf{plain,left=1}{v3,v4,v3}
\fmf{plain,right=0.5}{v1,v2,v1}
\fmfdot{v1,v2,v3,v4}
\end{fmfgraph*}
\end{center}}
\quad
\begin{tabular}{@{}c}
${\scs \mbox{\#10}}$ \\
$55296$ \\ 
${\scs ( 0, 0 , 2 , 0 ; 4 )}$
\end{tabular}
\parbox{13mm}{\begin{center}
\begin{fmfgraph*}(10,12.5)
\setval
\fmfforce{1/4w,1/5h}{v1}
\fmfforce{3/4w,1/5h}{v2}
\fmfforce{1/4w,4/5h}{v3}
\fmfforce{3/4w,4/5h}{v4}
\fmf{plain}{v1,v2}
\fmf{plain,left=1}{v1,v2,v1}
\fmf{plain}{v3,v4}
\fmf{plain,left=1}{v3,v4,v3}
\fmf{plain,left=0.5}{v1,v3}
\fmf{plain,right=0.5}{v2,v4}
\fmfdot{v1,v2,v3,v4}
\end{fmfgraph*}
\end{center}}
\quad
\begin{tabular}{@{}c}
${\scs \mbox{\#11}}$ \\
$497664$ \\ 
${\scs ( 1, 2  , 0 , 0 ; 2 )}$
\end{tabular}
\parbox{11mm}{\begin{center}
\begin{fmfgraph*}(8,13)
\setval
\fmfforce{0.5w,0h}{v1}
\fmfforce{0.5w,8/13h}{v2}
\fmfforce{0.0669873w,0.46154h}{v3}
\fmfforce{0.933w,0.46154h}{v4}
\fmfforce{0.5w,1h}{v5}
\fmf{plain,left=1}{v1,v2,v1}
\fmf{plain,left=1}{v2,v5,v2}
\fmf{plain}{v1,v3}
\fmf{plain}{v3,v4}
\fmf{plain}{v1,v4}
\fmfdot{v1,v2,v3,v4}
\end{fmfgraph*}\end{center}}
\quad
\begin{tabular}{@{}c}
${\scs \mbox{\#12}}$ \\
$165888$ \\ 
${\scs ( 2 , 0 , 1 , 0 ; 2 )}$
\end{tabular}
\parbox{15mm}{\begin{center}
\begin{fmfgraph*}(8,13)
\setval
\fmfforce{0w,0.3h}{v1}
\fmfforce{1w,0.3h}{v2}
\fmfforce{0.2w,0.55h}{v3}
\fmfforce{0.8w,0.55h}{v4}
\fmf{plain,left=1}{v1,v2,v1}
\fmf{plain,left=0.4}{v1,v2,v1}
\fmfi{plain}{reverse fullcircle scaled 0.625w shifted (1w,0.7h)}
\fmfi{plain}{reverse fullcircle scaled 0.625w shifted (0w,0.7h)}
\fmfdot{v1,v2,v3,v4}
\end{fmfgraph*}\end{center}}
\quad
\\
&\begin{tabular}{@{}c}
${\scs \mbox{\#13}}$ \\
$248832$ \\ 
${\scs ( 2,1, 0 , 0 ; 4 )}$
\end{tabular}
\parbox{10.5mm}{\begin{center}
\begin{fmfgraph*}(7.5,17.5)
\setval
\fmfforce{0w,0.5h}{v1}
\fmfforce{1w,0.5h}{v2}
\fmfforce{0.5w,h}{v3}
\fmfforce{0.5w,1h}{v4}
\fmfforce{0.5w,0h}{v5}
\fmfforce{0.5w,0.2857h}{v6}
\fmfforce{0.5w,0.71429h}{v7}
\fmf{plain,left=1}{v5,v6,v5}
\fmf{plain,left=1}{v7,v4,v7}
\fmf{plain,left=1}{v1,v2,v1}
\fmf{plain,left=0.4}{v1,v2,v1}
\fmfdot{v1,v2,v6,v7}
\end{fmfgraph*}\end{center}} 
\quad
\begin{tabular}{@{}c}
${\scs \mbox{\#14}}$ \\
$165888$ \\ 
${\scs ( 1,1, 1 , 0 ; 2 )}$
\end{tabular}
\parbox{10.5mm}{\begin{center}
\begin{fmfgraph*}(7.5,17.5)
\setval
\fmfforce{0w,0.21429h}{v1}
\fmfforce{1w,0.21429h}{v2}
\fmfforce{0.5w,0.42857h}{v3}
\fmfforce{0.5w,0.71429h}{v4}
\fmfforce{0.5w,1h}{v5}
\fmf{plain,left=1}{v1,v2,v1}
\fmf{plain,left=1}{v3,v4,v3}
\fmf{plain,left=1}{v4,v5,v4}
\fmf{plain,left=0.4}{v1,v2,v1}
\fmfdot{v1,v2,v3,v4}
\end{fmfgraph*}\end{center}} 
\quad
\begin{tabular}{@{}c}
${\scs \mbox{\#15}}$ \\
$248832$ \\ 
${\scs ( 3,1, 0 , 0 ; 2 )}$
\end{tabular}
\parbox{18mm}{\begin{center}
\begin{fmfgraph*}(15,20)
\setval
\fmfforce{1/2w,1/4h}{v1}
\fmfforce{1/2w,1/2h}{v2}
\fmfforce{1/2w,3/4h}{v3}
\fmfforce{1/2w,1h}{v4}
\fmfforce{0.36w,0.3125h}{v5}
\fmfforce{0.64w,0.3125h}{v6}
\fmf{plain,left=1}{v1,v2,v1}
\fmf{plain,left=1}{v2,v3,v2}
\fmf{plain,left=1}{v3,v4,v3}
\fmfi{plain}{reverse fullcircle scaled 0.333333w shifted (0.21132w,0.25h)}
\fmfi{plain}{reverse fullcircle scaled 0.333333w shifted (0.78868w,0.25h)}
\fmfdot{v2,v3,v5,v6}
\end{fmfgraph*}\end{center}} 
\quad
\begin{tabular}{@{}c}
${\scs\mbox{\#16}}$ \\
$62208$ \\ 
${\scs ( 4,0 , 0 , 0 ; 8 )}$
\end{tabular}
\parbox{18mm}{\begin{center}
\begin{fmfgraph*}(15,15)
\setval
\fmfforce{1/3w,1/2h}{v1}
\fmfforce{2/3w,1/2h}{v2}
\fmfforce{1/2w,1/3h}{v3}
\fmfforce{1/2w,2/3h}{v4}
\fmfforce{1/2w,1h}{v5}
\fmfforce{1/2w,0h}{v6}
\fmfforce{0w,0.5h}{v7}
\fmfforce{1w,0.5h}{v8}
\fmf{plain,left=1}{v3,v4,v3}
\fmf{plain,left=1}{v4,v5,v4}
\fmf{plain,left=1}{v3,v6,v3}
\fmf{plain,left=1}{v1,v7,v1}
\fmf{plain,left=1}{v2,v8,v2}
\fmfdot{v1,v2,v3,v4}
\end{fmfgraph*}\end{center}} 
\quad
\begin{tabular}{@{}c}
${\scs \mbox{\#17}}$ \\
$124416$ \\ 
${\scs ( 2,3, 0 , 0 ; 2 )}$
\end{tabular}
\parbox{28mm}{\begin{center}
\begin{fmfgraph*}(25,5)
\setval
\fmfleft{i1}
\fmfright{o1}
\fmf{plain,left=1}{i1,v1,i1}
\fmf{plain,left=1}{v1,v2,v1}
\fmf{plain,left=1}{v2,v3,v2}
\fmf{plain,left=1}{v3,v4,v3}
\fmf{plain,left=1}{o1,v4,o1}
\fmfdot{v1,v2,v3,v4}
\end{fmfgraph*}\end{center}}
\end{tabular}
\end{center}
\caption{\la{P} Connected vacuum diagrams and their 
multiplicities of the $\phi^4$-theory
up to five loops. Each diagram is characterized by the
vector $(S,D,T,F;N$) whose components specify the number of self-, double,
triple and fourfold connections, and of the identical vertex permutations,
respectively.}
\end{table}
\end{fmffile}
\newpage
\begin{fmffile}{graph3}
\vspace*{-2cm}
\begin{table}[t]
\begin{center}
\begin{tabular}{|c|c|}
\,\,\,$p$\,\,\,
&
\begin{tabular}{@{}c}
$\mbox{}$\\
$\mbox{}$
\end{tabular}
$\fullg_{12}^{(p)}$
\\
\hline
$1$ &
\hspace{-10pt}
\rule[-10pt]{0pt}{26pt}
\begin{tabular}{@{}c}
$\mbox{}$\\
${\scs \mbox{\#1.1}}$ \\
$12$\\ 
${\scs ( 1, 0, 0 ; 2 )}$\\
$\mbox{}$\\
\end{tabular}
\parbox{8mm}{\begin{center}
\begin{fmfgraph}(5,5)
\setval
\fmfstraight
\fmfforce{0w,0h}{i1}
\fmfforce{0.5w,0h}{v1}
\fmfforce{1w,0h}{o1}
\fmfforce{0.5w,1h}{v2}
\fmf{plain}{i1,v1}
\fmf{plain}{v1,o1}
\fmf{plain,left=1}{v1,v2,v1}
\fmfdot{v1}
\end{fmfgraph}
\end{center}}
\\
\hline
$2$ &
\hspace{-10pt}
\rule[-10pt]{0pt}{26pt}
\begin{tabular}{@{}c}
$\mbox{}$\\
${\scs \mbox{\#2.1}}$ \\
$192$\\ 
${\scs ( 0, 0, 1 ; 2 )}$\\
$\mbox{}$\\
\end{tabular}
\parbox{15mm}{\begin{center}
\begin{fmfgraph}(5,5)
\setval
\fmfforce{-1/2w,0.5h}{i1}
\fmfforce{0w,0.5h}{v1}
\fmfforce{1w,0.5h}{v2}
\fmfforce{1.5w,0.5h}{o1}
\fmfforce{0.5w,0h}{v3}
\fmfforce{0.5w,1h}{v4}
\fmf{plain}{i1,o1}
\fmf{plain,left=1}{v3,v4,v3}
\fmfdot{v1,v2}
\end{fmfgraph}
\end{center}}
\quad \,\,
\begin{tabular}{@{}c}
${\scs \mbox{\#3.1}}$ \\
$288$\\ 
${\scs ( 1, 1, 0 ; 2 )}$
\end{tabular}
\parbox{8mm}{\begin{center}
\begin{fmfgraph}(5,10)
\setval
\fmfforce{0w,0h}{i1}
\fmfforce{0.5w,0h}{v1}
\fmfforce{1w,0h}{o1}
\fmfforce{0.5w,0.5h}{v2}
\fmfforce{0.5w,1h}{v3}
\fmf{plain}{i1,v1}
\fmf{plain}{v1,o1}
\fmf{plain,left=1}{v1,v2,v1}
\fmf{plain,left=1}{v2,v3,v2}
\fmfdot{v1,v2}
\end{fmfgraph}
\end{center}}
\quad \,\, 
\begin{tabular}{@{}c}
${\scs \mbox{\#3.2}}$ \\
$288$\\ 
${\scs ( 2, 0, 0 ; 2 )}$
\end{tabular}
\parbox{15mm}{\begin{center}
\begin{fmfgraph}(12,5)
\setval
\fmfforce{0w,0h}{i1}
\fmfforce{1/5w,0h}{v1}
\fmfforce{1/5w,1h}{v2}
\fmfforce{4/5w,0h}{v3}
\fmfforce{4/5w,1h}{v4}
\fmfforce{1w,0h}{o1}
\fmf{plain}{i1,o1}
\fmf{plain,left=1}{v1,v2,v1}
\fmf{plain,left=1}{v3,v4,v3}
\fmfdot{v1,v3}
\end{fmfgraph}
\end{center}}
\\
\hline
$3$ &
\hspace{-10pt}
\rule[-10pt]{0pt}{26pt}
\begin{tabular}{@{}c}
$\mbox{}$
${\scs \mbox{\#4.1}}$ \\
$20736$\\ 
${\scs ( 0, 2, 0 ; 2 )}$\\
$\mbox{}$
\end{tabular}
\parbox{11mm}{\begin{center}
\begin{fmfgraph*}(8,8)
\setval
\fmfforce{-0.245513w,0.25h}{i1}
\fmfforce{1.24551w,0.25h}{o1}
\fmfforce{0.5w,0h}{v1}
\fmfforce{0.5w,1h}{v2}
\fmfforce{0.066987w,0.25h}{v3}
\fmfforce{0.93301w,0.25h}{v4}
\fmf{plain,left=1}{v1,v2,v1}
\fmf{plain}{i1,v3}
\fmf{plain}{o1,v4}
\fmf{plain}{v2,v3}
\fmf{plain}{v2,v4}
\fmfdot{v2,v3,v4}
\end{fmfgraph*}
\end{center}}
\quad \,\,
\begin{tabular}{@{}c}
${\scs \mbox{\#5.1}}$ \\
$6912$\\ 
${\scs ( 0, 0, 1 ; 4 )}$
\end{tabular}
\parbox{11mm}{\begin{center}
\begin{fmfgraph*}(8,8)
\setval
\fmfforce{0w,0.5h}{v1}
\fmfforce{1w,0.5h}{v2}
\fmfforce{0.5w,0h}{v4}
\fmfforce{0.1875w,0h}{i1}
\fmfforce{0.8125w,0h}{o1}
\fmf{plain,left=1}{v1,v2,v1}
\fmf{plain,left=0.4}{v1,v2,v1}
\fmf{plain,left=1}{v3,v4,v3}
\fmf{plain}{i1,o1}
\fmfdot{v1,v2,v3}
\end{fmfgraph*}\end{center}} 
\quad \,\,
\begin{tabular}{@{}c}
${\scs \mbox{\#5.2}}$ \\
$20736$\\ 
${\scs ( 1, 1, 0 ; 2 )}$
\end{tabular}
\parbox{13mm}{\begin{center}
\begin{fmfgraph*}(5,10)
\setval
\fmfforce{-1/2w,0.25h}{i1}
\fmfforce{3/2w,0.25h}{o1}
\fmfforce{0w,0.25h}{v1}
\fmfforce{1w,0.25h}{v2}
\fmfforce{0.5w,0.5h}{v3}
\fmfforce{0.5w,1h}{v4}
\fmf{plain}{i1,o1}
\fmf{plain,left=1}{v1,v2,v1}
\fmf{plain,left=1}{v3,v4,v3}
\fmfdot{v1,v2,v3}
\end{fmfgraph*}\end{center}} 
\quad \,\,
\begin{tabular}{@{}c}
${\scs \mbox{\#5.3}}$ \\
$13824$\\ 
${\scs ( 1, 0, 1 ; 1 )}$
\end{tabular}
\parbox{20mm}{\begin{center}
\begin{fmfgraph}(15,5)
\setval
\fmfforce{0w,0.5h}{i1}
\fmfforce{1/6w,0.5h}{v1}
\fmfforce{1/2w,0.5h}{v2}
\fmfforce{5/6w,0.5h}{v3}
\fmfforce{5/6w,1.5h}{v4}
\fmfforce{1w,0.5h}{o1}
\fmf{plain}{i1,o1}
\fmf{plain,left=1}{v3,v4,v3}
\fmf{plain,left=1}{v1,v2,v1}
\fmfdot{v1,v2,v3}
\end{fmfgraph}
\end{center}}
\\ & 
\quad \,\,
\begin{tabular}{@{}c}
$\mbox{}$\\
${\scs \mbox{\#6.1}}$ \\
$10368$\\ 
${\scs ( 2, 0, 0 ; 4 )}$\\
$\mbox{}$
\end{tabular}
\parbox{18mm}{\begin{center}
\begin{fmfgraph*}(15,15)
\setval
\fmfforce{1/3w,1/3h}{i1}
\fmfforce{2/3w,1/3h}{o1}
\fmfforce{1/2w,1/3h}{v1}
\fmfforce{1/2w,2/3h}{v2}
\fmfforce{1/2w,1h}{v3}
\fmfforce{0.355662432w,0.5833333333h}{v4}
\fmfforce{0.64433568w,0.5833333333h}{v5}
\fmfforce{0.067w,3/4h}{v6}
\fmfforce{0.933w,3/4h}{v7}
\fmf{plain,left=1}{v1,v2,v1}
\fmf{plain}{i1,v1}
\fmf{plain}{v1,o1}
\fmf{plain,left=1}{v4,v6,v4}
\fmf{plain,left=1}{v5,v7,v5}
\fmfdot{v1,v4,v5}
\end{fmfgraph*}\end{center}} 
\quad \,\,
\begin{tabular}{@{}c}
${\scs \mbox{\#6.2}}$ \\
$10368$\\ 
${\scs ( 3, 0, 0 ; 2 )}$
\end{tabular}
\parbox{23mm}{\begin{center}
\begin{fmfgraph}(20,5)
\setval
\fmfforce{0w,0h}{i1}
\fmfforce{1/8w,0h}{v1}
\fmfforce{1/8w,1h}{v2}
\fmfforce{1/2w,0h}{v3}
\fmfforce{1/2w,1h}{v4}
\fmfforce{7/8w,0h}{v5}
\fmfforce{7/8w,1h}{v6}
\fmfforce{1w,0h}{o1}
\fmf{plain}{i1,o1}
\fmf{plain,left=1}{v1,v2,v1}
\fmf{plain,left=1}{v3,v4,v3}
\fmf{plain,left=1}{v5,v6,v5}
\fmfdot{v1,v3,v5}
\end{fmfgraph}
\end{center}}
\quad \,\,
\begin{tabular}{@{}c}
${\scs \mbox{\#7.1}}$ \\
$10368$\\ 
${\scs ( 1, 2, 0 ; 2 )}$
\end{tabular}
\parbox{8mm}{\begin{center}
\begin{fmfgraph}(5,15)
\setval
\fmfforce{0w,0h}{i1}
\fmfforce{0.5w,0h}{v1}
\fmfforce{1w,0h}{o1}
\fmfforce{0.5w,1/3h}{v2}
\fmfforce{0.5w,2/3h}{v3}
\fmfforce{0.5w,1h}{v4}
\fmf{plain}{i1,v1}
\fmf{plain}{v1,o1}
\fmf{plain,left=1}{v1,v2,v1}
\fmf{plain,left=1}{v2,v3,v2}
\fmf{plain,left=1}{v3,v4,v3}
\fmfdot{v1,v2,v3}
\end{fmfgraph}
\end{center}}
\quad \,\,
\begin{tabular}{@{}c}
${\scs \mbox{\#7.2}}$ \\
$20736$\\ 
${\scs ( 2, 1, 0 ; 1 )}$
\end{tabular}
\parbox{15mm}{\begin{center}
\begin{fmfgraph}(12.5,10)
\setval
\fmfforce{0w,0h}{i1}
\fmfforce{1/5w,0h}{v1}
\fmfforce{1/5w,0.5h}{v2}
\fmfforce{1/5w,1h}{v5}
\fmfforce{4/5w,0h}{v3}
\fmfforce{4/5w,0.5h}{v4}
\fmfforce{1w,0h}{o1}
\fmf{plain}{i1,o1}
\fmf{plain,left=1}{v1,v2,v1}
\fmf{plain,left=1}{v3,v4,v3}
\fmf{plain,left=1}{v2,v5,v2}
\fmfdot{v1,v2,v3}
\end{fmfgraph}
\end{center}}
\\
\hline
$4$ &
\hspace{-10pt}
\rule[-10pt]{0pt}{26pt}
\begin{tabular}{@{}c}
$\mbox{}$\\
${\scs \mbox{\#8.1}}$ \\
$995328$\\ 
${\scs ( 0, 3, 0 ; 2 )}$\\
$\mbox{}$\\
\end{tabular}
\parbox{11mm}{\begin{center}
\begin{fmfgraph*}(8,8)
\setval
\fmfforce{-0.1035534w,0.1464466h}{i1}
\fmfforce{0.1464466w,0.1464466h}{v1}
\fmfforce{0.1464466w,0.8535534h}{v2}
\fmfforce{0.8535534w,0.8535534h}{v3}
\fmfforce{0.8535534w,0.1464466h}{v4}
\fmfforce{1.103554w,0.1464466h}{o1}
\fmfforce{1/2w,0h}{v5}
\fmfforce{1/2w,1h}{v6}
\fmf{plain,left=1}{v5,v6,v5}
\fmf{plain}{v1,v2}
\fmf{plain}{v2,v3}
\fmf{plain}{v3,v4}
\fmf{plain}{i1,v1}
\fmf{plain}{v4,o1}
\fmfdot{v1,v2,v3,v4}
\end{fmfgraph*}
\end{center}}
\hspace*{0.3cm} \,\,
\begin{tabular}{@{}c}
${\scs \mbox{\#9.1}}$ \\
$1990656$\\ 
${\scs ( 0, 1, 0 ; 4 )}$
\end{tabular}
\parbox{13mm}{\begin{center}
\begin{fmfgraph*}(10,10)
\setval
\fmfforce{-1/4w,1/2h}{i1}
\fmfforce{5/4w,1/2h}{o1}
\fmfforce{0w,1/2h}{v1}
\fmfforce{1w,1/2h}{v2}
\fmfforce{1/2w,1/4h}{v3}
\fmfforce{1/2w,3/4h}{v4}
\fmf{plain}{v1,i1}
\fmf{plain}{v2,o1}
\fmf{plain,left=1}{v1,v2}
\fmf{plain,left=1}{v3,v4,v3}
\fmf{plain,right=0.5}{v1,v2,v1}
\fmfdot{v1,v2,v3,v4}
\end{fmfgraph*}
\end{center}}
\hspace*{0.3cm}
\begin{tabular}{@{}c}
${\scs \mbox{\#9.2}}$ \\
$1990656$\\ 
${\scs ( 0, 2, 0 ; 2 )}$
\end{tabular}
\parbox{23mm}{\begin{center}
\begin{fmfgraph*}(20,10)
\setval
\fmfforce{0w,0.5h}{i1}
\fmfforce{1/8w,0.5h}{v1}
\fmfforce{3/8w,0.5h}{v2}
\fmfforce{5/8w,0.5h}{v3}
\fmfforce{7/8w,0.5h}{v4}
\fmfforce{1w,0.5h}{o1}
\fmf{plain}{i1,v1}
\fmf{plain}{v4,o1}
\fmf{plain,left=1}{v1,v2,v1}
\fmf{plain}{v2,v3}
\fmf{plain,left=1}{v1,v3}
\fmf{plain,right=1}{v2,v4}
\fmf{plain,left=1}{v3,v4,v3}
\fmf{plain}{v4,o}
\fmfdot{v1,v2,v3,v4}
\end{fmfgraph*}
\end{center}}
\hspace*{0.3cm}
\begin{tabular}{@{}c}
${\scs \mbox{\#10.1}}$ \\
$221184$\\ 
${\scs ( 0, 0, 2 ; 2 )}$
\end{tabular}
\parbox{23mm}{\begin{center}
\begin{fmfgraph*}(20,5)
\setval
\fmfforce{0w,0.5h}{i1}
\fmfforce{1/8w,0.5h}{v1}
\fmfforce{3/8w,0.5h}{v2}
\fmfforce{5/8w,0.5h}{v3}
\fmfforce{7/8w,0.5h}{v4}
\fmfforce{1w,0.5h}{o1}
\fmf{plain}{i1,o1}
\fmf{plain,left=1}{v1,v2,v1}
\fmf{plain,left=1}{v3,v4,v3}
\fmfdot{v1,v2,v3,v4}
\end{fmfgraph*}
\end{center}}
\hspace*{0.3cm}
\begin{tabular}{@{}c}
${\scs \mbox{\#10.2}}$ \\
$663552$\\ 
${\scs ( 0, 1, 1 ; 2 )}$
\end{tabular}
\parbox{13mm}{\begin{center}
\begin{fmfgraph*}(10,12.5)
\setval
\fmfforce{1/4w,1/5h}{v1}
\fmfforce{3/4w,1/5h}{v2}
\fmfforce{1/4w,4/5h}{v3}
\fmfforce{3/4w,4/5h}{v4}
\fmfforce{0w,1/5h}{i1}
\fmfforce{1w,1/5h}{o1}
\fmf{plain}{i1,v1}
\fmf{plain}{v2,o1}
\fmf{plain,left=1}{v1,v2,v1}
\fmf{plain}{v3,v4}
\fmf{plain,left=1}{v3,v4,v3}
\fmf{plain,left=0.5}{v1,v3}
\fmf{plain,right=0.5}{v2,v4}
\fmfdot{v1,v2,v3,v4}
\end{fmfgraph*}
\end{center}}
\\ & 
\begin{tabular}{@{}c}
${\scs \mbox{\#11.1}}$ \\
$995328$\\ 
${\scs ( 0, 2, 0 ; 4 )}$
\end{tabular}
\parbox{11mm}{\begin{center}
\begin{fmfgraph*}(8,8)
\setval
\fmfforce{0.1875w,0h}{i1}
\fmfforce{0.8125w,0h}{o1}
\fmfforce{0.5w,0h}{v1}
\fmfforce{0.5w,1h}{v2}
\fmfforce{0.066987w,0.25h}{v3}
\fmfforce{0.93301w,0.25h}{v4}
\fmf{plain,left=1}{v1,v2,v1}
\fmf{plain}{i1,o1}
\fmf{plain}{v2,v3}
\fmf{plain}{v2,v4}
\fmf{plain}{v3,v4}
\fmfdot{v1,v2,v3,v4}
\end{fmfgraph*}
\end{center}}
\hspace*{0.3cm}
\begin{tabular}{@{}c}
${\scs \mbox{\#11.2}}$ \\
$1990656$\\ 
${\scs ( 1, 2, 0 ; 1 )}$
\end{tabular}
\hspace*{0.6cm}
\parbox{20mm}{\begin{center}
\begin{fmfgraph*}(8,8)
\setval
\fmfforce{-1.12051w,0.25h}{i1}
\fmfforce{-0.80801w,0.25h}{v1}
\fmfforce{-0.80801w,0.875h}{v5}
\fmfforce{1.24551w,0.25h}{o1}
\fmfforce{0.5w,1h}{v2}
\fmfforce{0.066987w,0.25h}{v3}
\fmfforce{0.93301w,0.25h}{v4}
\fmf{plain,right=0.55}{v2,v3}
\fmf{plain,left=0.55}{v2,v4}
\fmf{plain,left=1}{v1,v5,v1}
\fmf{plain}{i1,o1}
\fmf{plain}{v2,v3}
\fmf{plain}{v2,v4}
\fmf{plain}{v3,v4}
\fmfdot{v1,v2,v3,v4}
\end{fmfgraph*}
\end{center}}
\hspace*{0.3cm}
\begin{tabular}{@{}c}
${\scs \mbox{\#11.3}}$ \\
$995328$\\ 
${\scs ( 1, 2, 0 ; 2 )}$
\end{tabular}
\parbox{16mm}{\begin{center}
\begin{fmfgraph*}(8,13)
\setval
\fmfforce{0.5w,0h}{v1}
\fmfforce{0.5w,8/13h}{v2}
\fmfforce{0.0669873w,0.46154h}{v3}
\fmfforce{0.933w,0.46154h}{v4}
\fmfforce{0.5w,1h}{v5}
\fmfforce{-0.2455127w,0.46154h}{i1}
\fmfforce{1.2455w,0.46154h}{o1}
\fmf{plain,left=1}{v1,v2,v1}
\fmf{plain,left=1}{v2,v5,v2}
\fmf{plain}{v1,v3}
\fmf{plain}{i1,v3}
\fmf{plain}{o1,v4}
\fmf{plain}{v1,v4}
\fmfdot{v1,v2,v3,v4}
\end{fmfgraph*}\end{center}}
\hspace*{0.3cm}
\begin{tabular}{@{}c}
${\scs \mbox{\#11.4}}$ \\
$3981312$\\ 
${\scs ( 1, 1, 0 ; 1 )}$
\end{tabular}
\parbox{16mm}{\begin{center}
\begin{fmfgraph*}(13,7)
\setval
\fmfforce{0w,0h}{i1}
\fmfforce{2.5/13w,0h}{v1}
\fmfforce{10.5/13w,0h}{v2}
\fmfforce{1w,0h}{o1}
\fmfforce{0.23w,0.58h}{v3}
\fmfforce{6.5/13w,0.84h}{v4}
\fmf{plain,left=1.5}{v1,v2}
\fmf{plain}{i1,o1}
\fmf{plain}{v1,v4}
\fmf{plain}{v2,v4}
\fmfi{plain}{reverse fullcircle scaled 0.35w shifted (0.08w,0.78h)}
\fmfdot{v1,v2,v3,v4}
\end{fmfgraph*}\end{center}}
\quad \,
\begin{tabular}{@{}c}
${\scs \mbox{\#12.1}}$ \\
$995328$\\ 
${\scs ( 2, 1, 0 ; 2 )}$
\end{tabular}
\parbox{15mm}{\begin{center}
\begin{fmfgraph*}(8,13)
\setval
\fmfforce{0w,0.3h}{v1}
\fmfforce{1w,0.3h}{v2}
\fmfforce{-0.3125w,0.3h}{i1}
\fmfforce{1.3125w,0.3h}{o1}
\fmfforce{0.2w,0.55h}{v3}
\fmfforce{0.8w,0.55h}{v4}
\fmf{plain}{i1,v1}
\fmf{plain}{v2,o1}
\fmf{plain,left=1}{v1,v2,v1}
\fmf{plain}{v1,v2}
\fmfi{plain}{reverse fullcircle scaled 0.625w shifted (1w,0.7h)}
\fmfi{plain}{reverse fullcircle scaled 0.625w shifted (0w,0.7h)}
\fmfdot{v1,v2,v3,v4}
\end{fmfgraph*}\end{center}} \quad
\\ & 
\begin{tabular}{@{}c}
${\scs \mbox{\#12.2}}$ \\
$331776$\\ 
${\scs ( 2, 0, 1 ; 2 )}$
\end{tabular}
\hspace*{0.3cm}
\parbox{20mm}{\begin{center}
\begin{fmfgraph}(15,5)
\setval
\fmfforce{-1/3w,0.5h}{i1}
\fmfforce{1/6w,0.5h}{v1}
\fmfforce{1/2w,0.5h}{v2}
\fmfforce{5/6w,0.5h}{v3}
\fmfforce{5/6w,1.5h}{v4}
\fmfforce{-1/6w,0.5h}{v5}
\fmfforce{-1/6w,1.5h}{v6}
\fmfforce{1w,0.5h}{o1}
\fmf{plain}{i1,o1}
\fmf{plain,left=1}{v3,v4,v3}
\fmf{plain,left=1}{v1,v2,v1}
\fmf{plain,left=1}{v5,v6,v5}
\fmfdot{v1,v2,v3,v5}
\end{fmfgraph}
\end{center}}
\hspace*{0.3cm}
\begin{tabular}{@{}c}
${\scs \mbox{\#12.3}}$ \\
$663552$\\ 
${\scs ( 2, 0, 1 ; 1 )}$
\end{tabular}
\parbox{27mm}{\begin{center}
\begin{fmfgraph}(22.5,5)
\setval
\fmfforce{0w,0.5h}{i1}
\fmfforce{1/9w,0.5h}{v1}
\fmfforce{3/9w,0.5h}{v2}
\fmfforce{5/9w,0.5h}{v3}
\fmfforce{5/9w,1.5h}{v4}
\fmfforce{8/9w,0.5h}{v5}
\fmfforce{8/9w,1.5h}{v6}
\fmfforce{1w,0.5h}{o1}
\fmf{plain}{i1,o1}
\fmf{plain,left=1}{v1,v2,v1}
\fmf{plain,left=1}{v3,v4,v3}
\fmf{plain,left=1}{v5,v6,v5}
\fmfdot{v1,v2,v3,v5}
\end{fmfgraph}
\end{center}}
\hspace*{0.3cm}
\begin{tabular}{@{}c}
${\scs \mbox{\#12.4}}$ \\
$663552$\\ 
${\scs ( 1, 0, 1 ; 2 )}$
\end{tabular}
\parbox{15mm}{\begin{center}
\begin{fmfgraph}(8,8)
\setval
\fmfforce{0.1875w,0h}{i1}
\fmfforce{0.8125w,0h}{o1}
\fmfforce{0.5w,0h}{v1}
\fmfforce{0.5w,1h}{v2}
\fmfforce{0.1w,0.2h}{v3}
\fmfforce{0.9w,0.8h}{v4}
\fmfforce{0.97w,0.35h}{v5}
\fmf{plain}{i1,o1}
\fmf{plain,left=1}{v1,v2,v1}
\fmf{plain,left=0.4}{v3,v4,v3}
\fmfi{plain}{reverse fullcircle scaled 0.625w shifted (1.25w,0.2h)}
\fmfdot{v1,v3,v4,v5}
\end{fmfgraph}\end{center}}
\hspace*{0.3cm}
\begin{tabular}{@{}c}
${\scs \mbox{\#13.1}}$ \\
$995328$\\ 
${\scs ( 2, 0, 0 ; 4 )}$
\end{tabular}
\parbox{15mm}{\begin{center}
\begin{fmfgraph*}(10,15)
\setval
\fmfforce{0w,0.5h}{i1}
\fmfforce{1w,0.5h}{o1}
\fmfforce{1/4w,0.5h}{v1}
\fmfforce{3/4w,0.5h}{v2}
\fmfforce{0.5w,1h}{v4}
\fmfforce{0.5w,0h}{v5}
\fmfforce{0.5w,1/3h}{v6}
\fmfforce{0.5w,2/3h}{v7}
\fmf{plain,left=1}{v5,v6,v5}
\fmf{plain,left=1}{v7,v4,v7}
\fmf{plain,left=1}{v1,v2,v1}
\fmf{plain}{i1,o1}
\fmfdot{v1,v2,v6,v7}
\end{fmfgraph*}\end{center}} 
\hspace*{0.3cm}
\begin{tabular}{@{}c}
${\scs \mbox{\#13.2}}$ \\
$995328$\\ 
${\scs ( 1, 1, 0 ; 4 )}$
\end{tabular}
\parbox{10.5mm}{\begin{center}
\begin{fmfgraph*}(7.5,17.5)
\setval
\fmfforce{1/6w,0.2857h}{i1}
\fmfforce{5/6w,0.2857h}{o1}
\fmfforce{0w,0.5h}{v1}
\fmfforce{1w,0.5h}{v2}
\fmfforce{0.5w,h}{v3}
\fmfforce{0.5w,1h}{v4}
\fmfforce{0.5w,0.2857h}{v6}
\fmfforce{0.5w,0.71429h}{v7}
\fmf{plain}{i1,o1}
\fmf{plain,left=1}{v5,v6,v5}
\fmf{plain,left=1}{v7,v4,v7}
\fmf{plain,left=1}{v1,v2,v1}
\fmf{plain,left=0.4}{v1,v2,v1}
\fmfdot{v1,v2,v6,v7}
\end{fmfgraph*}\end{center}} 
\\ & 
\begin{tabular}{@{}c}
${\scs \mbox{\#13.3}}$ \\
$1990656$\\ 
${\scs ( 2, 1, 0 ; 1 )}$
\end{tabular}
\parbox{18mm}{\begin{center}
\begin{fmfgraph*}(15,10)
\setval
\fmfforce{0w,0.25h}{i1}
\fmfforce{1w,0.25h}{o1}
\fmfforce{1/6w,0.25h}{v1}
\fmfforce{1/2w,0.25h}{v2}
\fmfforce{1/3w,0.5h}{v3}
\fmfforce{1/3w,1h}{v4}
\fmfforce{5/6w,0.25h}{v5}
\fmfforce{5/6w,0.75h}{v6}
\fmf{plain}{i1,o1}
\fmf{plain,left=1}{v1,v2,v1}
\fmf{plain,left=1}{v3,v4,v3}
\fmf{plain,left=1}{v5,v6,v5}
\fmfdot{v1,v2,v3,v5}
\end{fmfgraph*}\end{center}} 
\hspace*{0.3cm}
\begin{tabular}{@{}c}
${\scs \mbox{\#14.1}}$ \\
$995328$\\ 
${\scs ( 1, 2, 0 ; 2 )}$\
\end{tabular}
\parbox{18mm}{\begin{center}
\begin{fmfgraph*}(10,15)
\setval
\fmfforce{0w,1/6h}{i1}
\fmfforce{1w,1/6h}{o1}
\fmfforce{1/4w,1/6h}{v1}
\fmfforce{3/4w,1/6h}{v2}
\fmfforce{1/2w,1/3h}{v3}
\fmfforce{1/2w,2/3h}{v4}
\fmfforce{1/2w,1h}{v5}
\fmf{plain}{i1,o1}
\fmf{plain,left=1}{v1,v2,v1}
\fmf{plain,left=1}{v3,v4,v3}
\fmf{plain,left=1}{v4,v5,v4}
\fmfdot{v1,v2,v3,v4}
\end{fmfgraph*}\end{center}} 
\hspace*{0.3cm}
\begin{tabular}{@{}c}
${\scs \mbox{\#14.2}}$ \\
$663552$\\ 
${\scs ( 1, 1, 1 ; 1 )}$
\end{tabular}
\parbox{20mm}{\begin{center}
\begin{fmfgraph}(15,10)
\setval
\fmfforce{0w,1/4h}{i1}
\fmfforce{1/6w,1/4h}{v1}
\fmfforce{1/2w,1/4h}{v2}
\fmfforce{5/6w,1/4h}{v3}
\fmfforce{5/6w,3/4h}{v4}
\fmfforce{5/6w,5/4h}{v5}
\fmfforce{1w,1/4h}{o1}
\fmf{plain}{i1,o1}
\fmf{plain,left=1}{v3,v4,v3}
\fmf{plain,left=1}{v1,v2,v1}
\fmf{plain,left=1}{v4,v5,v4}
\fmfdot{v1,v2,v3,v4}
\end{fmfgraph}
\end{center}}
\hspace*{0.3cm}
\begin{tabular}{@{}c}
${\scs \mbox{\#14.3}}$ \\
$663552$\\ 
${\scs ( 1, 0, 1 ; 2 )}$
\end{tabular}
\parbox{25mm}{\begin{center}
\begin{fmfgraph*}(8,8)
\setval
\fmfforce{-0.625w,0h}{v5}
\fmfforce{-0.625w,5/8h}{v6}
\fmfforce{0w,0.5h}{v1}
\fmfforce{1w,0.5h}{v2}
\fmfforce{0.5w,0h}{v4}
\fmfforce{-0.9375w,0h}{i1}
\fmfforce{0.8125w,0h}{o1}
\fmf{plain,left=1}{v1,v2,v1}
\fmf{plain,left=0.4}{v1,v2,v1}
\fmf{plain,left=1}{v3,v4,v3}
\fmf{plain,left=1}{v5,v6,v5}
\fmf{plain}{i1,o1}
\fmfdot{v1,v2,v3,v5}
\end{fmfgraph*}\end{center}} 
\hspace*{-0.3cm}
\begin{tabular}{@{}c}
${\scs \mbox{\#14.4}}$ \\
$331776$\\ 
${\scs ( 0, 1, 1 ; 4 )}$
\end{tabular}
\parbox{11mm}{\begin{center}
\begin{fmfgraph*}(8,13)
\setval
\fmfforce{0.5w,0h}{v5}
\fmfforce{0.1875w,0h}{i1}
\fmfforce{0.8125w,0h}{o1}
\fmfforce{0w,9/13h}{v1}
\fmfforce{1w,9/13h}{v2}
\fmfforce{0.5w,5/13h}{v4}
\fmf{plain}{i1,o1}
\fmf{plain,left=1}{v1,v2,v1}
\fmf{plain,left=0.4}{v1,v2,v1}
\fmf{plain,left=1}{v3,v4,v3}
\fmf{plain,left=1}{v5,v4,v5}
\fmf{plain}{i1,o1}
\fmfdot{v1,v2,v3,v5}
\end{fmfgraph*}\end{center}} 
\\ & 
\hspace*{-0.7cm}
\begin{tabular}{@{}c}
${\scs \mbox{\#15.1}}$ \\
$995328$\\
${\scs ( 3, 1, 0 ; 1 )}$
\end{tabular}
\parbox{22mm}{\begin{center}
\begin{fmfgraph}(20,10)
\setval
\fmfforce{0w,0h}{i1}
\fmfforce{1/8w,0h}{v1}
\fmfforce{1/8w,0.5h}{v2}
\fmfforce{1/8w,1h}{v5}
\fmfforce{1/2w,0h}{v3}
\fmfforce{1/2w,0.5h}{v4}
\fmfforce{7/8w,0h}{v7}
\fmfforce{7/8w,0.5h}{v8}
\fmfforce{1w,0h}{o1}
\fmf{plain}{i1,o1}
\fmf{plain,left=1}{v1,v2,v1}
\fmf{plain,left=1}{v3,v4,v3}
\fmf{plain,left=1}{v2,v5,v2}
\fmf{plain,left=1}{v7,v8,v7}
\fmfdot{v1,v2,v3,v7}
\end{fmfgraph}
\end{center}}
\hspace*{0.1cm}
\begin{tabular}{@{}c}
${\scs \mbox{\#15.2}}$ \\
$497664$\\ 
${\scs ( 3, 1, 0 ; 2 )}$
\end{tabular}
\parbox{22mm}{\begin{center}
\begin{fmfgraph}(20,10)
\setval
\fmfforce{0w,0h}{i1}
\fmfforce{1/8w,0h}{v1}
\fmfforce{1/8w,1/2h}{v2}
\fmfforce{1/2w,0h}{v3}
\fmfforce{1/2w,1/2h}{v4}
\fmfforce{1/2w,1h}{v7}
\fmfforce{7/8w,0h}{v5}
\fmfforce{7/8w,1/2h}{v6}
\fmfforce{1w,0h}{o1}
\fmf{plain}{i1,o1}
\fmf{plain,left=1}{v1,v2,v1}
\fmf{plain,left=1}{v3,v4,v3}
\fmf{plain,left=1}{v5,v6,v5}
\fmf{plain,left=1}{v4,v7,v4}
\fmfdot{v1,v3,v5,v4}
\end{fmfgraph}
\end{center}}
\hspace*{0.1cm}
\begin{tabular}{@{}c}
${\scs \mbox{\#15.3}}$ \\
$497664$\\ 
${\scs ( 2, 1, 0 ; 4 )}$
\end{tabular}
\parbox{17mm}{\begin{center}
\begin{fmfgraph*}(15,20)
\setval
\fmfforce{1/3w,1/4h}{i1}
\fmfforce{2/3w,1/4h}{o1}
\fmfforce{1/2w,1/4h}{v1}
\fmfforce{1/2w,1/2h}{v8}
\fmfforce{1/2w,3/4h}{v2}
\fmfforce{1/2w,1h}{v3}
\fmfforce{0.355662432w,0.6975h}{v4}
\fmfforce{0.64433568w,0.6975h}{v5}
\fmf{plain,left=1}{v8,v2,v8}
\fmf{plain}{i1,v1}
\fmf{plain}{v1,o1}
\fmfi{plain}{reverse fullcircle scaled 0.33w shifted (0.225w,0.765h)}
\fmfi{plain}{reverse fullcircle scaled 0.33w shifted (0.775w,0.765h)}
\fmf{plain,left=1}{v1,v8,v1}
\fmfdot{v1,v4,v5,v8}
\end{fmfgraph*}\end{center}} 
\begin{tabular}{@{}c}
${\scs \mbox{\#15.4}}$ \\
$995328$\\ 
${\scs ( 2, 1, 0 ; 2 )}$
\end{tabular}
\hspace*{0.2cm}
\parbox{18mm}{\begin{center}
\begin{fmfgraph*}(15,15)
\setval
\fmfforce{1/3w,1/3h}{i1}
\fmfforce{2/3w,1/3h}{o1}
\fmfforce{1/2w,1/3h}{v1}
\fmfforce{1/2w,2/3h}{v2}
\fmfforce{1/2w,1h}{v3}
\fmfforce{0.355662432w,0.5833333333h}{v4}
\fmfforce{0.64433568w,0.5833333333h}{v5}
\fmfforce{0.067w,3/4h}{v6}
\fmfforce{0.933w,3/4h}{v7}
\fmfforce{-0.22166w,0.9166666666666h}{v8}
\fmf{plain,left=1}{v6,v8,v6}
\fmf{plain,left=1}{v1,v2,v1}
\fmf{plain}{i1,v1}
\fmf{plain}{v1,o1}
\fmf{plain,left=1}{v4,v6,v4}
\fmf{plain,left=1}{v5,v7,v5}
\fmfdot{v1,v4,v5,v6}
\end{fmfgraph*}\end{center}} 
\begin{tabular}{@{}c}
${\scs \mbox{\#15.5}}$ \\
$995328$\\ 
${\scs ( 3, 0, 0 ; 2 )}$
\end{tabular}
\parbox{17mm}{\begin{center}
\begin{fmfgraph*}(15,15)
\setval
\fmfforce{1/3w,1/3h}{i1}
\fmfforce{4/3w,1/3h}{o1}
\fmfforce{7/6w,1/3h}{v8}
\fmfforce{7/6w,2/3h}{v9}
\fmfforce{1/2w,1/3h}{v1}
\fmfforce{1/2w,2/3h}{v2}
\fmfforce{1/2w,1h}{v3}
\fmfforce{0.355662432w,0.5833333333h}{v4}
\fmfforce{0.64433568w,0.5833333333h}{v5}
\fmfforce{0.067w,3/4h}{v6}
\fmfforce{0.933w,3/4h}{v7}
\fmf{plain,left=1}{v1,v2,v1}
\fmf{plain,left=1}{v8,v9,v8}
\fmf{plain}{i1,v1}
\fmf{plain}{v1,o1}
\fmf{plain,left=1}{v4,v6,v4}
\fmf{plain,left=1}{v5,v7,v5}
\fmfdot{v1,v4,v5,v8}
\end{fmfgraph*}\end{center}} 
\\ & 
\begin{tabular}{@{}c}
${\scs\mbox{\#16.1}}$ \\
$497664$ \\ 
${\scs ( 3, 0, 0 ; 4 )}$\\
$\mbox{}$
\end{tabular}
\parbox{18mm}{\begin{center}
\begin{fmfgraph*}(15,10)
\setval
\fmfforce{1/3w,0h}{i1}
\fmfforce{2/3w,0h}{o1}
\fmfforce{1/3w,1/4h}{v1}
\fmfforce{2/3w,1/4h}{v2}
\fmfforce{1/2w,0h}{v3}
\fmfforce{1/2w,1/2h}{v4}
\fmfforce{1/2w,1h}{v5}
\fmfforce{0w,1/4h}{v7}
\fmfforce{1w,1/4h}{v8}
\fmf{plain}{i1,o1}
\fmf{plain,left=1}{v3,v4,v3}
\fmf{plain,left=1}{v4,v5,v4}
\fmf{plain,left=1}{v3,v6,v3}
\fmf{plain,left=1}{v1,v7,v1}
\fmf{plain,left=1}{v2,v8,v2}
\fmfdot{v1,v2,v3,v4}
\end{fmfgraph*}\end{center}} 
\hspace*{0.3cm}
\begin{tabular}{@{}c}
${\scs \mbox{\#16.2}}$ \\
$497664$\\ 
${\scs ( 4, 0, 0 ; 2 )}$
\end{tabular}
\parbox{31mm}{\begin{center}
\begin{fmfgraph}(27.5,5)
\setval
\fmfforce{0w,0h}{i1}
\fmfforce{1/11w,0h}{v1}
\fmfforce{1/11w,1h}{v2}
\fmfforce{4/11w,0h}{v3}
\fmfforce{4/11w,1h}{v4}
\fmfforce{7/11w,0h}{v5}
\fmfforce{7/11w,1h}{v6}
\fmfforce{10/11w,0h}{v7}
\fmfforce{10/11w,1h}{v8}
\fmfforce{1w,0h}{o1}
\fmf{plain}{i1,o1}
\fmf{plain,left=1}{v1,v2,v1}
\fmf{plain,left=1}{v3,v4,v3}
\fmf{plain,left=1}{v5,v6,v5}
\fmf{plain,left=1}{v7,v8,v7}
\fmfdot{v1,v3,v5,v7}
\end{fmfgraph}
\end{center}}
\hspace*{0.3cm}
\begin{tabular}{@{}c}
${\scs \mbox{\#17.1}}$ \\
$497664$\\ 
${\scs ( 1, 3, 0 ; 2 )}$
\end{tabular}
\parbox{8mm}{\begin{center}
\begin{fmfgraph}(5,20)
\setval
\fmfforce{0w,0h}{i1}
\fmfforce{0.5w,0h}{v1}
\fmfforce{1w,0h}{o1}
\fmfforce{0.5w,1/4h}{v2}
\fmfforce{0.5w,1/2h}{v3}
\fmfforce{0.5w,3/4h}{v4}
\fmfforce{0.5w,1h}{v5}
\fmf{plain}{i1,v1}
\fmf{plain}{v1,o1}
\fmf{plain,left=1}{v1,v2,v1}
\fmf{plain,left=1}{v2,v3,v2}
\fmf{plain,left=1}{v3,v4,v3}
\fmf{plain,left=1}{v5,v4,v5}
\fmfdot{v1,v2,v3,v4}
\end{fmfgraph}
\end{center}}
\hspace*{0.3cm}
\begin{tabular}{@{}c}
${\scs \mbox{\#17.2}}$ \\
$995328$\\ 
${\scs ( 2, 2, 0 ; 1 )}$\\
\end{tabular}
\parbox{16mm}{\begin{center}
\begin{fmfgraph}(12.5,15)
\setval
\fmfforce{0w,0h}{i1}
\fmfforce{1/5w,0h}{v1}
\fmfforce{1w,0h}{o1}
\fmfforce{1/5w,1/3h}{v2}
\fmfforce{1/5w,2/3h}{v3}
\fmfforce{1/5w,1h}{v4}
\fmfforce{4/5w,0h}{v5}
\fmfforce{4/5w,1/3h}{v6}
\fmf{plain}{i1,v1}
\fmf{plain}{v1,o1}
\fmf{plain,left=1}{v1,v2,v1}
\fmf{plain,left=1}{v2,v3,v2}
\fmf{plain,left=1}{v3,v4,v3}
\fmf{plain,left=1}{v5,v6,v5}
\fmfdot{v1,v2,v3,v5}
\end{fmfgraph}
\end{center}}
\hspace*{0.3cm}
\begin{tabular}{@{}c}
${\scs \mbox{\#17.3}}$ \\
$497664$\\ 
${\scs ( 2, 2, 0 ; 2 )}$\\
\end{tabular}
\parbox{15mm}{\begin{center}
\begin{fmfgraph}(12,10)
\setval
\fmfforce{0w,0h}{i1}
\fmfforce{1/5w,0h}{v1}
\fmfforce{1/5w,0.5h}{v2}
\fmfforce{1/5w,1h}{v5}
\fmfforce{4/5w,0h}{v3}
\fmfforce{4/5w,0.5h}{v4}
\fmfforce{4/5w,1h}{v6}
\fmfforce{1w,0h}{o1}
\fmf{plain}{i1,o1}
\fmf{plain,left=1}{v1,v2,v1}
\fmf{plain,left=1}{v3,v4,v3}
\fmf{plain,left=1}{v2,v5,v2}
\fmf{plain,left=1}{v4,v6,v4}
\fmfdot{v1,v2,v3,v4}
\end{fmfgraph}
\end{center}}
\end{tabular}
\end{center}
\caption{\la{TWO} Connected diagrams of the two-point function and their 
multiplicities of the $\phi^4$-theory
up to four loops. Each diagram is characterized by the
vector $(S,D,T;N$) whose components specify the number of self-, double,
triple connections, and of the identical vertex permutations,
respectively.}
\end{table}
\end{fmffile}
\newpage
\begin{fmffile}{graph4}
\hspace*{-2cm}
\vspace*{-1cm}
\setlength{\unitlength}{1mm}
\begin{table}[t]
\begin{center}
\begin{tabular}{|c|c|}
\,\,\,$p$\,\,\,
&
\begin{tabular}{@{}c}
$\mbox{}$\\
$\mbox{}$
\end{tabular}
$\fullg_{1234}^{{\rm c},(p)}$
\\
\hline
$1$ &
\hspace{-10pt}
\rule[-10pt]{0pt}{26pt}
\begin{tabular}{@{}c}
$\mbox{}$\\
${\scs \#1.1.1}$ \\
${\scs 24}$\\ 
${\scs ( 0, 0, 0 ; 24 )}$\\
$\mbox{}$\\
\end{tabular}
\parbox{8mm}{\begin{center}
\begin{fmfgraph}(5,5)
\setval
\fmfstraight
\fmfforce{0w,0h}{i1}
\fmfforce{0w,1h}{i2}
\fmfforce{1w,0h}{o1}
\fmfforce{1w,1h}{o2}
\fmfforce{0.5w,0.5h}{v1}
\fmf{plain}{i1,o2}
\fmf{plain}{i2,o1}
\fmfdot{v1}
\end{fmfgraph}
\end{center}}
\\ \hline
$2$ &
\hspace{-10pt}
\rule[-10pt]{0pt}{26pt}
\begin{tabular}{@{}c}
$\mbox{}$\\
${\scs \#2.1.1, \#3.1.1}$ \\
${\scs 1152,576}$\\ 
${\scs 1728}$\\
${\scs ( 0, 1, 0 ; 8 )}$\\
$\mbox{}$\\
\end{tabular}
\parbox{13mm}{\begin{center}
\begin{fmfgraph}(10,5)
\setval
\fmfstraight
\fmfforce{0w,0h}{i1}
\fmfforce{0w,1h}{i2}
\fmfforce{1w,0h}{o1}
\fmfforce{1w,1h}{o2}
\fmfforce{1/4w,0.5h}{v1}
\fmfforce{3/4w,0.5h}{v2}
\fmf{plain}{i1,v1}
\fmf{plain}{v2,o1}
\fmf{plain}{i2,v1}
\fmf{plain}{v2,o2}
\fmf{plain,left=1}{v1,v2,v1}
\fmfdot{v1,v2}
\end{fmfgraph}
\end{center}}
\quad
\begin{tabular}{@{}c}
$\mbox{}$\\
${\scs \#3.1.2, \#3.2.1}$ \\
${\scs 1152,1152}$\\ 
${\scs 2304}$\\
${\scs ( 1, 0, 0 ; 6 )}$\\
$\mbox{}$\\
\end{tabular}
\parbox{13mm}{\begin{center}
\begin{fmfgraph}(10,10)
\setval
\fmfstraight
\fmfforce{0w,1/2h}{i1}
\fmfforce{1/4w,3/4h}{i2}
\fmfforce{1/4w,1/4h}{i3}
\fmfforce{1w,1/2h}{o1}
\fmfforce{3/4w,1/2h}{v1}
\fmfforce{3/4w,1h}{v2}
\fmfforce{1/4w,1/2h}{v3}
\fmf{plain}{i1,o1}
\fmf{plain}{i2,v3}
\fmf{plain}{i3,v3}
\fmf{plain,left=1}{v1,v2,v1}
\fmfdot{v1,v3}
\end{fmfgraph}
\end{center}}
\\ \hline
$3$ &
\hspace{-10pt}
\rule[-10pt]{0pt}{26pt}
\begin{tabular}{@{}c}
$\mbox{}$\\
${\scs \#4.1.1, \#7.1.1}$ \\
${\scs 41472,20736}$\\ 
${\scs 62208}$\\
${\scs ( 0, 2, 0 ; 8 )}$\\
$\mbox{}$\\
\end{tabular}
\parbox{18mm}{\begin{center}
\begin{fmfgraph}(15,5)
\setval
\fmfstraight
\fmfforce{0w,0h}{i1}
\fmfforce{0w,1h}{i2}
\fmfforce{1w,0h}{o1}
\fmfforce{1w,1h}{o2}
\fmfforce{1/6w,0.5h}{v1}
\fmfforce{1/2w,0.5h}{v2}
\fmfforce{5/6w,0.5h}{v3}
\fmf{plain}{i1,v1}
\fmf{plain}{v3,o1}
\fmf{plain}{i2,v1}
\fmf{plain}{v3,o2}
\fmf{plain,left=1}{v1,v2,v1}
\fmf{plain,left=1}{v2,v3,v2}
\fmfdot{v1,v2,v3}
\end{fmfgraph}
\end{center}}
\quad
\begin{tabular}{@{}c}
$\mbox{}$\\
${\scs \#4.1.2, \#5.1.1, \#5.2.1}$ \\
${\scs 165888,41472,41472}$\\ 
${\scs 248832}$\\
${\scs ( 0, 1, 0 ; 4 )}$\\
$\mbox{}$\\
\end{tabular}
\parbox{13mm}{\begin{center}
\begin{fmfgraph}(10,5)
\setval
\fmfstraight
\fmfforce{0w,1/2h}{i1}
\fmfforce{1w,1/2h}{o1}
\fmfforce{1/4w,1/2h}{v1}
\fmfforce{3/4w,1/2h}{v2}
\fmfforce{1/2w,1h}{v3}
\fmfforce{0.75w,1.5h}{i2}
\fmfforce{0.25w,1.5h}{o2}
\fmf{plain}{i1,o1}
\fmf{plain}{i2,v3}
\fmf{plain}{o2,v3}
\fmf{plain,left=1}{v1,v2,v1}
\fmfdot{v1,v2,v3}
\end{fmfgraph}
\end{center}}
\quad
\begin{tabular}{@{}c}
$\mbox{}$\\
${\scs \#5.1.2, \#5.3.2}$ \\
${\scs 27648,27648}$\\ 
${\scs 55296}$\\
${\scs ( 0, 0, 1 ; 6 )}$\\
$\mbox{}$\\
\end{tabular}
\parbox{15.5mm}{\begin{center}
\begin{fmfgraph}(12.5,5)
\setval
\fmfstraight
\fmfforce{0w,1/2h}{i1}
\fmfforce{1/5w,0h}{i2}
\fmfforce{1/5w,1h}{i3}
\fmfforce{1w,1/2h}{o1}
\fmfforce{1/5w,1/2h}{v1}
\fmfforce{2/5w,1/2h}{v2}
\fmfforce{4/5w,1/2h}{v3}
\fmf{plain}{i1,o1}
\fmf{plain}{i2,v1}
\fmf{plain}{i3,v1}
\fmf{plain,left=1}{v2,v3,v2}
\fmfdot{v1,v3,v2}
\end{fmfgraph}
\end{center}}
\\
&
\hspace{-10pt}
\rule[-10pt]{0pt}{26pt}
\begin{tabular}{@{}c}
$\mbox{}$\\
${\scs \#5.2.2, \#6.1.1}$ \\
${\scs 82944,41472}$\\ 
${\scs 124416}$\\
${\scs ( 1, 0, 0 ; 8 )}$\\
$\mbox{}$\\
\end{tabular}
\parbox{13mm}{\begin{center}
\begin{fmfgraph}(10,10)
\setval
\fmfstraight
\fmfforce{0w,0h}{i1}
\fmfforce{0w,1/2h}{i2}
\fmfforce{1w,0h}{o1}
\fmfforce{1w,1/2h}{o2}
\fmfforce{1/4w,1/4h}{v1}
\fmfforce{3/4w,1/4h}{v2}
\fmfforce{1/2w,1/2h}{v3}
\fmfforce{1/2w,1h}{v4}
\fmf{plain}{i1,v1}
\fmf{plain}{v2,o1}
\fmf{plain}{i2,v1}
\fmf{plain}{v2,o2}
\fmf{plain,left=1}{v1,v2,v1}
\fmf{plain,left=1}{v3,v4,v3}
\fmfdot{v1,v2,v3}
\end{fmfgraph}
\end{center}}
\quad
\begin{tabular}{@{}c}
$\mbox{}$\\
${\scs \#5.2.3, \#5.3.1, \#7.1.2, \#7.2.1}$ \\
${\scs 82944,82944,41472,41472}$\\ 
${\scs 248832}$\\
${\scs ( 1, 1, 0 ; 2 )}$\\
$\mbox{}$\\
\end{tabular}
\parbox{18mm}{\begin{center}
\begin{fmfgraph*}(12.5,5)
\setval
\fmfstraight
\fmfforce{0w,0h}{i1}
\fmfforce{1w,0h}{o1}
\fmfforce{0w,3/2h}{i2}
\fmfforce{2/5w,3/2h}{o2}
\fmfforce{1/5w,0h}{v1}
\fmfforce{1/5w,1h}{v2}
\fmfforce{4/5w,0h}{v3}
\fmfforce{4/5w,1h}{v4}
\fmf{plain,left=1}{v1,v2,v1}
\fmf{plain,left=1}{v3,v4,v3}
\fmf{plain}{i1,o1}
\fmf{plain}{i2,v2}
\fmf{plain}{o2,v2}
\fmfdot{v1,v2,v3}
\end{fmfgraph*}
\end{center}}
\quad 
\begin{tabular}{@{}c}
$\mbox{}$\\
${\scs \#6.1.2, \#6.2.2, \#7.2.2}$ \\
${\scs 20736,20736,82944}$\\ 
${\scs 124416}$\\
${\scs ( 2, 0, 0 ; 4 )}$\\
$\mbox{}$\\
\end{tabular}
\parbox{15.5mm}{\begin{center}
\begin{fmfgraph}(12.5,5)
\setval
\fmfstraight
\fmfforce{0w,0h}{i1}
\fmfforce{1/2w,1/2h}{i2}
\fmfforce{1/2w,-1/2h}{o1}
\fmfforce{1w,0h}{o2}
\fmfforce{1/5w,0h}{v1}
\fmfforce{1/5w,1h}{v2}
\fmfforce{4/5w,0h}{v3}
\fmfforce{4/5w,1h}{v4}
\fmfforce{1/2w,0h}{v5}
\fmf{plain}{i1,o2}
\fmf{plain}{i2,v5}
\fmf{plain}{o1,v5}
\fmf{plain,left=1}{v1,v2,v1}
\fmf{plain,left=1}{v3,v4,v3}
\fmfdot{v1,v3,v5}
\end{fmfgraph}
\end{center}}
\\
&
\hspace{-10pt}
\rule[-10pt]{0pt}{26pt}
\begin{tabular}{@{}c}
$\mbox{}$\\
${\scs \#6.1.3, \#6.2.1}$ \\
${\scs 41472,41472}$\\ 
${\scs 82944}$\\
${\scs ( 2, 0, 0 ; 6 )}$\\
$\mbox{}$\\
\end{tabular}
\parbox{20.5mm}{\begin{center}
\begin{fmfgraph}(17.5,10)
\setval
\fmfstraight
\fmfforce{0w,1/2h}{i1}
\fmfforce{1/7w,3/4h}{i2}
\fmfforce{1/7w,1/4h}{i3}
\fmfforce{1w,1/2h}{o1}
\fmfforce{3/7w,1/2h}{v1}
\fmfforce{3/7w,1h}{v2}
\fmfforce{1/7w,1/2h}{v3}
\fmfforce{6/7w,1/2h}{v4}
\fmfforce{6/7w,1h}{v5}
\fmf{plain}{i1,o1}
\fmf{plain}{i2,v3}
\fmf{plain}{i3,v3}
\fmf{plain,left=1}{v1,v2,v1}
\fmf{plain,left=1}{v4,v5,v4}
\fmfdot{v1,v3,v4}
\end{fmfgraph}
\end{center}}
\quad
\begin{tabular}{@{}c}
$\mbox{}$\\
${\scs \#7.1.3, \#7.2.3}$ \\
${\scs 41472,41472}$\\ 
${\scs 82944}$\\
${\scs ( 1, 1, 0 ; 6 )}$\\
$\mbox{}$\\
\end{tabular}
\parbox{13mm}{\begin{center}
\begin{fmfgraph}(10,15)
\setval
\fmfstraight
\fmfforce{0w,1/3h}{i1}
\fmfforce{1/4w,1/2h}{i2}
\fmfforce{1/4w,1/6h}{i3}
\fmfforce{1w,1/3h}{o1}
\fmfforce{3/4w,1/3h}{v1}
\fmfforce{3/4w,2/3h}{v2}
\fmfforce{1/4w,1/3h}{v3}
\fmfforce{3/4w,1h}{v4}
\fmf{plain}{i1,o1}
\fmf{plain}{i2,v3}
\fmf{plain}{i3,v3}
\fmf{plain,left=1}{v1,v2,v1}
\fmf{plain,left=1}{v4,v2,v4}
\fmfdot{v1,v2,v3}
\end{fmfgraph}
\end{center}}
\\ \hline
$4$ &
\hspace{-10pt}
\rule[-10pt]{0pt}{26pt}
\begin{tabular}{@{}c}
$\mbox{}$\\
${\scs \#8.1.1, \#17.1.1}$ \\
${\scs 1990656,995328}$\\ 
${\scs 2985984}$\\
${\scs ( 0, 3, 0 ; 8 )}$\\
$\mbox{}$\\
\end{tabular}
\parbox{23mm}{\begin{center}
\begin{fmfgraph}(20,5)
\setval
\fmfstraight
\fmfforce{0w,0h}{i1}
\fmfforce{0w,1h}{i2}
\fmfforce{1w,0h}{o1}
\fmfforce{1w,1h}{o2}
\fmfforce{1/8w,0.5h}{v1}
\fmfforce{3/8w,0.5h}{v2}
\fmfforce{5/8w,0.5h}{v3}
\fmfforce{7/8w,0.5h}{v4}
\fmf{plain}{i1,v1}
\fmf{plain}{v4,o1}
\fmf{plain}{i2,v1}
\fmf{plain}{v4,o2}
\fmf{plain,left=1}{v1,v2,v1}
\fmf{plain,left=1}{v2,v3,v2}
\fmf{plain,left=1}{v4,v3,v4}
\fmfdot{v1,v2,v3,v4}
\end{fmfgraph}
\end{center}}
\quad
\begin{tabular}{@{}c}
$\mbox{}$\\
${\scs \#8.1.2, \#9.2.1, \#10.2.1}$ \\
${\scs 3981312,3981312,3981312}$\\ 
${\scs 11943936}$\\
${\scs ( 0, 2, 0 ; 4 )}$\\
$\mbox{}$\\
\end{tabular}
\parbox{15.5mm}{\begin{center}
\begin{fmfgraph}(12.5,5)
\setval
\fmfstraight
\fmfforce{0w,0h}{i1}
\fmfforce{0w,1h}{i2}
\fmfforce{1w,0h}{o1}
\fmfforce{1w,1h}{o2}
\fmfforce{1/5w,0h}{v1}
\fmfforce{1/5w,1h}{v2}
\fmfforce{4/5w,0h}{v3}
\fmfforce{4/5w,1h}{v4}
\fmf{plain}{i1,o1}
\fmf{plain}{i2,o2}
\fmf{plain,left=1}{v1,v2,v1}
\fmf{plain,left=1}{v4,v3,v4}
\fmfdot{v1,v2,v3,v4}
\end{fmfgraph}
\end{center}}
\quad
\begin{tabular}{@{}c}
$\mbox{}$\\
${\scs \#8.1.3, \#11.1.2, \#11.3.1}$ \\
${\scs 7962624,1990656,1990656}$\\ 
${\scs 11943936}$\\ 
${\scs ( 0, 2, 0 ; 4 )}$\\
$\mbox{}$\\
\end{tabular}
\parbox{18mm}{\begin{center}
\begin{fmfgraph}(15,7.5)
\setval
\fmfstraight
\fmfforce{0w,1/3h}{i1}
\fmfforce{1w,1/3h}{o1}
\fmfforce{1/6w,1/3h}{v1}
\fmfforce{1/2w,1/3h}{v2}
\fmfforce{5/6w,1/3h}{v3}
\fmfforce{1/2w,1h}{v4}
\fmfforce{1/3w,4/3h}{i2}
\fmfforce{2/3w,4/3h}{o2}
\fmf{plain}{i1,v1}
\fmf{plain}{v3,o1}
\fmf{plain}{i2,v4}
\fmf{plain}{o2,v4}
\fmf{plain,left=1}{v1,v2,v1}
\fmf{plain,left=1}{v2,v3,v2}
\fmf{plain,left=1}{v1,v3}
\fmfdot{v1,v2,v3,v4}
\end{fmfgraph}
\end{center}}
\\ 
&
\hspace{-10pt}
\rule[-10pt]{0pt}{26pt}
\begin{tabular}{@{}c}
$\mbox{}$\\
${\scs \#9.1.1, \#13.2.1}$ \\
${\scs 3981312,1990656}$\\
${\scs 5971968}$\\
${\scs ( 0, 1, 0 ; 16)}$\\
$\mbox{}$\\
\end{tabular}
\parbox{14mm}{\begin{center}
\begin{fmfgraph*}(8,8)
\setval
\fmfforce{-0.3w,0.2h}{i1}
\fmfforce{-0.3w,0.8h}{i2}
\fmfforce{1.3w,0.2h}{o1}
\fmfforce{1.3w,0.8h}{o2}
\fmfforce{0w,1/2h}{v1}
\fmfforce{1/2w,0h}{v2}
\fmfforce{1/2w,1h}{v3}
\fmfforce{1w,1/2h}{v4}
\fmf{plain}{i1,v1}
\fmf{plain}{i2,v1}
\fmf{plain}{o1,v4}
\fmf{plain}{o2,v4}
\fmf{plain,left=1}{v1,v4,v1}
\fmf{plain,left=0.4}{v2,v3,v2}
\fmfdot{v1,v2,v3,v4}
\end{fmfgraph*}\end{center}} 
\quad
\begin{tabular}{@{}c}
$\mbox{}$\\
${\scs \#9.1.2}$ \\
${\scs 7962624}$\\
${\scs ( 0, 0, 0 ; 24 )}$\\
$\mbox{}$\\
\end{tabular}
\parbox{14mm}{\begin{center}
\begin{fmfgraph}(8,8)
\setval
\fmfforce{-0.3w,1/2h}{i1}
\fmfforce{1/2w,-0.3h}{i2}
\fmfforce{1/2w,1.3h}{o1}
\fmfforce{1.3w,1/2h}{o2}
\fmfforce{0w,1/2h}{v1}
\fmfforce{1/2w,0h}{v2}
\fmfforce{1/2w,1h}{v3}
\fmfforce{1w,1/2h}{v4}
\fmf{plain}{i1,v1}
\fmf{plain}{i2,v2}
\fmf{plain}{o1,v3}
\fmf{plain}{o2,v4}
\fmf{plain}{v1,v4}
\fmf{plain}{v2,v3}
\fmf{plain,left=1}{v1,v4,v1}
\fmfdot{v1,v2,v3,v4}
\end{fmfgraph}\end{center}}
\quad
\begin{tabular}{@{}c}
$\mbox{}$\\
${\scs \#9.1.3, \#9.2.3, \#11.1.1, \#11.4.1}$ \\
${\scs 15925248,15925248,7962624,7962624}$\\
${\scs 47775744}$\\
${\scs ( 0, 1, 0 ; 2)}$\\
$\mbox{}$\\
\end{tabular}
\parbox{23mm}{\begin{center}
\begin{fmfgraph*}(20,10)
\setval
\fmfforce{0w,0.5h}{i1}
\fmfforce{1/8w,0.25h}{i2}
\fmfforce{1/8w,0.5h}{v1}
\fmfforce{3/8w,0.5h}{v2}
\fmfforce{5/8w,0.5h}{v3}
\fmfforce{7/8w,0.5h}{v4}
\fmfforce{1w,0.5h}{o1}
\fmfforce{2/8w,0.25h}{o2}
\fmf{plain}{i1,v1}
\fmf{plain}{i2,v1}
\fmf{plain}{v4,o1}
\fmf{plain}{v2,o2}
\fmf{plain}{v1,v2}
\fmf{plain}{v2,v3}
\fmf{plain,left=1}{v1,v3}
\fmf{plain,right=1}{v2,v4}
\fmf{plain,left=1}{v3,v4,v3}
\fmf{plain}{v4,o}
\fmfdot{v1,v2,v3,v4}
\end{fmfgraph*}
\end{center}}
\\ 
&
\hspace{-10pt}
\rule[-10pt]{0pt}{26pt}
\begin{tabular}{@{}c}
$\mbox{}$\\
${\scs \#9.2.2, \#14.1.1, \#14.4.3}$ \\
${\scs 7962624,1990656,1990656}$\\ 
${\scs 11943936}$\\
${\scs ( 0, 2, 0 ; 4 )}$\\
$\mbox{}$\\
\end{tabular}
\parbox{13mm}{\begin{center}
\begin{fmfgraph}(10,10)
\setval
\fmfstraight
\fmfforce{0w,1/4h}{i1}
\fmfforce{1w,1/4h}{o1}
\fmfforce{1/4w,1/4h}{v1}
\fmfforce{3/4w,1/4h}{v2}
\fmfforce{1/2w,1/2h}{v3}
\fmfforce{1/2w,1h}{v4}
\fmfforce{0.75w,1.25h}{i2}
\fmfforce{0.25w,1.25h}{o2}
\fmf{plain}{i1,o1}
\fmf{plain}{i2,v4}
\fmf{plain}{o2,v4}
\fmf{plain,left=1}{v1,v2,v1}
\fmf{plain,left=1}{v3,v4,v3}
\fmfdot{v1,v2,v3,v4}
\end{fmfgraph}
\end{center}}
\quad
\begin{tabular}{@{}c}
$\mbox{}$\\
${\scs \#10.1.1, \#10.2.3, \# 14.2.1, \#14.4.2}$ \\
${\scs 2654208,2654208,1327104,1327104}$\\
${\scs 7962624}$\\
${\scs ( 0, 1,1; 2)}$\\
$\mbox{}$\\
\end{tabular}
\parbox{18mm}{\begin{center}
\begin{fmfgraph}(15,5)
\setval
\fmfstraight
\fmfforce{0w,0h}{i1}
\fmfforce{1w,0h}{o1}
\fmfforce{1/6w,0h}{v1}
\fmfforce{1/6w,1h}{v2}
\fmfforce{0w,3/2h}{i2}
\fmfforce{2/6w,3/2h}{o2}
\fmfforce{3/6w,0h}{v3}
\fmfforce{5/6w,0h}{v4}
\fmf{plain}{i1,o1}
\fmf{plain}{v2,i2}
\fmf{plain}{v2,o2}
\fmf{plain,left=1}{v1,v2,v1}
\fmf{plain,left=1}{v3,v4,v3}
\fmfdot{v1,v2,v3,v4}
\end{fmfgraph}
\end{center}}
\quad
\begin{tabular}{@{}c}
$\mbox{}$\\
${\scs \#10.2.2, \#12.4.1}$ \\
${\scs 2654208,1327104}$\\ 
${\scs 3981312}$\\
${\scs ( 0, 0, 1 ; 8 )}$\\
$\mbox{}$\\
\end{tabular}
\parbox{14mm}{\begin{center}
\begin{fmfgraph}(8,8)
\setval
\fmfforce{0w,0.5h}{v1}
\fmfforce{1w,0.5h}{v2}
\fmfforce{0.2w,0.9h}{v3}
\fmfforce{0.8w,0.9h}{v4}
\fmfforce{-0.2w,0.9h}{i1}
\fmfforce{0.2w,1.3h}{i2}
\fmfforce{1.2w,0.9h}{o1}
\fmfforce{0.8w,1.3h}{o2}
\fmfdot{v1,v2}
\fmf{plain,left=1}{v1,v2,v1}
\fmf{plain,left=0.4}{v1,v2,v1}
\fmf{plain}{i1,v3}
\fmf{plain}{i2,v3}
\fmf{plain}{o1,v4}
\fmf{plain}{o2,v4}
\fmfdot{v1,v2,v3,v4}
\end{fmfgraph}\end{center}}
\\ 
&
\hspace{-10pt}
\rule[-10pt]{0pt}{26pt}
\begin{tabular}{@{}c}
$\mbox{}$\\
${\scs \#11.1.3, \#11.2.1}$ \\
${\scs 3981312,3981312}$\\
${\scs 7962624}$\\
${\scs ( 0, 2, 0 ; 6)}$\\
$\mbox{}$\\
\end{tabular}
\hspace*{0.5cm}
\parbox{15mm}{\begin{center}
\begin{fmfgraph*}(8,8)
\setval
\fmfforce{-0.8w,0.25h}{i1}
\fmfforce{-0.5w,-0.05h}{i2}
\fmfforce{-0.5w,0.55h}{i3}
\fmfforce{-0.5w,0.25h}{v1}
\fmfforce{1.24551w,0.25h}{o1}
\fmfforce{0.5w,1h}{v2}
\fmfforce{0.066987w,0.25h}{v3}
\fmfforce{0.93301w,0.25h}{v4}
\fmf{plain,right=0.55}{v2,v3}
\fmf{plain,left=0.55}{v2,v4}
\fmf{plain,left=1}{v1,v5,v1}
\fmf{plain}{i2,i3}
\fmf{plain}{i1,o1}
\fmf{plain}{v2,v3}
\fmf{plain}{v2,v4}
\fmf{plain}{v3,v4}
\fmfdot{v1,v2,v3,v4}
\end{fmfgraph*}
\end{center}}
\quad
\begin{tabular}{@{}c}
$\mbox{}$\\
${\scs \#11.2.2, \#11.4.3, \#14.1.2, \#14.3.3}$ \\
${\scs 7962624,7962624,3981312,3981312}$\\
${\scs 23887872}$\\ 
${\scs ( 1, 1, 0 ; 2)}$\\
$\mbox{}$\\
\end{tabular}
\parbox{13mm}{\begin{center}
\begin{fmfgraph}(12.5,10)
\setval
\fmfstraight
\fmfforce{0w,1/4h}{i1}
\fmfforce{4/5w,1/4h}{o1}
\fmfforce{1/5w,1/4h}{v1}
\fmfforce{3/5w,1/4h}{v2}
\fmfforce{2/5w,1/2h}{v3}
\fmfforce{4/5w,1/2h}{v5}
\fmfforce{4/5w,1h}{v6}
\fmfforce{1/5w,1/2h}{i2}
\fmfforce{1w,1/2h}{o2}
\fmf{plain}{i1,o1}
\fmf{plain}{i2,o2}
\fmf{plain,left=1}{v1,v2,v1}
\fmf{plain,left=1}{v5,v6,v5}
\fmfdot{v1,v2,v3,v5}
\end{fmfgraph}
\end{center}}
\quad
\begin{tabular}{@{}c}
$\mbox{}$\\
${\scs \#11.2.3, \#11.4.2, \#13.2.2, \#13.3.1}$ \\
${\scs 7962624,7962624,3981312,3981312}$\\
${\scs 23887872}$\\ 
${\scs ( 1, 1, 0 ; 2)}$\\
$\mbox{}$\\
\end{tabular}
\parbox{18mm}{\begin{center}
\begin{fmfgraph}(15,7.5)
\setval
\fmfstraight
\fmfforce{0w,1/3h}{i1}
\fmfforce{1w,1/3h}{o1}
\fmfforce{1/6w,1/3h}{v1}
\fmfforce{1/2w,1/3h}{v2}
\fmfforce{1/3w,2/3h}{v3}
\fmfforce{5/6w,1/3h}{v4}
\fmfforce{5/6w,1h}{v5}
\fmfforce{1/6w,1h}{i2}
\fmfforce{1/2w,1h}{o2}
\fmf{plain}{i1,o1}
\fmf{plain}{i2,v3}
\fmf{plain}{o2,v3}
\fmf{plain,left=1}{v1,v2,v1}
\fmf{plain,left=1}{v4,v5,v4}
\fmfdot{v1,v2,v3,v4}
\end{fmfgraph}
\end{center}}
\\ 
&
\hspace{-10pt}
\rule[-10pt]{0pt}{26pt}
\begin{tabular}{@{}c}
$\mbox{}$\\
${\scs \#11.2.4, \#11.3.2, \#17.1.2, \#17.2.1}$ \\
${\scs 3981312,3981312,1990656,1990656}$\\
${\scs 11943936}$\\
${\scs ( 1, 2, 0 ; 2)}$\\
$\mbox{}$\\
\end{tabular}
\parbox{18mm}{\begin{center}
\begin{fmfgraph*}(12.5,10)
\setval
\fmfstraight
\fmfforce{0w,0h}{i1}
\fmfforce{1w,0h}{o1}
\fmfforce{0w,5/4h}{i2}
\fmfforce{2/5w,5/4h}{o2}
\fmfforce{1/5w,0h}{v1}
\fmfforce{1/5w,1/2h}{v2}
\fmfforce{4/5w,0h}{v3}
\fmfforce{4/5w,1/2h}{v4}
\fmfforce{1/5w,1h}{v5}
\fmf{plain,left=1}{v1,v2,v1}
\fmf{plain,left=1}{v3,v4,v3}
\fmf{plain,left=1}{v2,v5,v2}
\fmf{plain}{i1,o1}
\fmf{plain}{i2,v5}
\fmf{plain}{o2,v5}
\fmfdot{v1,v2,v3,v5}
\end{fmfgraph*}
\end{center}}
\quad
\begin{tabular}{@{}c}
$\mbox{}$\\
${\scs \#11.3.3, \#11.4.4, \#12.1.1, \#12.4.5}$ \\
${\scs 7962624,7962624,3981312,3981312}$\\
${\scs 23887872}$\\
${\scs ( 1, 1, 0 ; 2)}$\\
$\mbox{}$\\
\end{tabular}
\parbox{13mm}{\begin{center}
\begin{fmfgraph}(13,8)
\setval
\fmfstraight
\fmfforce{0w,1/2h}{i1}
\fmfforce{1w,1/2h}{o1}
\fmfforce{2.5/13w,1/2h}{v1}
\fmfforce{10.5/13w,1/2h}{v2}
\fmfforce{4/13w,0.87h}{v3}
\fmfforce{9/13w,0.87h}{v4}
\fmfforce{0.05w,0.87h}{i2}
\fmfforce{4/13w,1.3h}{o2}
\fmf{plain}{i1,o1}
\fmf{plain}{i2,v3}
\fmf{plain}{o2,v3}
\fmfi{plain}{reverse fullcircle scaled 0.38w shifted (10.5/13w,1.1h)}
\fmf{plain,left=1}{v1,v2,v1}
\fmfdot{v1,v2,v3,v4}
\end{fmfgraph}
\end{center}}
\quad

\begin{tabular}{@{}c}
$\mbox{}$\\
${\scs \#11.4.5, \#15.3.1, \#15.4.1}$ \\
${\scs 7962624,1990656,1990656}$\\
${\scs 11943936}$\\
${\scs ( 1, 1, 0 ; 4)}$\\
$\mbox{}$\\
\end{tabular}
\quad
\parbox{18mm}{\begin{center}
\begin{fmfgraph}(15,10)
\setval
\fmfstraight
\fmfforce{0w,0h}{i1}
\fmfforce{0w,1/2h}{i2}
\fmfforce{1w,0h}{o1}
\fmfforce{1w,1/2h}{o2}
\fmfforce{1/6w,1/4h}{v1}
\fmfforce{1/2w,1/4h}{v2}
\fmfforce{5/6w,1/4h}{v3}
\fmfforce{4/6w,1/2h}{v4}
\fmfforce{4/6w,1h}{v5}
\fmf{plain}{i1,v1}
\fmf{plain}{v3,o1}
\fmf{plain}{i2,v1}
\fmf{plain}{v3,o2}
\fmf{plain,left=1}{v1,v2,v1}
\fmf{plain,left=1}{v2,v3,v2}
\fmf{plain,left=1}{v4,v5,v4}
\fmfdot{v1,v2,v3,v4}
\end{fmfgraph}
\end{center}}
\\ 
&
\hspace{-10pt}
\rule[-10pt]{0pt}{26pt}
\begin{tabular}{@{}c}
$\mbox{}$\\
${\scs \#11.4.6, \#13.1.1, \#13.2.3}$ \\
${\scs 15925248,3981312,3981312}$\\
${\scs 23887872}$\\
${\scs ( 1, 0, 0 ; 4)}$\\
$\mbox{}$\\
\end{tabular}
\parbox{13mm}{\begin{center}
\begin{fmfgraph}(10,10)
\setval
\fmfstraight
\fmfforce{0w,3/4h}{i1}
\fmfforce{1w,3/4h}{o1}
\fmfforce{1/4w,3/4h}{v1}
\fmfforce{3/4w,3/4h}{v2}
\fmfforce{1/2w,1h}{v3}
\fmfforce{1/2w,1/2h}{v4}
\fmfforce{1/2w,0h}{v5}
\fmfforce{3/4w,5/4h}{i2}
\fmfforce{1/4w,5/4h}{o2}
\fmf{plain}{i1,o1}
\fmf{plain}{i2,v3}
\fmf{plain}{o2,v3}
\fmf{plain,left=1}{v1,v2,v1}
\fmf{plain,left=1}{v4,v5,v4}
\fmfdot{v1,v2,v3,v4}
\end{fmfgraph}
\end{center}}
\quad
\begin{tabular}{@{}c}
$\mbox{}$\\
${\scs \#12.1.2, \#12.2.2, \#13.3.3, \#17.2.2}$ \\
${\scs 1990656,1990656,3981312,3981312}$\\
${\scs 11943936}$\\
${\scs ( 2, 1, 0 ; 2)}$\\
$\mbox{}$\\
\end{tabular}
\parbox{23mm}{\begin{center}
\begin{fmfgraph}(20,7.5)
\setval
\fmfstraight
\fmfforce{0w,1/3h}{i1}
\fmfforce{1w,1/3h}{o1}
\fmfforce{3/8w,0h}{i2}
\fmfforce{5/8w,0h}{o2}
\fmfforce{1/8w,1/3h}{v1}
\fmfforce{1/8w,1h}{v5}
\fmfforce{3/8w,1/3h}{v2}
\fmfforce{5/8w,1/3h}{v3}
\fmfforce{7/8w,1/3h}{v4}
\fmfforce{7/8w,1h}{v6}
\fmf{plain}{i1,v2}
\fmf{plain}{v3,o1}
\fmf{plain}{i2,v2}
\fmf{plain}{o2,v3}
\fmf{plain,left=1}{v1,v5,v1}
\fmf{plain,left=1}{v2,v3,v2}
\fmf{plain,left=1}{v4,v6,v4}
\fmfdot{v1,v3,v2,v4}
\end{fmfgraph}
\end{center}}
\quad
\begin{tabular}{@{}c}
$\mbox{}$\\
${\scs \#12.1.3, \#16.1.2}$ \\
${\scs 3981312,1990656}$\\
${\scs 5971968}$\\
${\scs ( 2, 0, 0 ; 8)}$\\
$\mbox{}$\\
\end{tabular}
\parbox{13mm}{\begin{center}
\begin{fmfgraph}(13,8)
\setval
\fmfstraight
\fmfforce{0w,0.25h}{i1}
\fmfforce{0w,0.75h}{i2}
\fmfforce{1w,0.25h}{o1}
\fmfforce{1w,0.75h}{o2}
\fmfforce{2.5/13w,1/2h}{v1}
\fmfforce{10.5/13w,1/2h}{v2}
\fmfforce{4/13w,0.87h}{v3}
\fmfforce{9/13w,0.87h}{v4}
\fmf{plain}{i1,v1}
\fmf{plain}{i2,v1}
\fmf{plain}{o1,v2}
\fmf{plain}{o2,v2}
\fmfi{plain}{reverse fullcircle scaled 0.38w shifted (2.5/13w,1.1h)}
\fmfi{plain}{reverse fullcircle scaled 0.38w shifted (10.5/13w,1.1h)}
\fmf{plain,left=1}{v1,v2,v1}
\fmfdot{v1,v2,v3,v4}
\end{fmfgraph}
\end{center}}
\\ 
&
\hspace{-10pt}
\rule[-10pt]{0pt}{26pt}
\begin{tabular}{@{}c}
$\mbox{}$\\
${\scs \#12.1.4, \#12.3.3, \#15.1.1, \#15.3.2}$ \\
${\scs 3981312,3981312,1990656,1990656}$\\
${\scs 11943936}$\\
${\scs ( 2, 1, 0 ; 2)}$\\
$\mbox{}$\\
\end{tabular}
\parbox{23mm}{\begin{center}
\begin{fmfgraph*}(20,5)
\setval
\fmfstraight
\fmfforce{0w,0h}{i1}
\fmfforce{1w,0h}{o1}
\fmfforce{0w,3/2h}{i2}
\fmfforce{2/8w,3/2h}{o2}
\fmfforce{1/8w,0h}{v1}
\fmfforce{1/8w,1h}{v2}
\fmfforce{4/8w,0h}{v3}
\fmfforce{4/8w,1h}{v4}
\fmfforce{7/8w,0h}{v5}
\fmfforce{7/8w,1h}{v6}
\fmf{plain,left=1}{v1,v2,v1}
\fmf{plain,left=1}{v3,v4,v3}
\fmf{plain,left=1}{v5,v6,v5}
\fmf{plain}{i1,o1}
\fmf{plain}{i2,v2}
\fmf{plain}{o2,v2}
\fmfdot{v1,v2,v3,v5}
\end{fmfgraph*}
\end{center}}
\begin{tabular}{@{}c}
$\mbox{}$\\
${\scs \#12.2.1, \#12.4.2}$ \\
${\scs 1327104,1327104}$\\
${\scs 2654208}$\\
${\scs ( 1, 0, 1 ; 6)}$\\
$\mbox{}$\\
\end{tabular}
\parbox{20.5mm}{\begin{center}
\begin{fmfgraph}(17.5,7.55)
\setval
\fmfstraight
\fmfforce{0w,1/3h}{i1}
\fmfforce{1/7w,0h}{i2}
\fmfforce{1/7w,2/3h}{i3}
\fmfforce{1w,1/3h}{o1}
\fmfforce{1/7w,1/3h}{v1}
\fmfforce{2/7w,1/3h}{v2}
\fmfforce{4/7w,1/3h}{v3}
\fmfforce{6/7w,1/3h}{v4}
\fmfforce{6/7w,1h}{v5}
\fmf{plain}{i1,o1}
\fmf{plain}{i2,v1}
\fmf{plain}{i3,v1}
\fmf{plain,left=1}{v2,v3,v2}
\fmf{plain,left=1}{v4,v5,v4}
\fmfdot{v1,v3,v2,v4}
\end{fmfgraph}
\end{center}}
\begin{tabular}{@{}c}
$\mbox{}$\\
${\scs \#12.3.2, \#12.4.3, \#14.2.2, \#14.3.2}$ \\
${\scs 1327104,1327104,2654208,2654208}$\\
${\scs 7962624}$\\
${\scs ( 1, 0, 1 ; 2)}$\\
$\mbox{}$\\
\end{tabular}
\parbox{20.5mm}{\begin{center}
\begin{fmfgraph}(17.5,7.5)
\setval
\fmfstraight
\fmfforce{0w,1/3h}{i1}
\fmfforce{1w,1/3h}{o1}
\fmfforce{3/7w,2/3h}{i2}
\fmfforce{3/7w,0h}{o2}
\fmfforce{1/7w,1/3h}{v1}
\fmfforce{1/7w,1h}{v5}
\fmfforce{3/7w,1/3h}{v2}
\fmfforce{4/7w,1/3h}{v3}
\fmfforce{6/7w,1/3h}{v4}
\fmf{plain}{i1,o1}
\fmf{plain}{i2,o2}
\fmf{plain,left=1}{v4,v3,v4}
\fmf{plain,left=1}{v1,v5,v1}
\fmfdot{v1,v3,v2,v4}
\end{fmfgraph}
\end{center}}
\\ 
&
\hspace{-10pt}
\rule[-10pt]{0pt}{26pt}
\begin{tabular}{@{}c}
$\mbox{}$\\
${\scs \#12.3.1, \#12.4.4}$ \\
${\scs 1327104,1327104}$\\
${\scs 2654208}$\\
${\scs ( 1, 0, 1 ; 6)}$\\
$\mbox{}$\\
\end{tabular}
\parbox{23mm}{\begin{center}
\begin{fmfgraph}(20,7.5)
\setval
\fmfstraight
\fmfforce{0w,1/3h}{i1}
\fmfforce{1/8w,0h}{i2}
\fmfforce{1/8w,2/3h}{i3}
\fmfforce{1w,1/3h}{o1}
\fmfforce{1/8w,1/3h}{v1}
\fmfforce{3/8w,1/3h}{v2}
\fmfforce{5/8w,1/3h}{v3}
\fmfforce{7/8w,1/3h}{v4}
\fmfforce{3/8w,1h}{v5}
\fmf{plain}{i1,o1}
\fmf{plain}{i2,v1}
\fmf{plain}{i3,v1}
\fmf{plain,left=1}{v3,v4,v3}
\fmf{plain,left=1}{v2,v5,v2}
\fmfdot{v1,v3,v2,v4}
\end{fmfgraph}
\end{center}}
\quad
\begin{tabular}{@{}c}
$\mbox{}$\\
${\scs \#13.1.2, \#13.3.4, \#15.4.2, \#15.5.1}$ \\
${\scs 7962624,7962624,3981312,3981312}$\\
${\scs 23887872}$\\
${\scs ( 2, 0, 0 ; 2)}$\\
$\mbox{}$\\
\end{tabular}
\quad
\parbox{15mm}{\begin{center}
\begin{fmfgraph}(12.5,10)
\setval
\fmfforce{0w,0h}{i1}
\fmfforce{1/5w,0h}{v1}
\fmfforce{1/5w,0.5h}{v2}
\fmfforce{1/5w,1h}{v5}
\fmfforce{4/5w,0h}{v3}
\fmfforce{4/5w,0.5h}{v4}
\fmfforce{0w,0.25h}{v6}
\fmfforce{1w,0h}{o1}
\fmfforce{-1/5w,0h}{i2}
\fmfforce{-1/5w,1/2h}{o2}
\fmf{plain}{i1,o1}
\fmf{plain}{i2,v6}
\fmf{plain}{o2,v6}
\fmf{plain,left=1}{v1,v2,v1}
\fmf{plain,left=1}{v3,v4,v3}
\fmf{plain,left=1}{v2,v5,v2}
\fmfdot{v1,v2,v3,v6}
\end{fmfgraph}
\end{center}}
\quad
\begin{tabular}{@{}c}
$\mbox{}$\\
${\scs \#13.1.3, \#16.1.1}$ \\
${\scs 1990656,995328}$\\
${\scs 2985984}$\\
${\scs ( 2, 0, 0 ; 16)}$\\
$\mbox{}$\\
\end{tabular}
\parbox{13mm}{\begin{center}
\begin{fmfgraph}(10,15)
\setval
\fmfstraight
\fmfforce{0w,1/3h}{i1}
\fmfforce{0w,2/3h}{i2}
\fmfforce{1w,2/3h}{o1}
\fmfforce{1w,1/3h}{o2}
\fmfforce{1/2w,0h}{v1}
\fmfforce{1/2w,1/3h}{v2}
\fmfforce{1/2w,2/3h}{v3}
\fmfforce{1/2w,1h}{v4}
\fmfforce{1/4w,1/2h}{v5}
\fmfforce{3/4w,1/2h}{v6}
\fmf{plain}{i1,v5}
\fmf{plain}{v6,o1}
\fmf{plain}{i2,v5}
\fmf{plain}{v6,o2}
\fmf{plain,left=1}{v1,v2,v1}
\fmf{plain,left=1}{v2,v3,v2}
\fmf{plain,left=1}{v3,v4,v3}
\fmfdot{v2,v3,v5,v6}
\end{fmfgraph}
\end{center}}
\\ 
&
\hspace{-10pt}
\rule[-10pt]{0pt}{26pt}
\begin{tabular}{@{}c}
$\mbox{}$\\
${\scs \#13.2.4, \#13.3.5}$ \\
${\scs 3981312,3981312}$\\
${\scs 7962624}$\\
${\scs ( 1, 1, 0 ; 6)}$\\
$\mbox{}$\\
\end{tabular}
\hspace*{0.3cm}
\parbox{15.5mm}{\begin{center}
\begin{fmfgraph*}(12.5,10)
\setval
\fmfforce{0w,1/4h}{i1}
\fmfforce{1/5w,1/2h}{i2}
\fmfforce{1/5w,0h}{o1}
\fmfforce{1w,1/4h}{o2}
\fmfforce{2/5w,1/4h}{v1}
\fmfforce{4/5w,1/4h}{v2}
\fmfforce{3/5w,0.5h}{v4}
\fmfforce{3/5w,1h}{v7}
\fmfforce{1/5w,1/4h}{v6}
\fmf{plain}{i1,o2}
\fmf{plain}{i2,o1}
\fmf{plain}{v2,o2}
\fmf{plain,left=1}{v7,v4,v7}
\fmf{plain,left=1}{v1,v2,v1}
\fmfdot{v1,v2,v4,v6}
\end{fmfgraph*}\end{center}} 
\quad
\begin{tabular}{@{}c}
$\mbox{}$\\
${\scs \#13.3.2, \#15.2.1, \#15.3.3}$ \\
${\scs 3981312,995328,995328}$\\
${\scs 5971968}$\\
${\scs( 2, 1, 0 ; 4)}$\\
$\mbox{}$\\
\end{tabular}
\parbox{23mm}{\begin{center}
\begin{fmfgraph*}(20,5)
\setval
\fmfstraight
\fmfforce{0w,0h}{i1}
\fmfforce{1/8w,0h}{v1}
\fmfforce{1/8w,1h}{v2}
\fmfforce{4/8w,0h}{v3}
\fmfforce{4/8w,1h}{v4}
\fmfforce{7/8w,0h}{v5}
\fmfforce{7/8w,1h}{v6}
\fmfforce{1w,0h}{o1}
\fmfforce{3/8w,3/2h}{i2}
\fmfforce{5/8w,3/2h}{o2}
\fmf{plain,left=1}{v1,v2,v1}
\fmf{plain,left=1}{v3,v4,v3}
\fmf{plain,left=1}{v5,v6,v5}
\fmf{plain}{i1,o1}
\fmf{plain}{i2,v4}
\fmf{plain}{o2,v4}
\fmfdot{v1,v3,v4,v5}
\end{fmfgraph*}
\end{center}}
\quad
\begin{tabular}{@{}c}
$\mbox{}$\\
${\scs \#14.1.3, \#14.2.3, \#17.1.3, \#17.3.1}$ \\
${\scs 3981312,3981312,1990656,1990656}$\\
${\scs 11943936}$\\
${\scs( 1, 2, 0 ; 2)}$\\
$\mbox{}$\\
\end{tabular}
\parbox{18mm}{\begin{center}
\begin{fmfgraph*}(12.5,10)
\setval
\fmfstraight
\fmfforce{0w,0h}{i1}
\fmfforce{1w,0h}{o1}
\fmfforce{0w,3/4h}{i2}
\fmfforce{2/5w,3/4h}{o2}
\fmfforce{1/5w,0h}{v1}
\fmfforce{1/5w,1/2h}{v2}
\fmfforce{4/5w,0h}{v3}
\fmfforce{4/5w,1/2h}{v4}
\fmfforce{4/5w,1h}{v5}
\fmf{plain,left=1}{v1,v2,v1}
\fmf{plain,left=1}{v3,v4,v3}
\fmf{plain,left=1}{v4,v5,v4}
\fmf{plain}{i1,o1}
\fmf{plain}{i2,v2}
\fmf{plain}{o2,v2}
\fmfdot{v1,v2,v3,v4}
\end{fmfgraph*}
\end{center}}
\\ 
&
\hspace{-10pt}
\rule[-10pt]{0pt}{26pt}
\begin{tabular}{@{}c}
$\mbox{}$\\
${\scs \#14.1.4, \#15.4.4}$ \\
${\scs 3981312,1990656}$\\
${\scs 5971968}$\\
${\scs ( 1, 1, 0 ; 8)}$\\
$\mbox{}$\\
\end{tabular}
\parbox{13mm}{\begin{center}
\begin{fmfgraph}(10,15)
\setval
\fmfstraight
\fmfforce{0w,0h}{i1}
\fmfforce{0w,1/3h}{i2}
\fmfforce{1w,0h}{o1}
\fmfforce{1w,1/3h}{o2}
\fmfforce{1/4w,1/6h}{v1}
\fmfforce{3/4w,1/6h}{v2}
\fmfforce{1/2w,1/3h}{v3}
\fmfforce{1/2w,2/3h}{v4}
\fmfforce{1/2w,1h}{v5}
\fmf{plain}{i1,v1}
\fmf{plain}{v2,o1}
\fmf{plain}{i2,v1}
\fmf{plain}{v2,o2}
\fmf{plain,left=1}{v1,v2,v1}
\fmf{plain,left=1}{v3,v4,v3}
\fmf{plain,left=1}{v5,v4,v5}
\fmfdot{v1,v2,v3,v4}
\end{fmfgraph}
\end{center}}
\quad
\begin{tabular}{@{}c}
$\mbox{}$\\
${\scs \#14.3.1, \#14.4.1}$ \\
${\scs 1327104,1327104}$\\
${\scs 2654208}$\\
${\scs ( 0, 0, 1 ; 12)}$\\
$\mbox{}$\\
\end{tabular}

\hspace*{0.3cm}
\parbox{11mm}{\begin{center}
\begin{fmfgraph*}(8,8)
\setval
\fmfforce{0w,0.5h}{v1}
\fmfforce{1w,0.5h}{v2}
\fmfforce{0.5w,0h}{v4}
\fmfforce{-0.5w,0h}{i1}
\fmfforce{-0.2w,0.3h}{i2}
\fmfforce{-0.2w,-0.3h}{o2}
\fmfforce{0.8125w,0h}{o1}
\fmfforce{-0.2w,0h}{v5}
\fmf{plain,left=1}{v1,v2,v1}
\fmf{plain,left=0.4}{v1,v2,v1}
\fmf{plain,left=1}{v3,v4,v3}
\fmf{plain}{i1,o1}
\fmf{plain}{i2,o2}
\fmfdot{v1,v2,v3,v5}
\end{fmfgraph*}\end{center}} 
\quad
\begin{tabular}{@{}c}
$\mbox{}$\\
${\scs \#15.1.2, \#15.5.3, \#16.1.3, \#16.2.2}$ \\
${\scs 3981312,3981312,1990656,1990656}$\\
${\scs 11943936}$\\
${\scs ( 3, 0, 0 ; 2)}$\\
$\mbox{}$\\
\end{tabular}
\parbox{25.5mm}{\begin{center}
\begin{fmfgraph}(22.5,5)
\setval
\fmfstraight
\fmfforce{0w,0h}{i1}
\fmfforce{3/9w,1/2h}{i2}
\fmfforce{3/9w,-1/2h}{o1}
\fmfforce{1w,0h}{o2}
\fmfforce{1/9w,0h}{v1}
\fmfforce{1/9w,1h}{v2}
\fmfforce{5/9w,0h}{v3}
\fmfforce{5/9w,1h}{v4}
\fmfforce{3/9w,0h}{v5}
\fmfforce{8/9w,0h}{v6}
\fmfforce{8/9w,1h}{v7}
\fmf{plain}{i1,o2}
\fmf{plain}{i2,v5}
\fmf{plain}{o1,v5}
\fmf{plain,left=1}{v1,v2,v1}
\fmf{plain,left=1}{v3,v4,v3}
\fmf{plain,left=1}{v6,v7,v6}
\fmfdot{v1,v3,v5,v6}
\end{fmfgraph}
\end{center}}
\\ 
&
\hspace{-10pt}
\rule[-10pt]{0pt}{26pt}
\begin{tabular}{@{}c}
$\mbox{}$\\
${\scs \#15.1.3, \#15.4.3, \#17.2.3, \#17.3.2}$ \\
${\scs 1990656,1990656,3981312,3981312}$\\
${\scs 11943936}$\\
${\scs ( 2, 1, 0 ; 2)}$\\
$\mbox{}$\\
\end{tabular}
\parbox{15.5mm}{\begin{center}
\begin{fmfgraph}(12.5,10)
\setval
\fmfstraight
\fmfforce{0w,0h}{i1}
\fmfforce{1/2w,1/4h}{i2}
\fmfforce{1/2w,-1/4h}{o1}
\fmfforce{1w,0h}{o2}
\fmfforce{1/5w,0h}{v1}
\fmfforce{1/5w,1/2h}{v2}
\fmfforce{4/5w,0h}{v3}
\fmfforce{4/5w,1/2h}{v4}
\fmfforce{1/2w,0h}{v5}
\fmfforce{4/5w,1h}{v6}
\fmf{plain}{i1,o2}
\fmf{plain}{i2,v5}
\fmf{plain}{o1,v5}
\fmf{plain,left=1}{v1,v2,v1}
\fmf{plain,left=1}{v3,v4,v3}
\fmf{plain,left=1}{v6,v4,v6}
\fmfdot{v1,v3,v5,v4}
\end{fmfgraph}
\end{center}}
\quad
\begin{tabular}{@{}c}
$\mbox{}$\\
${\scs \#15.1.4, \#15.4.5}$ \\
${\scs 1990656,1990656}$\\
${\scs 3981312}$\\
${\scs ( 2, 1, 0 ; 6)}$\\
$\mbox{}$\\
\end{tabular}
\parbox{20.5mm}{\begin{center}
\begin{fmfgraph}(17.5,10)
\setval
\fmfstraight
\fmfforce{0w,0h}{i1}
\fmfforce{1/7w,1/4h}{i2}
\fmfforce{1/7w,-1/4h}{i3}
\fmfforce{1w,0h}{o1}
\fmfforce{3/7w,0h}{v1}
\fmfforce{3/7w,1/2h}{v2}
\fmfforce{1/7w,0h}{v3}
\fmfforce{6/7w,0h}{v4}
\fmfforce{6/7w,1/2h}{v5}
\fmfforce{6/7w,1h}{v6}
\fmf{plain}{i1,o1}
\fmf{plain}{i2,v3}
\fmf{plain}{i3,v3}
\fmf{plain,left=1}{v1,v2,v1}
\fmf{plain,left=1}{v4,v5,v4}
\fmf{plain,left=1}{v6,v5,v6}
\fmfdot{v1,v3,v4,v5}
\end{fmfgraph}
\end{center}}
\quad
\begin{tabular}{@{}c}
$\mbox{}$\\
${\scs \#15.2.2, \#15.5.2}$ \\
${\scs 1990656,1990656}$\\
${\scs 3981312}$\\
${\scs ( 3, 0, 0 ; 6)}$\\
$\mbox{}$\\
\end{tabular}
\parbox{18mm}{\begin{center}
\begin{fmfgraph}(15,12.5)
\setval
\fmfstraight
\fmfforce{0w,3/5h}{i1}
\fmfforce{3/6w,4/5h}{i2}
\fmfforce{1w,3/5h}{o1}
\fmfforce{3/6w,0h}{o2}
\fmfforce{1/6w,3/5h}{v1}
\fmfforce{1/6w,1h}{v2}
\fmfforce{3/6w,3/5h}{v3}
\fmfforce{3/6w,1/5h}{v4}
\fmfforce{5/6w,1/5h}{v5}
\fmfforce{5/6w,3/5h}{v6}
\fmfforce{5/6w,1h}{v7}
\fmf{plain}{i1,o1}
\fmf{plain}{i2,o2}
\fmf{plain,left=1}{v1,v2,v1}
\fmf{plain,left=1}{v4,v5,v4}
\fmf{plain,left=1}{v6,v7,v6}
\fmfdot{v1,v3,v4,v6}
\end{fmfgraph}
\end{center}}
\\ 
&
\hspace{-10pt}
\rule[-10pt]{0pt}{26pt}
\begin{tabular}{@{}c}
$\mbox{}$\\
${\scs \#15.2.3, \#15.4.6}$ \\
${\scs 1990656,1990656}$\\
${\scs 3981312}$\\
${\scs ( 2, 1, 0 ; 6)}$\\
$\mbox{}$\\
\end{tabular}
\parbox{20.5mm}{\begin{center}
\begin{fmfgraph}(17.5,10)
\setval
\fmfstraight
\fmfforce{0w,0h}{i1}
\fmfforce{1/7w,1/4h}{i2}
\fmfforce{1/7w,-1/4h}{i3}
\fmfforce{1w,0h}{o1}
\fmfforce{3/7w,0h}{v1}
\fmfforce{3/7w,1/2h}{v2}
\fmfforce{1/7w,0h}{v3}
\fmfforce{6/7w,0h}{v4}
\fmfforce{6/7w,1/2h}{v5}
\fmfforce{3/7w,1h}{v6}
\fmf{plain}{i1,o1}
\fmf{plain}{i2,v3}
\fmf{plain}{i3,v3}
\fmf{plain,left=1}{v1,v2,v1}
\fmf{plain,left=1}{v4,v5,v4}
\fmf{plain,left=1}{v6,v2,v6}
\fmfdot{v1,v3,v4,v2}
\end{fmfgraph}
\end{center}}
\quad
\begin{tabular}{@{}c}
$\mbox{}$\\
${\scs \#15.3.4, \#15.5.4}$ \\
${\scs 1990656,1990656}$\\
${\scs 3981312}$\\
${\scs ( 2, 0, 0 ; 12)}$\\
$\mbox{}$\\
\end{tabular}
\parbox{20mm}{\begin{center}
\begin{fmfgraph*}(15,15)
\setval
\fmfforce{-1/6w,1/3h}{i1}
\fmfforce{0w,1/2h}{i2}
\fmfforce{2/3w,1/3h}{o1}
\fmfforce{0w,1/6h}{o2}
\fmfforce{1/2w,1/3h}{v1}
\fmfforce{1/2w,2/3h}{v2}
\fmfforce{1/2w,1h}{v3}
\fmfforce{0.355662432w,0.5833333333h}{v4}
\fmfforce{0.64433568w,0.5833333333h}{v5}
\fmfforce{0.067w,3/4h}{v6}
\fmfforce{0.933w,3/4h}{v7}
\fmfforce{0w,1/3h}{v8}
\fmf{plain,left=1}{v1,v2,v1}
\fmf{plain}{i1,v1}
\fmf{plain}{i2,o2}
\fmf{plain}{v1,o1}
\fmf{plain,left=1}{v4,v6,v4}
\fmf{plain,left=1}{v5,v7,v5}
\fmfdot{v1,v4,v5,v8}
\end{fmfgraph*}\end{center}} 
\quad
\begin{tabular}{@{}c}
$\mbox{}$\\
${\scs \#16.1.4, \#16.2.1}$ \\
${\scs 1990656,1990656}$\\
${\scs 3981312}$\\
${\scs ( 3, 0, 0 ; 6)}$\\
$\mbox{}$\\
\end{tabular}
\parbox{28mm}{\begin{center}
\begin{fmfgraph}(25,10)
\setval
\fmfstraight
\fmfforce{0w,1/2h}{i1}
\fmfforce{1/10w,3/4h}{i2}
\fmfforce{1/10w,1/4h}{i3}
\fmfforce{1w,1/2h}{o1}
\fmfforce{3/10w,1/2h}{v1}
\fmfforce{3/10w,1h}{v2}
\fmfforce{1/10w,1/2h}{v3}
\fmfforce{6/10w,1/2h}{v4}
\fmfforce{6/10w,1h}{v5}
\fmfforce{9/10w,1/2h}{v6}
\fmfforce{9/10w,1h}{v7}
\fmf{plain}{i1,o1}
\fmf{plain}{i2,v3}
\fmf{plain}{i3,v3}
\fmf{plain,left=1}{v1,v2,v1}
\fmf{plain,left=1}{v4,v5,v4}
\fmf{plain,left=1}{v6,v7,v6}
\fmfdot{v1,v3,v4,v6}
\end{fmfgraph}
\end{center}}
\\ 
&
\hspace{-10pt}
\rule[-10pt]{0pt}{26pt}
\begin{tabular}{@{}c}
$\mbox{}$\\
${\scs \#17.1.4, \#17.2.4}$ \\
${\scs 1990656,1990656}$\\
${\scs 3981312}$\\
${\scs ( 1, 2, 0 ; 6)}$\\
$\mbox{}$\\
\end{tabular}
\parbox{13mm}{\begin{center}
\begin{fmfgraph}(10,20)
\setval
\fmfstraight
\fmfforce{0w,1/4h}{i1}
\fmfforce{1/4w,3/8h}{i2}
\fmfforce{1/4w,1/8h}{i3}
\fmfforce{1w,1/4h}{o1}
\fmfforce{3/4w,1/4h}{v1}
\fmfforce{3/4w,2/4h}{v2}
\fmfforce{1/4w,1/4h}{v3}
\fmfforce{3/4w,3/4h}{v4}
\fmfforce{3/4w,1h}{v5}
\fmf{plain}{i1,o1}
\fmf{plain}{i2,v3}
\fmf{plain}{i3,v3}
\fmf{plain,left=1}{v1,v2,v1}
\fmf{plain,left=1}{v4,v2,v4}
\fmf{plain,left=1}{v4,v5,v4}
\fmfdot{v1,v2,v3,v4}
\end{fmfgraph}
\end{center}}
\end{tabular}
\end{center}
\caption{\la{FOUR} Connected diagrams of the four-point function 
and their multiplicities of the $\phi^4$-theory
up to three loops. Each diagram is characterized by the
vector $(S,D,T;N$) whose components specify the number of self-, double,
triple connections, and of the identical vertex permutations,
respectively.}
\end{table}
\end{fmffile}
\newpage
\begin{fmffile}{graph5}
\begin{table}[t]
\begin{center}
\begin{tabular}{|c|c|}
\,\,\,$p$\,\,\,
&
\begin{tabular}{@{}c}
$\mbox{}$\\
$\mbox{}$
\end{tabular}
$W^{(p)}$
\\
\hline
$1$ &
\rule[-10pt]{0pt}{26pt}
\begin{tabular}{@{}c}
$\mbox{}$\\
${\scs \mbox{\#1.1}}$ \\
$2$ \\ 
${\scs ( 0, 1; 2 )}$\\
$\mbox{}$
\end{tabular}
\parbox{10mm}{\centerline{
\begin{fmfgraph}(7,5)
\setval
\fmfleft{v1}
\fmfright{v2}
\fmf{boson}{v1,v2}
\fmf{plain,left=0.7}{v2,v1}
\fmf{plain,left=0.7}{v1,v2}
\fmfdot{v1,v2}
\end{fmfgraph}
}}
\begin{tabular}{@{}c}
${\scs \mbox{\#1.2}}$ \\
$1$ \\ 
${\scs ( 2, 0; 2 )}$
\end{tabular}
\parbox{19mm}{\begin{center}
\begin{fmfgraph}(15,7)
\setval
\fmfforce{0.33w,0.5h}{v1}
\fmfforce{0.66w,0.5h}{v2}
\fmf{boson}{v1,v2}
\fmfi{plain}{reverse fullcircle scaled 0.33w shifted (0.165w,0.5h)}
\fmfi{plain}{fullcircle rotated 180 scaled 0.33w shifted (0.825w,0.5h)}
\fmfdot{v1,v2}
\end{fmfgraph}
\end{center}}
\\
\hline
$2$ &
\hspace{-10pt}
\rule[-10pt]{0pt}{26pt}
\quad
\begin{tabular}{@{}c}
${\scs \mbox{\#2.1}}$ \\
$48$ \\ ${\scs ( 0, 0; 8 )}$
\end{tabular}
\parbox{14mm}{\begin{center}
\begin{fmfgraph}(5,5)
\setval
\fmfleft{i2,i1}
\fmfright{o2,o1}
\fmf{plain,left=0.5}{i1,o1,o2,i2,i1}
\fmf{boson}{i1,o2}
\fmf{boson}{i2,o1}
\fmfdot{i1,i2,o1,o2}
\end{fmfgraph}
\end{center}}
\quad
\begin{tabular}{@{}c}
$\mbox{}$\\
${\scs \mbox{\#2.2}}$ \\
$24$ \\ ${\scs ( 0, 2; 4 )}$\\
$\mbox{}$
\end{tabular}
%4
\parbox{18mm}{\begin{center}
\begin{fmfgraph}(8,5)
\setval
\fmfleft{i2,i1}
\fmfright{o2,o1}
\fmf{plain,left=1}{i1,i2,i1}
\fmf{boson}{i1,o1}
\fmf{boson}{i2,o2}
\fmf{plain,left=1}{o1,o2,o1}
\fmfdot{i1,i2,o1,o2}
\end{fmfgraph}
\end{center}} 
\quad
\begin{tabular}{@{}c}
${\scs \mbox{\#3.1}}$ \\
$96$ \\ ${\scs ( 0, 0; 4 )}$
\end{tabular}
\parbox{14mm}{\begin{center}
\begin{fmfgraph}(5,5)
\setval
\fmfleft{i2,i1}
\fmfright{o2,o1}
\fmf{plain,left=0.5}{i1,o1,o2,i2,i1}
\fmf{boson,left=0.4}{i1,i2}
\fmf{boson,right=0.4}{o1,o2}
\fmfdot{i1,i2,o1,o2}
\end{fmfgraph}
\end{center}}
\quad
\begin{tabular}{@{}c}
${\scs \mbox{\#3.2}}$ \\
$96$ \\ ${\scs ( 1, 0; 2 )}$
\end{tabular}
\parbox{18mm}{\begin{center}
\begin{fmfgraph}(13,5)
\setval
\fmfforce{0.385w,0.5h}{v1}
\fmfforce{0.615w,0.5h}{v2}
\fmfforce{0.192w,1h}{v3}
\fmfforce{0.192w,0h}{v4}
\fmf{plain,left=0.5}{v3,v1,v4}
\fmf{plain,left=1}{v4,v3}
\fmf{boson}{v3,v4}
\fmf{boson}{v1,v2}
\fmfi{plain}{fullcircle rotated 180 scaled 0.384w shifted (0.808w,0.5h)}
\fmfdot{v1,v2,v3,v4}
\end{fmfgraph}
\end{center}}
\quad
\begin{tabular}{@{}c}
${\scs \mbox{\#3.3}}$ \\
$24$ \\ ${\scs ( 2, 1; 2 )}$
\end{tabular}
\parbox{26mm}{\begin{center}
\begin{fmfgraph}(21,5)
\setval
\fmfforce{0.238w,0.5h}{v1}
\fmfforce{0.381w,0.5h}{v2}
\fmfforce{0.619w,0.5h}{v3}
\fmfforce{0.762w,0.5h}{v4}
\fmf{plain,left=1}{v2,v3,v2}
\fmf{boson}{v1,v2}
\fmf{boson}{v3,v4}
\fmfi{plain}{reverse fullcircle scaled 0.238w shifted (0.119w,0.5h)}
\fmfi{plain}{fullcircle rotated 180 scaled 0.238w shifted (0.881w,0.5h)}
\fmfdot{v1,v2,v3,v4}
\end{fmfgraph}
\end{center}}
\\
\hline
$3$ &
\hspace{-10pt}
\rule[-10pt]{0pt}{26pt}
\quad
\begin{tabular}{@{}c}
${\scs \mbox{\#4.1}}$ \\
$3840$ \\ ${\scs ( 0, 0; 12 )}$
\end{tabular}
\parbox{11mm}{\begin{center}
\begin{fmfgraph}(7,7)
\setval
\fmfforce{0w,0.5h}{v1}
\fmfforce{0.25w,0.933h}{v2}
\fmfforce{0.75w,0.933h}{v3}
\fmfforce{1w,0.5h}{v4}
\fmfforce{0.75w,0.067h}{v5}
\fmfforce{0.25w,0.067h}{v6}
\fmf{plain,left=0.3}{v1,v2,v3,v4,v5,v6,v1}
\fmf{boson}{v1,v4}
\fmf{boson}{v2,v5}
\fmf{boson}{v3,v6}
\fmfdot{v1,v2,v3,v4,v5,v6}
\end{fmfgraph}
\end{center}}
\quad
\begin{tabular}{@{}c}
${\scs \mbox{\#4.2}}$ \\
$11520$ \\ ${\scs ( 0, 0; 4 )}$
\end{tabular}
\parbox{11mm}{\begin{center}
\begin{fmfgraph}(7,7)
\setval
\fmfforce{0w,0.5h}{v1}
\fmfforce{0.25w,0.933h}{v2}
\fmfforce{0.75w,0.933h}{v3}
\fmfforce{1w,0.5h}{v4}
\fmfforce{0.75w,0.067h}{v5}
\fmfforce{0.25w,0.067h}{v6}
\fmf{plain,left=0.3}{v1,v2,v3,v4,v5,v6,v1}
\fmf{boson}{v1,v4}
\fmf{boson}{v2,v6}
\fmf{boson}{v3,v5}
\fmfdot{v1,v2,v3,v4,v5,v6}
\end{fmfgraph}
\end{center}} 
\quad
\begin{tabular}{@{}c}
${\scs \mbox{\#4.3}}$ \\
$3840$ \\ ${\scs ( 0, 0; 12 )}$
\end{tabular}
\parbox{18mm}{\begin{center}
\begin{fmfgraph}(13,5)
\setval
\fmfforce{0.192w,1h}{i1}
\fmfforce{0.385w,0.5h}{i2}
\fmfforce{0.192w,0h}{i3}
\fmfforce{0.808w,1h}{o1}
\fmfforce{0.808w,0h}{o2}
\fmfforce{0.615w,0.5h}{o3}
\fmf{plain,left=0.5}{i1,i2,i3}
\fmf{plain,left=1}{i3,i1}
\fmf{boson}{i1,o1}
\fmf{boson}{i2,o3}
\fmf{boson}{i3,o2}
\fmf{plain,left=1}{o1,o2}
\fmf{plain,left=0.5}{o2,o3,o1}
\fmfdot{i1,i2,i3,o1,o2,o3}
\end{fmfgraph}
\end{center}}
\quad
\begin{tabular}{@{}c}
${\scs \mbox{\#4.4}}$ \\
$960$ \\ ${\scs ( 0, 3; 6 )}$
\end{tabular}
\parbox{20mm}{\begin{center}
\begin{fmfgraph}(10,5)
\setval
\fmfforce{0w,1h}{i1}
\fmfforce{0w,0h}{i2}
\fmfforce{0.33w,0h}{v1}
\fmfforce{0.66w,0h}{v2}
\fmfforce{1w,1h}{o1}
\fmfforce{1w,0h}{o2}
\fmf{plain,left=1}{i1,i2,i1}
\fmf{plain,left=1}{o1,o2,o1}
\fmf{boson}{i1,o1}
\fmf{boson}{i2,v1}
\fmf{plain,left=1}{v1,v2,v1}
\fmf{boson}{v2,o2}
\fmfdot{i1,i2,v1,v2,o1,o2}
\end{fmfgraph}
\end{center}}
\quad
\begin{tabular}{@{}c}
${\scs \mbox{\#4.5}}$ \\
$5760$ \\ ${\scs ( 0, 1; 4 )}$
\end{tabular}
\parbox{18mm}{\begin{center}
\begin{fmfgraph}(13,5)
\setval
\fmfforce{0w,0.5h}{i1}
\fmfforce{0.192w,1h}{i2}
\fmfforce{0.385w,0.5h}{i3}
\fmfforce{0.192w,0h}{i4}
\fmfforce{0.808w,1h}{o1}
\fmfforce{0.808w,0h}{o2}
\fmf{plain,left=0.5}{i1,i2,i3,i4,i1}
\fmf{boson}{i1,i3}
\fmf{boson}{i2,o1}
\fmf{boson}{i4,o2}
\fmf{plain,left=1}{o1,o2,o1}
\fmfdot{i1,i2,i3,i4,o1,o2}
\end{fmfgraph}
\end{center}} 
\\  &
\quad
\begin{tabular}{@{}c}
${\scs \mbox{\#5.1}}$ \\
$11520$ \\ ${\scs ( 0, 1; 2 )}$
\end{tabular}
\parbox{18mm}{\begin{center}
\begin{fmfgraph}(13,5)
\setval
\fmfforce{0.192w,1h}{i1}
\fmfforce{0.192w,0h}{i2}
\fmfforce{0.026w,0.75h}{i3}
\fmfforce{0.026w,0.25h}{i4}
\fmfforce{0.808w,1h}{o1}
\fmfforce{0.808w,0h}{o2}
\fmf{plain,left=1}{i1,i2}
\fmf{plain,left=1/3}{i2,i4,i3,i1}
\fmf{boson,left=0.8}{i3,i4}
\fmf{boson}{i1,o1}
\fmf{boson}{i2,o2}
\fmf{plain,left=1}{o1,o2,o1}
\fmfdot{i1,i2,i3,i4,o1,o2}
\end{fmfgraph}
\end{center}}
\quad
\begin{tabular}{@{}c}
${\scs \mbox{\#5.2}}$ \\
$23040$ \\ ${\scs ( 0, 0; 2 )}$
\end{tabular}
\parbox{11mm}{\begin{center}
\begin{fmfgraph}(7,7)
\setval
\fmfforce{0w,0.5h}{v1}
\fmfforce{0.25w,0.933h}{v2}
\fmfforce{0.75w,0.933h}{v3}
\fmfforce{1w,0.5h}{v4}
\fmfforce{0.75w,0.067h}{v5}
\fmfforce{0.25w,0.067h}{v6}
\fmf{plain,left=0.3}{v1,v2,v3,v4,v5,v6,v1}
\fmf{boson,right=0.7}{v2,v3}
\fmf{boson,right=0.2}{v4,v6}
\fmf{boson,right=0.2}{v5,v1}
\fmfdot{v1,v2,v3,v4,v5,v6}
\end{fmfgraph}
\end{center}} 
\begin{tabular}{@{}c}
${\scs \mbox{\#5.3}}$ \\
$11520$ \\ ${\scs ( 1, 0; 2 )}$
\end{tabular}
\parbox{17mm}{\begin{center}
\begin{fmfgraph}(13,5)
\setval
\fmfforce{0.252w,0.976h}{i2}
\fmfforce{0.037w,0.794h}{i3}
\fmfforce{0.037w,0.206h}{i4}
\fmfforce{0.252w,0.024h}{i5}
\fmfforce{0.385w,0.5h}{i1}
\fmfforce{0.615w,0.5h}{o1}
\fmf{plain,right=0.3}{i1,i2,i3,i4,i5,i1}
\fmf{boson}{i3,i5}
\fmf{boson}{i2,i4}
\fmf{boson}{i1,o1}
\fmfi{plain}{fullcircle rotated 180 scaled 0.385w shifted (0.808w,0.5h)}
\fmfdot{i1,i2,i3,i4,i5,o1}
\end{fmfgraph}
\end{center}}
\quad
\begin{tabular}{@{}c}
${\scs \mbox{\#5.4}}$ \\
$5760$ \\ ${\scs ( 1, 1; 2 )}$
\end{tabular}
\parbox{25mm}{\begin{center}
\begin{fmfgraph}(21,5)
\setval
\fmfforce{0.238w,0.5h}{i1}
\fmfforce{0.381w,0.5h}{v1}
\fmfforce{0.5w,1h}{v2}
\fmfforce{0.5w,0h}{v3}
\fmfforce{0.881w,1h}{o1}
\fmfforce{0.881w,0h}{o2}
\fmfi{plain}{reverse fullcircle scaled 0.238w shifted (0.119w,0.5h)}
\fmf{boson}{i1,v1}
\fmf{plain,right=0.5}{v2,v1,v3}
\fmf{plain,right=1}{v3,v2}
\fmf{boson}{v2,o1}
\fmf{boson}{v3,o2}
\fmf{plain,right=1}{o1,o2,o1}
\fmfdot{i1,v1,v2,v3,o1,o2}
\end{fmfgraph}
\end{center}}
\quad
\begin{tabular}{@{}c}
${\scs \mbox{\#6.1}}$ \\
$7680$ \\ ${\scs ( 0, 0;6 )}$
\end{tabular}
\parbox{11mm}{\begin{center}
\begin{fmfgraph}(7,7)
\setval
\fmfforce{0w,0.5h}{v1}
\fmfforce{0.25w,0.933h}{v2}
\fmfforce{0.75w,0.933h}{v3}
\fmfforce{1w,0.5h}{v4}
\fmfforce{0.75w,0.067h}{v5}
\fmfforce{0.25w,0.067h}{v6}
\fmf{plain,left=0.3}{v1,v2,v3,v4,v5,v6,v1}
\fmf{boson,right=0.7}{v2,v3}
\fmf{boson,right=0.7}{v4,v5}
\fmf{boson,right=0.7}{v6,v1}
\fmfdot{v1,v2,v3,v4,v5,v6}
\end{fmfgraph}
\end{center}}
\\ &
\begin{tabular}{@{}c}
${\scs \mbox{\#6.2}}$ \\
$11520$ \\ ${\scs ( 1, 0; 2 )}$
\end{tabular}
\parbox{17mm}{\begin{center}
\begin{fmfgraph}(13,5)
\setval
\fmfforce{0.252w,0.976h}{i2}
\fmfforce{0.037w,0.794h}{i3}
\fmfforce{0.037w,0.206h}{i4}
\fmfforce{0.252w,0.024h}{i5}
\fmfforce{0.385w,0.5h}{i1}
\fmfforce{0.615w,0.5h}{o1}
\fmf{plain,right=0.3}{i1,i2,i3,i4,i5,i1}
\fmf{boson,left=0.7}{i2,i3}
\fmf{boson,left=0.7}{i4,i5}
\fmf{boson}{i1,o1}
\fmfi{plain}{fullcircle rotated 180 scaled 0.385w shifted (0.808w,0.5h)}
\fmfdot{i1,i2,i3,i4,i5,o1}
\end{fmfgraph}
\end{center}}
\hspace*{0.1cm}
\begin{tabular}{@{}c}
${\scs \mbox{\#6.3}}$ \\
$5760$ \\ ${\scs ( 2, 0; 2 )}$
\end{tabular}
\parbox{25mm}{\begin{center}
\begin{fmfgraph}(21,5)
\setval
\fmfforce{0.238w,0.5h}{v1}
\fmfforce{0.381w,0.5h}{v2}
\fmfforce{0.5595w,0.933h}{v4}
\fmfforce{0.4405w,0.933h}{v3}
\fmfforce{0.619w,0.5h}{v5}
\fmfforce{0.762w,0.5h}{v6}
\fmfi{plain}{reverse fullcircle scaled 0.238w shifted (0.119w,0.5h)}
\fmfi{plain}{fullcircle rotated 180 scaled 0.238w shifted (0.881w,0.5h)}
\fmf{boson}{v1,v2}
\fmf{plain,left=0.33}{v2,v3,v4,v5}
\fmf{plain,left=1}{v5,v2}
\fmf{boson,right=0.8}{v3,v4}
\fmf{boson}{v5,v6}
\fmfdot{v1,v2,v3,v4,v5,v6}
\end{fmfgraph}
\end{center}}
\hspace*{0.1cm}
\begin{tabular}{@{}c}
${\scs \mbox{\#6.4}}$ \\
$960$ \\ ${\scs ( 3, 0; 6 )}$
\end{tabular}
\parbox{25mm}{\begin{center}
\begin{fmfgraph}(21,13)
\setval
\fmfforce{0.238w,0.192h}{v1}
\fmfforce{0.381w,0.192h}{v2}
\fmfforce{0.5w,0.384h}{v4}
\fmfforce{0.5w,0.615h}{v3}
\fmfforce{0.619w,0.192h}{v5}
\fmfforce{0.762w,0.192h}{v6}
\fmfi{plain}{reverse fullcircle scaled 0.238w shifted (0.119w,0.192h)}
\fmfi{plain}{fullcircle rotated 180 scaled 0.238w shifted (0.881w,0.192h)}
\fmfi{plain}{fullcircle rotated 270 scaled 0.238w shifted (0.5w,0.808h)}
\fmf{boson}{v1,v2}
\fmf{plain,left=0.5}{v2,v4,v5}
\fmf{plain,left=1}{v5,v2}
\fmf{boson}{v3,v4}
\fmf{boson}{v5,v6}
\fmfdot{v1,v2,v3,v4,v5,v6}
\end{fmfgraph}
\end{center}}
\hspace*{0.1cm}
\begin{tabular}{@{}c}
${\scs \mbox{\#7.1}}$ \\
$11520$ \\ ${\scs ( 0, 0; 4 )}$
\end{tabular}
\parbox{11mm}{\begin{center}
\begin{fmfgraph}(7,7)
\setval
\fmfforce{0w,0.5h}{v1}
\fmfforce{0.25w,0.933h}{v2}
\fmfforce{0.75w,0.933h}{v3}
\fmfforce{1w,0.5h}{v4}
\fmfforce{0.75w,0.067h}{v5}
\fmfforce{0.25w,0.067h}{v6}
\fmf{plain,left=0.3}{v1,v2,v3,v4,v5,v6,v1}
\fmf{boson,right=0.7}{v2,v3}
\fmf{boson}{v1,v4}
\fmf{boson,right=0.7}{v5,v6}
\fmfdot{v1,v2,v3,v4,v5,v6}
\end{fmfgraph}
\end{center}}
\begin{tabular}{@{}c}
${\scs \mbox{\#7.2}}$ \\
$5760$ \\ ${\scs ( 0, 0; 8 )}$
\end{tabular}
\parbox{17mm}{\begin{center}
\begin{fmfgraph}(13,5)
\setval
\fmfforce{0.192w,1h}{i1}
\fmfforce{0.385w,0.5h}{i2}
\fmfforce{0.192w,0h}{i3}
\fmfforce{0.808w,1h}{o1}
\fmfforce{0.808w,0h}{o2}
\fmfforce{0.615w,0.5h}{o3}
\fmf{plain,left=0.5}{i1,i2,i3}
\fmf{plain,left=1}{i3,i1}
\fmf{boson}{i1,i3}
\fmf{boson}{i2,o3}
\fmf{boson}{o1,o2}
\fmf{plain,right=1}{o2,o1}
\fmf{plain,right=0.5}{o1,o3,o2}
\fmfdot{i1,i2,i3,o1,o2,o3}
\end{fmfgraph}
\end{center}} 
\\ &
\quad
\begin{tabular}{@{}c}
${\scs \mbox{\#7.3}}$ \\
$11520$ \\ ${\scs ( 1, 0; 2 )}$
\end{tabular}
\parbox{17mm}{\begin{center}
\begin{fmfgraph}(13,5)
\setval
\fmfforce{0.252w,0.976h}{i2}
\fmfforce{0.037w,0.794h}{i3}
\fmfforce{0.037w,0.206h}{i4}
\fmfforce{0.252w,0.024h}{i5}
\fmfforce{0.385w,0.5h}{i1}
\fmfforce{0.615w,0.5h}{o1}
\fmf{plain,right=0.3}{i1,i2,i3,i4,i5,i1}
\fmf{boson}{i2,i5}
\fmf{boson,left=0.7}{i3,i4}
\fmf{boson}{i1,o1}
\fmfi{plain}{fullcircle rotated 180 scaled 0.385w shifted (0.808w,0.5h)}
\fmfdot{i1,i2,i3,i4,i5,o1}
\end{fmfgraph}
\end{center}}
\quad
\begin{tabular}{@{}c}
${\scs \mbox{\#7.4}}$ \\
$5760$ \\ ${\scs ( 1, 1; 2 )}$
\end{tabular}
\parbox{25mm}{\begin{center}
\begin{fmfgraph}(21,5)
\setval
\fmfforce{0.238w,0.5h}{v1}
\fmfforce{0.381w,0.5h}{v2}
\fmfforce{0.119w,1h}{v4}
\fmfforce{0.119w,0h}{v3}
\fmfforce{0.619w,0.5h}{v5}
\fmfforce{0.762w,0.5h}{v6}
\fmfi{fermion}{reverse fullcircle scaled 0.238w shifted (0.119w,0.5h)}
\fmfi{fermion}{fullcircle rotated 180 scaled 0.238w shifted (0.881w,0.5h)}
\fmf{boson}{v1,v2}
\fmf{fermion,left=1}{v2,v5,v2}
\fmf{boson}{v3,v4}
\fmf{boson}{v5,v6}
\fmfdot{v1,v2,v3,v4,v5,v6}
\end{fmfgraph}
\end{center}}
\quad
\begin{tabular}{@{}c}
${\scs \mbox{\#7.5}}$ \\
$1440$ \\ ${\scs ( 2, 2; 2 )}$
\end{tabular}
\parbox{32mm}{\begin{center}
\begin{fmfgraph}(29,5)
\setval
\fmfforce{0.172w,0.5h}{v1}
\fmfforce{0.276w,0.5h}{v2}
\fmfforce{0.448w,0.5h}{v3}
\fmfforce{0.552w,0.5h}{v4}
\fmfforce{0.724w,0.5h}{v5}
\fmfforce{0.828w,0.5h}{v6}
\fmfi{plain}{reverse fullcircle scaled 0.172w shifted (0.086w,0.5h)}
\fmfi{plain}{fullcircle rotated 180 scaled 0.172w shifted (0.914w,0.5h)}
\fmf{boson}{v1,v2}
\fmf{plain,left=1}{v2,v3,v2}
\fmf{plain,left=1}{v4,v5,v4}
\fmf{boson}{v3,v4}
\fmf{boson}{v5,v6}
\fmfdot{v1,v2,v3,v4,v5,v6}
\end{fmfgraph}
\end{center}}
\quad
\begin{tabular}{@{}c}
${\scs \mbox{\#7.6}}$ \\
$2880$ \\ ${\scs ( 2, 0; 4 )}$
\end{tabular}
\parbox{25mm}{\begin{center}
\begin{fmfgraph}(21,5)
\setval
\fmfforce{0.238w,0.5h}{v1}
\fmfforce{0.381w,0.5h}{v2}
\fmfforce{0.5w,1h}{v3}
\fmfforce{0.5w,0h}{v4}
\fmfforce{0.619w,0.5h}{v5}
\fmfforce{0.762w,0.5h}{v6}
\fmfi{plain}{reverse fullcircle scaled 0.238w shifted (0.119w,0.5h)}
\fmfi{plain}{fullcircle rotated 180 scaled 0.238w shifted (0.881w,0.5h)}
\fmf{boson}{v1,v2}
\fmf{plain,left=0.5}{v2,v3,v5,v4,v2}
\fmf{boson}{v3,v4}
\fmf{boson}{v5,v6}
\fmfdot{v1,v2,v3,v4,v5,v6}
\end{fmfgraph}
\end{center}} 
\end{tabular}
\end{center}
\caption{\la{Y}Connected vacuum diagrams and their 
multiplicities of the $\phi^2 A$-theory
up to four loops. Each diagram is characterized by the
vector $(S,D;N)$ whose components specify the number of self- and double
connections as well as the identical vertex permutations,
respectively.}
\end{table}
\end{fmffile}
\newpage
\begin{table}
\begin{tabular}{@{}c}
$\mbox{}$ \\ $\mbox{}$ 
\begin{tabular}{|c||c|c|c|c|}
\hline
\multicolumn{5}{|c|}{$W^{(1)}$: 1 diagram}\\
\hline
$i$&1&\multicolumn{3}{c|}{}\\
$j$&1&\multicolumn{3}{c|}{}\\
\hline
\#&\multicolumn{1}{c|}{$M_{ij}$}&$\scriptstyle(S,D,T,F;N)$&$M$&$W^{-1}$\\
\hline
1&2&(2,1,0,0;1)&3&8\\
\hline
\end{tabular} 
\\  \\
\begin{tabular}{|c||c|c|c|c|c|}
\hline
\multicolumn{6}{|c|}{$W^{(2)}$: 2 diagrams}\\
\hline
$i$&1&22&\multicolumn{3}{c|}{}\\
$j$&1&12&\multicolumn{3}{c|}{}\\
\hline
\#&\multicolumn{2}{c|}{$M_{ij}$}&$\scriptstyle(S,D,T,F;N)$&$M$&$W^{-1}$\\
\hline
2&0&40&(0,0,0,1;2)&24&48\\
3&1&21&(2,1,0,0;2)&72&16\\
\hline
\end{tabular}
\\  \\
\begin{tabular}{|c||c|c|c|c|c|c|}
\hline
\multicolumn{7}{|c|}{$W^{(3)}$: 4 diagrams}\\
\hline
$i$&1&22&333&\multicolumn{3}{c|}{}\\
$j$&1&12&123&\multicolumn{3}{c|}{}\\
\hline
\#&\multicolumn{3}{c|}{$M_{ij}$}&$\scriptstyle(S,D,T,F;N)$&$M$&$W^{-1}$\\
\hline
5&0&11&310&(1,0,1,0;2)&3456&24\\
4&0&20&220&(0,3,0,0;6)&1728&48\\
7&0&21&201&(2,2,0,0;2)&2592&32\\
6&1&11&111&(3,0,0,0;6)&1728&48\\
\hline
\end{tabular}
\\  \\
\begin{tabular}{|c||c|c|c|c|c|c|c|}
\hline
\multicolumn{8}{|c|}{$W^{(4)}$: 10 diagrams}\\
\hline
$i$&1&22&333&4444&\multicolumn{3}{c|}{}\\
$j$&1&12&123&1234&\multicolumn{3}{c|}{}\\
\hline
\#&\multicolumn{4}{c|}{$M_{ij}$}&$\scriptstyle(S,D,T,F;N)$&$M$&$W^{-1}$\\
\hline
10&0&00&130&3100&(0,0,2,0;4)&55296&144\\
8&0&00&220&2200&(0,4,0,0;8)&62208&128\\
12&0&01&111&3100&(2,0,1,0;2)&165888&48\\
14&0&01&120&3010&(1,1,1,0;2)&165888&48\\
17&0&01&201&2200&(2,3,0,0;2)&124416&64\\
11&0&01&210&2110&(1,2,0,0;2)&497664&16\\
9&0&10&120&2110&(0,2,0,0;8)&248832&32\\
13&0&11&101&2110&(2,1,0,0;4)&248832&32\\
15&0&11&111&2001&(3,1,0,0;2)&248832&32\\
16&1&01&111&1101&(4,0,0,0;8)&62208&128\\
\hline
\end{tabular}
$\mbox{}$\\ $\mbox{}$
\end{tabular}
\caption{\la{DATA1} Unique matrix representation of all connected
vacuum diagrams of $\phi^4$-theory up to the order $p=4$.
The number in the first column corresponds to their graphical
representation in Table \ref{P}. 
The matrix elements $M_{ij}$ represent the
numbers of lines connecting two vertices $i$ and $j$, with omitting
$M_{i0}=0$ for simplicity. The running numbers
of the vertices are listed on top of each column in the first
two rows. The further columns contain the vector $(S,D,T,F;N)$
characterizing the topology of the diagram, the
multiplicity $M$ and the weight $W=M/[(4!)^p p!]$.}
\end{table}
\newpage
\tiny
\begin{table}
\begin{tabular}{@{}cc}
$\mbox{}$ & $\mbox{}$ \\
\begin{tabular}{@{}c}
\begin{tabular}{|c||c|c|c|c|}
\hline
\multicolumn{5}{|c|}{$\fullg_{12}^{(1)}$: 1 diagram}\\
\hline
$i$&11&\multicolumn{3}{c|}{}\\
$j$&01&\multicolumn{3}{c|}{}\\
\hline
\#&\multicolumn{1}{c|}{$M_{ij}$}&$\scriptstyle(S,D,T;N)$&$M$&$W^{-1}$\\
\hline
1.1&21&(1,0,0;2)&12&4\\
\hline
\end{tabular} 
\\ \\
\begin{tabular}{|c||c|c|c|c|c|}
\hline
\multicolumn{6}{|c|}{$\fullg_{12}^{(2)}$: 3 diagrams}\\
\hline
$i$&11&222&\multicolumn{3}{c|}{}\\
$j$&01&012&\multicolumn{3}{c|}{}\\
\hline
\#&\multicolumn{2}{c|}{$M_{ij}$}&$\scriptstyle(S,D,T;N)$&$M$&$W^{-1}$\\
\hline
3.1&01&220&(1,1,0;2)&288&8\\
2.1&10&130&(0,0,1;2)&192&12\\
3.2&11&111&(2,0,0;2)&288&8\\
\hline
\end{tabular} 
\\ \\
\begin{tabular}{|c||c|c|c|c|c|c|}
\hline
\multicolumn{7}{|c|}{$\fullg_{12}^{(3)}$: 8 diagrams}\\
\hline
$i$&11&222&3333&\multicolumn{3}{c|}{}\\
$j$&01&012&0123&\multicolumn{3}{c|}{}\\
\hline
\#&\multicolumn{3}{c|}{$M_{ij}$}&$\scriptstyle(S,D,T;N)$&$M$&$W^{-1}$\\
\hline
7.1&00&021&2200&(1,2,0;2)&10368&16\\
5.1&00&030&2110&(0,0,1;4)&6912&24\\
5.3&00&111&1300&(1,0,1;1)&13824&12\\
4.1&00&120&1210&(0,2,0;2)&20736&8\\
6.1&01&011&2110&(2,0,0;4)&10368&16\\
7.2&01&101&1210&(2,1,0;1)&20736&8\\
5.2&01&110&1120&(1,1,0;2)&20736&8\\
6.2&01&111&1101&(3,0,0;2)&10368&16\\
\hline
\end{tabular} 
\end{tabular} &
\begin{tabular}{|c||c|c|c|c|c|c|c|}
\hline
\multicolumn{8}{|c|}{$\fullg_{12}^{(4)}$: 30 diagrams}\\
\hline
$i$&11&222&3333&44444&\multicolumn{3}{c|}{}\\
$j$&01&012&0123&01234&\multicolumn{3}{c|}{}\\
\hline
\#&\multicolumn{4}{c|}{$M_{ij}$}&$\scriptstyle(S,D,T;N)$&$M$&$W^{-1}$\\
\hline
17.1&00&001&0220&22000&(1,3,0;2)&497664&32\\
12.4&00&001&0310&21100&(1,0,1;2)&663552&24\\
14.2&00&001&1120&13000&(1,1,1;1)&663552&24\\
11.3&00&001&1210&12100&(1,2,0;2)&995328&16\\
14.4&00&010&0130&22000&(0,1,1;4)&331776&48\\
11.1&00&010&0220&21100&(0,2,0;4)&995328&16\\
10.1&00&010&1030&13000&(0,0,2;2)&221184&72\\
9.2&00&010&1120&12100&(0,2,0;2)&1990656&8\\
15.3&00&011&0111&22000&(2,1,0;4)&497664&32\\
15.4&00&011&0201&21100&(2,1,0;2)&995328&16\\
13.2&00&011&0210&21010&(1,1,0;4)&995328&16\\
12.3&00&011&1011&13000&(2,0,1;1)&663552&24\\
13.3&00&011&1101&12100&(2,1,0;1)&1990656&8\\
11.4&00&011&1110&12010&(1,1,0;1)&3981312&4\\
11.2&00&020&1011&12100&(1,2,0;1)&1990656&8\\
8.1&00&020&1020&12010&(0,3,0;2)&995328&16\\
9.1&00&020&1110&11110&(0,1,0;4)&1990656&8\\
17.2&00&021&1001&12010&(2,2,0;1)&995328&16\\
14.1&00&021&1100&11020&(1,2,0;2)&995328&16\\
15.2&00&021&1101&11001&(3,1,0;2)&497664&32\\
14.3&00&030&1001&11110&(1,0,1;2)&663552&24\\
10.2&00&030&1010&11020&(0,1,1;2)&663552&24\\
12.2&00&030&1011&11001&(2,0,1;2)&331776&48\\
16.1&01&001&0111&21100&(3,0,0;4)&497664&32\\
15.1&01&001&1011&12100&(3,1,0;1)&995328&16\\
17.3&01&001&1020&12010&(2,2,0;2)&497664&32\\
13.1&01&001&1110&11110&(2,0,0;4)&995328&16\\
15.5&01&011&1001&11110&(3,0,0;2)&995328&16\\
12.1&01&011&1010&11020&(2,1,0;2)&995328&16\\
16.2&01&011&1011&11001&(4,0,0;2)&497664&32\\
\hline
\end{tabular}
\\ $\mbox{}$ & $\mbox{}$ 
\end{tabular}
\caption{\la{DATA2} Unique matrix representation of all connected
two-point function of $\phi^4$-theory up to the order $p=4$.
The numbers in the first column correspond to their graphical
representation in Table \ref{TWO}. 
The matrix elements $M_{ij}$ represent the
numbers of lines connecting two vertices $i$ and $j$, with omitting
$M_{i0}=0$ for simplicity. The running numbers
of the vertices are listed on top of each column in the first
two rows. The further columns contain the vector $(S,D,T;N)$
characterizing the topology of the diagram, the
multiplicity $M$ and the weight $W=M/[(4!)^p p!]$.}
\end{table}
\newpage
\vspace*{-2cm}
\begin{table}
\begin{tabular}{@{}c}
\begin{tabular}{@{}cc}
$\mbox{}$ & $\mbox{}$ \\
\begin{tabular}{@{}c}
\begin{tabular}{|c||c|c|c|c|}
\hline
\multicolumn{5}{|c|}{$\fullg_{1234}^{{\rm c},(1)}$: 1 diagram}\\
\hline
$i$&11&\multicolumn{3}{c|}{}\\
$j$&01&\multicolumn{3}{c|}{}\\
\hline
\#&\multicolumn{1}{c|}{$M_{ij}$}&$\scriptstyle(S,D,T;N)$&$M$&$W^{-1}$\\
\hline
1.1.1&40&(0,0,0;24)&24&24\\
\hline
\end{tabular}
\\ \\
\begin{tabular}{|c||c|c|c|c|c|}
\hline
\multicolumn{6}{|c|}{$\fullg_{1234}^{{\rm c},(2)}$: 2 diagrams}\\
\hline
$i$&11&222&\multicolumn{3}{c|}{}\\
$j$&01&012&\multicolumn{3}{c|}{}\\
\hline
\#&\multicolumn{2}{c|}{$M_{ij}$}&$\scriptstyle(S,D,T;N)$&$M$&$W^{-1}$\\
\hline
3.1.2, 3.2.1&11&310&(1,0,0;6)&2304&12\\
2.1.1, 3.1.1&20&220&(0,1,0;8)&1728&16\\
\hline
\end{tabular}
\end{tabular} \hspace*{0.5cm}  & \hspace*{0.5cm}
\begin{tabular}{|c||c|c|c|c|c|c|}
\hline
\multicolumn{7}{|c|}{$\fullg_{1234}^{{\rm c},(3)}$: 8 diagrams}\\
\hline
$i$&11&222&3333&\multicolumn{3}{c|}{}\\
$j$&01&012&0123&\multicolumn{3}{c|}{}\\
\hline
\#&\multicolumn{3}{c|}{$M_{ij}$}&$\scriptstyle(S,D,T;N)$&$M$&$W^{-1}$\\
\hline
5.1.2, 5.3.2&00&130&3100&(0,0,1;6)&55296&36\\
4.1.1, 7.1.1&00&220&2200&(0,2,0;8)&62208&32\\
6.1.3, 6.2.1&01&111&3100&(2,0,0;6)&82944&24\\
7.1.3, 7.2.3&01&120&3010&(1,1,0;6)&82944&24\\
5.2.2, 6.1.1&01&210&2110&(1,0,0;8)&124416&16\\
5.2.3, 5.3.1, 7.1.2, 7.2.1&10&111&2200&(1,1,0;2)&248832&8\\
4.1.2, 5.1.1, 5.2.1&10&120&2110&(0,1,0;4)&248832&8\\
6.1.2, 6.2.2, 7.2.2&11&101&2110&(2,0,0;4)&124416&16\\
\hline
\end{tabular} 
\end{tabular} \\ $\mbox{}$ \\ $\mbox{}$
\begin{tabular}{|c||c|c|c|c|c|c|c|}
\hline
\multicolumn{8}{|c|}{$\fullg_{1234}^{{\rm c},(4)}$: 37 diagrams}\\
\hline
$i$&11&222&3333&44444&\multicolumn{3}{c|}{}\\
$j$&01&012&0123&01234&\multicolumn{3}{c|}{}\\
\hline
\#&\multicolumn{4}{c|}{$M_{ij}$}&$\scriptstyle(S,D,T;N)$&$M$&$W^{-1}$\\
\hline
13.2.4, 13.3.5&00&011&1210&31000&(1,1,0;6)&7962624&24\\
12.3.1, 12.4.4&00&011&1300&30100&(1,0,1;6)&2654208&72\\
11.4.5, 15.3.1, 15.4.1&00&011&2110&22000&(1,1,0;4)&11943936&16\\
11.1.3, 11.2.1&00&020&1120&31000&(0,2,0;6)&7962624&24\\
8.1.1, 17.1.1&00&020&2020&22000&(0,3,0;8)&2985984&64\\
9.1.1, 13.2.1&00&020&2110&21100&(0,1,0;16)&5971968&32\\
15.2.3, 15.4.6&00&021&1101&31000&(2,1,0;6)&3981312&48\\
17.1.4, 17.2.4&00&021&1200&30010&(1,2,0;6)&3981312&48\\
14.1.4, 15.4.4&00&021&2100&21010&(1,1,0;8)&5971968&32\\
12.2.1, 12.4.2&00&030&1011&31000&(1,0,1;6)&2654208&72\\
14.3.1, 14.4.1&00&030&1110&30010&(0,0,1;12)&2654208&72\\
10.2.2, 12.4.1&00&030&2010&21010&(0,0,1;8)&3981312&48\\
11.2.4, 11.3.2, 17.1.2, 17.2.1&00&101&1210&22000&(1,2,0;2)&11943936&16\\
12.3.2, 12.4.3, 14.2.2, 14.3.2&00&101&1300&21100&(1,0,1;2)&7962624&24\\
9.2.2, 14.1.1, 14.4.3&00&110&1120&22000&(0,2,0;4)&11943936&16\\
9.1.3, 9.2.3, 11.1.1, 11.4.1&00&110&1210&21100&(0,1,0;2)&47775744&4\\
10.1.1, 10.2.3, 14.2.1, 14.4.2&00&110&1300&20200&(0,1,1;2)&7962624&24\\
13.3.2, 15.2.1, 15.3.3&00&111&1101&22000&(2,1,0;4)&5971968&32\\
11.2.3, 11.4.2, 13.2.2, 13.3.1&00&111&1200&21010&(1,1,0;2)&23887872&8\\
8.1.3, 11.1.2, 11.3.1&00&120&1200&20110&(0,2,0;4)&11943936&16\\
15.1.4, 15.4.5&01&001&1120&31000&(2,1,0;6)&3981312&48\\
13.1.3, 16.1.1&01&001&2110&21100&(2,0,0;16)&2985984&64\\
16.1.4, 16.2.1&01&011&1011&31000&(3,0,0;6)&3981312&48\\
15.3.4, 15.5.4&01&011&1110&30010&(2,0,0;12)&3981312&48\\
12.1.3, 16.1.2&01&011&2010&21010&(2,0,0;8)&5971968&32\\
11.3.3, 11.4.4, 12.1.1, 12.4.5&01&100&1120&21100&(1,1,0;2)&23887872&8\\
14.1.3, 14.2.3, 17.1.3, 17.3.1&01&100&1210&20200&(1,2,0;2)&11943936&16\\
15.1.2, 15.5.3, 16.1.3, 16.2.2&01&101&1101&21100&(3,0,0;2)&11943936&16\\
13.1.2, 13.3.4, 15.4.2, 15.5.1&01&101&1110&21010&(2,0,0;2)&23887872&8\\
15.1.3, 15.4.3, 17.2.3, 17.3.2&01&101&1200&20110&(2,1,0;2)&11943936&16\\
12.1.4, 12.3.3, 15.1.1, 15.3.2&01&110&1101&20200&(2,1,0;2)&11943936&16\\
11.4.6, 13.1.1, 13.2.3&01&110&1110&20110&(1,0,0;4)&23887872&8\\
8.1.2, 9.2.1, 10.2.1&10&100&1120&12100&(0,2,0;4)&11943936&16\\
12.1.2, 12.2.2, 13.3.3, 17.2.2&10&101&1101&12100&(2,1,0;2)&11943936&16\\
11.2.2, 11.4.3, 14.1.2, 14.3.3&10&101&1110&12010&(1,1,0;2)&23887872&8\\
9.1.2&10&110&1110&11110&(0,0,0;24)&7962624&24\\
15.2.2, 15.5.2&10&111&1101&11001&(3,0,0;6)&3981312&48\\
\hline
\end{tabular}
\\ $\mbox{}$
\end{tabular}
\caption{\la{DATA3} Unique matrix representation of all connected
two-point function of $\phi^4$-theory up to the order $p=4$.
The numbers in the first column correspond to their graphical
representation in Table \ref{FOUR}. 
The matrix elements $M_{ij}$ represent the
numbers of lines connecting two vertices $i$ and $j$, with omitting
$M_{i0}=0$ for simplicity. The running numbers
of the vertices are listed on top of each column in the first
two rows. The further columns contain the vector $(S,D,T;N)$
characterizing the topology of the diagram, the
multiplicity $M$ and the weight $W=M/[(4!)^p p!]$.}
\end{table}

\end{document}